\documentclass[reprint,amsmath,amssymb,aps,prd,superscriptaddress,nofootinbib,longbibliography]{revtex4-2}

\usepackage{silence}  
\usepackage[dvipsnames]{xcolor}
\usepackage{graphicx}
\usepackage{tikz}
\usepackage{bm}
\usepackage[
    separate-uncertainty=true,
    round-mode=uncertainty,
    round-precision=1,
]{siunitx}
\usepackage{booktabs}
\usepackage{float}  
\usepackage{multirow}
\usepackage{dcolumn}
\usepackage{xspace}
\usepackage{orcidlink}
\usepackage[normalem]{ulem}
\usepackage[export]{adjustbox}
\usepackage[colorinlistoftodos,prependcaption,textsize=tiny]{todonotes}
\usepackage{hyperref}

\definecolor{linkcolor}{HTML}{0b5394}

\graphicspath{{figures/}}

\newcommand{\cc}{\textsc{CaloHadronic}\xspace}
\newcommand{\as}{\textsc{AllShowers}\xspace}
\newcommand{\ct}{\textsc{CaloTrilogy}\xspace}
\newcommand{\ccc}{\textsc{CaloClouds3}\xspace}

\newcommand{\layerFM}{\textsc{LayerFM}\xspace}
\newcommand{\latentMF}{\textsc{Latent}\xspace \textsc{Mean}\xspace \textsc{Flow}\xspace \textsc{(LMF)}\xspace}
\newcommand{\LMF}{\textsc{LMF}\xspace}
\newcommand{\calochillname}{CaloCHILL}  
\newcommand{\calochillfull}{Calorimeter Hits In Learned Latents}  
\newcommand{\calochill}{\textsc{\calochillname}\xspace}  
\newcommand{\calochilldef}{\textsc{\calochillname} (\calochillfull)\xspace}  
\newcommand{\VAEdec}{\textsc{VAE}\xspace \textsc{decompressor}\xspace}
\newcommand{\VAEenc}{\textsc{VAE}\xspace \textsc{compressor}\xspace}

\usepackage{bold-extra}
\DeclareFontShape{OT1}{cmr}{m}{scit}{<->ssub*cmr/m/sc}{}

\hypersetup{
    colorlinks=true,
    linkcolor=linkcolor,
    citecolor=linkcolor,
    urlcolor=linkcolor
}

\newenvironment{widefigure}[1][tp]{\csname figure*\endcsname[tp]}{\csname endfigure*\endcsname}
\newenvironment{widetable}[1][tp]{\csname table*\endcsname[tp]}{\csname endtable*\endcsname}
\newcommand{\granularityfigwidth}{\linewidth}  
\newcommand{\ldabigfigwidth}{\linewidth}  
\newcommand{\ldafigwidth}{\linewidth}  
\newcommand{\linearityfigwidth}{\linewidth}  
\newcommand{\timefirsthitfigwidth}{0.8\linewidth}  
\newcommand{\calibfigwidth}{\linewidth}  
\newcommand{\modelspanelwidth}{0.27\linewidth}  
\newcommand{\modelslegendwidth}{0.85\linewidth}  

\newcommand{\timingfigwidth}{0.95\linewidth}  
\newcommand{\attnfigwidth}{\linewidth}  
\newcommand{\eventsthreedwidth}{\linewidth}  
\newcommand{\timingpanelwidth}{\linewidth}  
\usepackage{placeins}  
\newcommand{\prdlayout}{}  
\makeatletter\def\@fltstk{}\makeatother  

\begin{document}

\title{\boldmath Just \calochillname: fast hadronic shower generation in a learned latent space}

\author{Henry~Day-Hall~\orcidlink{0000-0002-9710-2980}}
\affiliation{Deutsches Elektronen-Synchrotron DESY, Notkestr.~85, 22607 Hamburg, Germany}
\author{Frank~Gaede~\orcidlink{0000-0002-7055-9200}}
\affiliation{Deutsches Elektronen-Synchrotron DESY, Notkestr.~85, 22607 Hamburg, Germany}
\author{Gregor~Kasieczka~\orcidlink{0000-0003-3457-2755}}
\affiliation{Institute for Experimental Physics, Universit\"at Hamburg, Luruper Chaussee 149, 22761 Hamburg, Germany}
\author{Katja~Kr\"uger~\orcidlink{0000-0002-1956-6608}}
\affiliation{Deutsches Elektronen-Synchrotron DESY, Notkestr.~85, 22607 Hamburg, Germany}
\author{Martina~Mozzanica~\orcidlink{0009-0002-1111-6247}}
\email{martina.mozzanica@uni-hamburg.de}
\affiliation{Institute for Experimental Physics, Universit\"at Hamburg, Luruper Chaussee 149, 22761 Hamburg, Germany}

\begin{abstract}
Calorimeter showers are the most expensive part of detector simulation, and highly granular calorimeters, in which a single hadronic shower leaves thousands of energy deposits, make them more expensive still. We present \calochilldef, a generative model that does not produce these deposits one by one: instead it generates a compact summary of the shower in a learned latent space and expands it to full resolution only at the end. The complete shower, including the arrival time of every hit, is built in three stages: a flow-matching model first sets the number of hits and the earliest hit time in each calorimeter layer, a mean-flow model then generates the latent summary, and the decompressor of a variational autoencoder turns it into the full point cloud, giving every hit its position, energy and time. Sparse attention along the shower axis keeps this compression affordable for showers with thousands of hits. On positive-pion showers in the calorimeter of the International Large Detector, a proposed detector for the FCC-ee, \calochill reproduces the main shower observables, including their time structure. This new model is \(\times 600\) faster than its predecessor, while improving on all accuracy metrics considered.

\end{abstract}

\maketitle

\section{Introduction}\label{sec:intro}

High Energy Physics (HEP) aims to advance our understanding of fundamental forces and fields by observing the outcomes of particle interactions at high energies in colliders.
Analysis in HEP compares observed events against simulated counterparts, and the gold standard for simulating detector interactions is the MC program \textsc{Geant4}~\cite{GEANT4:2002zbu,Allison:2006ve,Allison:2016lfl}.
Among all the steps involved, interactions with the calorimeter in the detector consume the most computational time~\cite{HEPSoftwareFoundation:2017ggl}.
Thus, some form of fast surrogate modeling is essential to generate sufficient simulations to prevent statistical uncertainties from dominating the analysis. Detectors planned for the future, such as those proposed for the FCC-ee, will feature significantly higher granularity~\cite{Thomson:2009rp,ILDConceptGroup:2020sfq}, presenting an even greater challenge for fast and accurate generative simulation. 

Substantial efforts have focused on developing solutions to meet this demand~\cite{Paganini:2017hrr,Paganini:2017dwg,Buhmann:2020pmy,Krause:2021ilc,Mikuni:2022xry,Amram:2023onf,Buhmann:2023bwk,Buhmann:2023kdg,ATLAS:2021pzo}, with flow matching~\cite{Favaro:2024rle} and diffusion models~\cite{Amram:2023onf} performing particularly well.
Methodologies explored just this year include tokenizer-based models for calorimeter simulation~\cite{Granger:2026vfd,Birk:2026udp,Cardona-Giraldo:2026rnn}, flow-based~\cite{Buss:2026yrf} and diffusion-based approaches~\cite{Jiang:2026zov}, transfer-learning techniques between detectors~\cite{Buss:2026tlx,Favaro:2026awn}, and even FPGA-specific optimizations~\cite{May:2026cnw}.
Comparative studies of earlier models appear in the second CaloChallenge paper~\cite{Krause:2024avx} and the first full-physics benchmark~\cite{Buss:2025bec}. Readers seeking a broader perspective on ongoing developments are directed to the simulation section of the HEP–ML Living Guide~\cite{HEPMLGuide:simulation} and to the review in ref.~\cite{Hashemi:2023rgo}.

To date, fast simulations have improved performance by replacing \textsc{Geant4} with surrogates that still generate every shower point individually~\cite{Buhmann:2023bwk,Buhmann:2023kdg,Buss:2025kiu,Buss:2025cyw,Buss:2026yrf}.
Point-cloud approaches normally encounter a trade-off between the desire to predict points at a higher granularity to reduce position related artifacts and the corresponding increase in inference time from predicting more points.
Generation in a learned latent space mitigates this trade-off, because the generative model acts on a compressed representation and only its expansion back into the full point cloud scales with the shower length.
Such approaches have also been explored.
\textsc{CaloLatent}~\cite{Madula:2023calolatent} and \textsc{CaloDREAM}~\cite{Favaro:2024rle} generate voxelised showers in the latent space of an autoencoder, following latent diffusion in image generation~\cite{rombach2022highresolutionimagesynthesislatent}, and the \textsc{CaloPointFlow} models~\cite{Schnake:2026iec,Schnake:2024mip} encode a point-cloud shower into a single global latent vector, from which the individual points are then sampled.
In this work the point cloud is instead compressed into an ordered sequence of $d_s$ latent vectors, each summarising a section of the shower along its axis, and the generator acts on this sequence rather than on the points.
A compressed representation reduces the amount of data to be generated rather than only accelerating its production, but the autoencoder must then compress a cloud of up to thousands of points into $d_s$ latent vectors and still reconstruct it faithfully.
Every observable therefore depends on what the latent keeps, not only on the generator.

Future hadronic calorimeters, such as the one designed for
International Large Detector (ILD)~\cite{ILDConceptGroup:2020sfq} 
may also offer high resolution timing information.
This is a valuable feature for mitigating pileup at higher luminosities~\cite{Butler:2019rpu,CMS:2017jpq}, and so far has only been included in a limited number of fast generative simulations; only two are known to us at the time of writing, both are for photon showers~\cite{CMS-DP-2025-016,Granger:2026vfd}.
This is the first work to includes the time predictions for hadronic showers for highly granular calorimeters and also timing information for both ECal and HCal.
Timing information on the hits is a further dimension, and since our approach compresses the latent space it is essential to ensure that enough information is retained to correctly reproduce this information. 
In this work, we present \calochilldef and demonstrate the feasibility of a compressed latent approach by generating complex hadronic showers in a high-granularity calorimeter, complete with detailed features and timing information.
\section{Dataset}\label{sec:dataset}

The showers used in this work are the \textsc{Geant4} sample of \cc~\cite{Buss:2025cyw}.
This sample also contains the arrival time of the energy deposits, which was not used in the previous work: the Geant4 global time of each energy-deposition step, measured from the creation of the primary particle at the particle gun, without simulation of the detector time response.
Single $\pi^{+}$ with energies drawn uniformly between
\SI{10}{\giga\electronvolt} and \SI{90}{\giga\electronvolt} enter the
calorimeter system of the ILD, and their detector response is simulated with \textsc{Geant4}~\cite{GEANT4:2002zbu}
version 11.2.2 using the \texttt{QGSP\_BERT} physics list.
The chosen ILD calorimeter system comprises a silicon--tungsten electromagnetic section (ECal) of 30 layers with
$5\times5$\,mm$^{2}$ cells and a scintillator--steel hadronic section (HCal) of
48 layers with $3\times3$\,cm$^{2}$ cells, 78 sampling layers in total~\cite{ILDConceptGroup:2020sfq}.

\begin{figure}[!ht]  
\centering
\includegraphics[width=\granularityfigwidth]{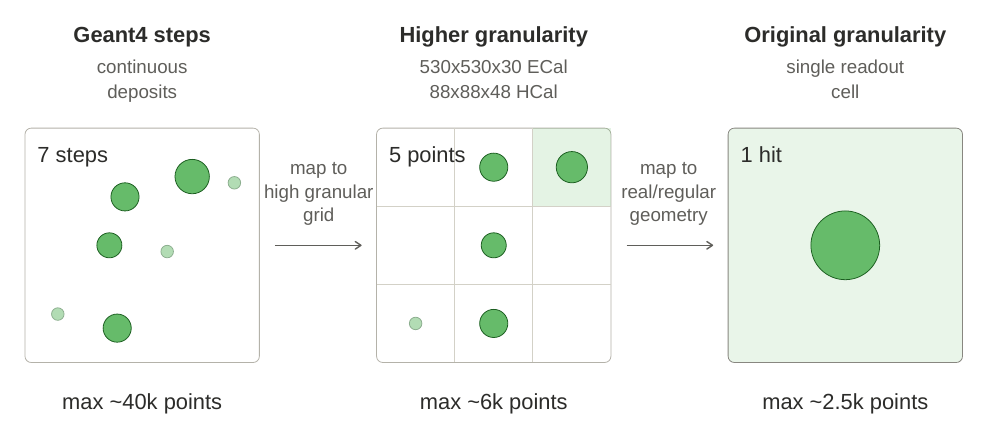}
\caption{Reduction of a \textsc{Geant4} shower to a point cloud, illustrated for
a single readout cell. Left: the energy deposits produced by \textsc{Geant4},
a quasi-continuous set of steps of which only a few are drawn. Centre: the
virtual grid of \cc~\cite{Buss:2025cyw}, about nine times finer than the
physical readout; this is the representation used here. Right: the physical readout granularity, recovered at
the analysis stage, where all deposits inside a cell are merged into a single
hit.}
\label{fig:granularity}
\end{figure}

A shower is represented as a point cloud rather than as a fixed voxel grid.
\textsc{Geant4} records the energy deposition as a quasi-continuous set of
steps, far more numerous than the channels the detector actually reads out;
figure~\ref{fig:granularity} shows how this is reduced. Deposits are retained
within $|x|<\SI{450}{\milli\meter}$ and $|z|<\SI{450}{\milli\meter}$ of the
shower axis. Following \cc~\cite{Buss:2025cyw}, the steps
inside that region are binned on a virtual grid about nine times finer than the
physical readout, with a transverse pitch of \SI{1.70}{\milli\meter} in the ECal
and \SI{10.18}{\milli\meter} in the HCal; the region and the pitch together give
$530\times530$ bins in each ECal layer and $88\times88$ in each HCal layer. The
physical cell granularity is recovered only at the analysis stage.

Each populated bin becomes one point, carrying the summed energy of its steps,
the bin-centre $x$ and $z$ coordinates, and the index of its layer as $y$, so
that the longitudinal coordinate is discrete by construction.

With the addition of time, the highly granular calorimeter captures five-dimensional information.
\cc~\cite{Buss:2025cyw} models the first four of those coordinates; in this work the
arrival time is generated as well. The energy deposition of a hadronic shower is not
instantaneous: a prompt electromagnetic core is followed by a slow component from
neutron scattering and capture and from nuclear de-excitation, extending over
hundreds of nanoseconds and measured in highly granular CALICE
prototypes~\cite{Simon:2011rm}.
Resolving it separates the neutron-rich, poorly
compensated part of the shower from the prompt one, which bears on software
compensation and hence on the hadronic energy resolution.
The reconstructed time also disentangles overlapping showers, rejects deposits from other bunch crossings.
Along with the momentum measured in the tracker,  arrival time allows calculation of a time-of-flight
that separates pions from kaons, which is important because the specific energy loss alone is ambiguous. 

Each cell is assigned the time of the earliest \textsc{Geant4} step falling into it,
so that the feature records when the shower front first reaches that cell
rather than when its energy is deposited on average, mirroring the readout.
Taking the time of the first energy deposit rather than average time in the cell is also better for separating the prompt electromagnetic front from the slow neutron tail.
Deposits arriving later
than \SI{200}{\nano\second} are discarded, as they fall outside the
bunch-crossing window of the ILC and would not be integrated by the calorimeter
readout.

Every shower is stored as a fixed-length array of 6000 points with five
features each, $(x, y, z, E, t)$.

\section{Model}\label{sec:model}

Generation in \calochill is factorised into three stages, shown in figure~\ref{fig:model}. 

\begin{widefigure}[!ht]
    \centering
    \includegraphics[width=0.95\linewidth]{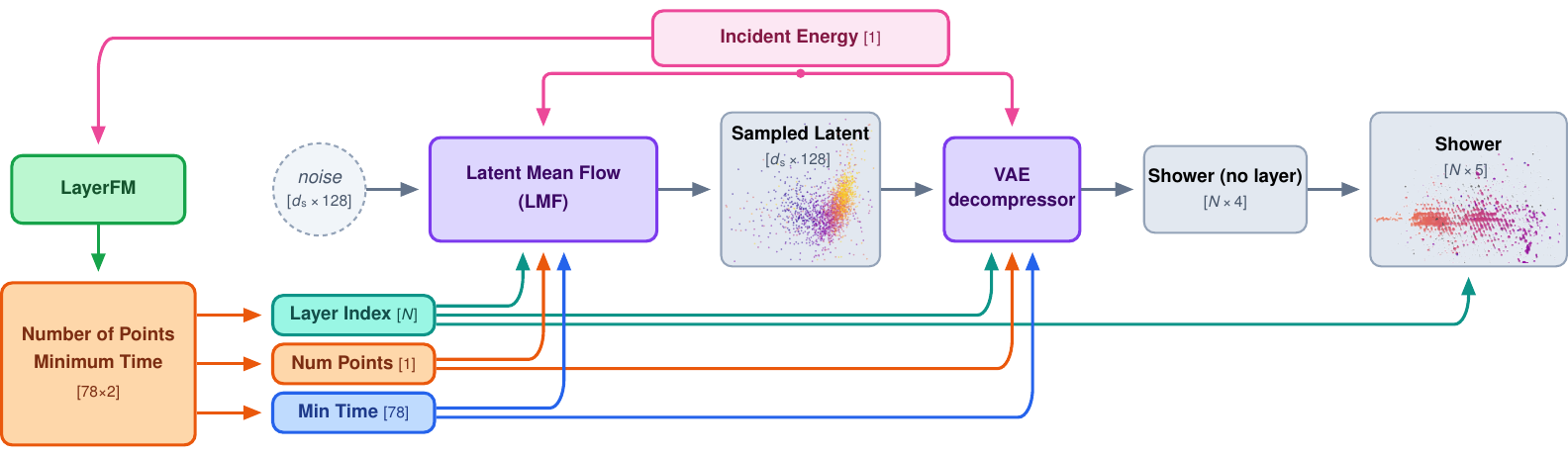}
    \caption{Generation pipeline of \calochill. Conditioned on the energy of the incident
    particle, a flow-matching model (\layerFM) first produces the per-layer
    observables: the number of points and the earliest hit time in each of the 78
    layers. Together with the incident energy, these condition both stages that
    follow. Starting from noise, the mean-flow model generates a latent code, which
    the \VAEdec expands into the full size point cloud carrying
    position, energy and time; the layer index of each point is attached last, 
    completing the five features of every point.
    }
    \label{fig:model}
\end{widefigure}

The first stage, \layerFM, generates quantities per layer, in this case the number of 
points per layer and the minimum time of point arrival per layer; the second, \latentMF generates a 
compressed description of the shower and the third, \VAEdec, expands that description back into a full 
resolution point cloud. 
The energy of the incident particle is used as a conditioning variable for all three 
stages, and the per-layer observables produced by the first stage are passed on to the second and
third stages as well.
They enter in two different forms: the minimum time per layer as a single
vector of 78 components, one per layer, and the number of points per layer through the layer
index carried by each point, together with the total number of points.
The resulting point cloud carries five features per point: the $x$ and $z$ coordinates, the layer index $y$, 
the energy $E$ and the time $t$. 

To use this model in a realistic analysis process conditioning on incident particle angle would also be needed.
At present the model does not support this; it retains the vertically directed particle gun of its predecessor, \cc{}'s, training set.
Additionally, it is possible to imagine training this model on a range of particle species, and conditioning on the required particle species, as was done with great success by the \as model.
We leave both these extensions for future work.

\subsection{\layerFM}\label{sec:stage_layerfm}
Predicting the number of points on each layer is an easier task than predicting the whole shower structure,
however, an accurate prediction of the macro features such as points per layer is very important for good population statistics.
Our architecture separates the generation of macro and micro information, with \layerFM generating macro features.
It is a flow-matching model over quantities defined layer by layer: for each of the 78 sampling layers, \layerFM generates the multiplicity of points and the earliest hit time.
Summing the multiplicity of points per layer gives the multiplicity of the shower
and expanding it gives the layer index of every point, so the longitudinal
outline is determined before the generation of individual points begins, as was first done in \as~\cite{Buss:2026yrf}.
These quantities then condition both of the stages that follow, together with the incident energy.
\layerFM is a development on the equivalent component from previous fast simulations: that of \cc~\cite{Buss:2025cyw} generated the point counts
alone, \as~\cite{Buss:2026yrf} widened its \emph{conditioning} to the particle
type and the direction of incidence, whereas in this case the  
\emph{output} is extended to also predict the hit times alongside the counts, with the energy per layer supported in the same way, as a generator such as that of
\ccc~\cite{Buss:2025kiu} would require. Appendix~\ref{sec:applayerfm} gives the
architecture, the preprocessing and the distillation that reduces sampling to a
single network evaluation.

\subsection{\textsc{VAE}: \textsc{compressor} and \textsc{decompressor}}\label{sec:stage_vae}
Generating $\mathcal{O}(10^3)$ points directly is expensive, and the points are strongly correlated,
so a variational autoencoder can be trained to compress the cloud into a much shorter sequence.
Compressing it efficiently is the central idea  of this work: additional speed comes from shrinking the length of the point cloud from which we draw samples.
The \VAEenc maps the full cloud onto $d_s$ latent vectors, and the \VAEdec expands
them back into one point per hit, conditioned on the incident energy and the
per-layer observables.
Only the latent sequence is produced by the generative stage, and scaled back to a shower by the \VAEdec.
In both the \VAEenc and the \VAEdec every point carries an embedding of its layer index, as in \as~\cite{Buss:2026yrf}.
The \VAEenc is used only in training --- at generation time there is no shower to compress, and the chain
starts from noise --- which is why figure~\ref{fig:model} shows the \VAEdec alone.
The \textsc{VAE} is trained on the usual sum of a reconstruction term and a
Kullback--Leibler term towards a standard normal prior. The reconstruction term
cannot be the usual point-wise one, because the output is an unordered set: there
is no one-to-one correspondence between the points of the decompressed cloud and
those of the input, so the loss has to compare the two sets as distributions.
The Chamfer and earth-mover distances, usual for point clouds, are
respectively too blunt and too expensive at these multiplicities, so the
comparison is made instead with a sliced Wasserstein distance, layer by layer.
Appendix~\ref{sec:appvae} gives the architecture and the reasoning.

\subsection{\latentMF}\label{sec:stage_lmf}
The generative stage proper, which produces that latent sequence, starts from Gaussian noise and is conditioned on the incident energy and the per-layer observables.
Its target is the posterior mean $\mathbf{z}_{\mu}$, rather than a draw from the posterior.
This is because a draw adds noise that is redrawn at every compression and so cannot be learned; appendix~\ref{sec:applda} shows that it roughly halves the energy information the latent carries.
It is a mean-flow model~\cite{Geng:2025meanflow}: it learns an \emph{average} velocity rather than
an instantaneous one, so only a single network evaluation is required to transform noise to data.
The objective was first brought to calorimeter simulation in
\ct~\cite{Jiang:2026zov}, which generates the shower itself in a single step; here it acts on the latent instead.
Appendix~\ref{sec:applmf} gives the parametrisation and the training details.

\subsection{Summarising and rebuilding a shower with local attention}\label{sec:attn}

\begin{figure}[!ht]
    \centering
    \includegraphics[width=\attnfigwidth]{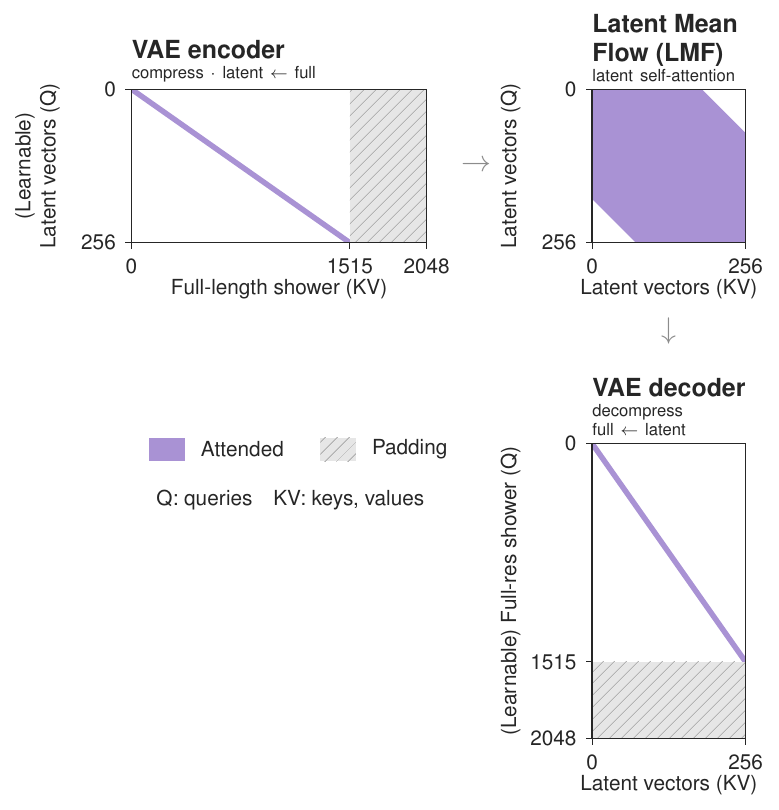}
    \caption{Sparse attention masks: the \VAEenc compressing a cloud onto the
    latent (left), the latent self-attention (centre), and the \VAEdec expanding
    it back (right). Rows are queries, columns keys and values, a filled entry is
    one that is evaluated, and the lighter regions are padding. The example is a
    toy one, 1515 points onto $d_s = 512$, sized for legibility rather than
    matching a configuration used in production.}
    \label{fig:attn}
\end{figure}

Attention over a full shower would significantly inflate the inference time:
a dense attention matrix grows as the square of the multiplicity, and at the granularity we require, showers can have \(\mathcal{O}(1000)\) points.
Most of that computation is wasted, because a hadronic
shower develops along its axis: points are strongly correlated with those a few
layers away on either side of them, and with very little beyond.
In \as~\cite{Buss:2026yrf} this is used to create a 
sparse mask with to flex attention~\cite{Dong:2024flexattention}.
The points are sorted by layer index, so that adjacency in the
sequence means adjacency in depth, and a point then interacts only with those at
most two layers away.
For \as, the shower is the only sequence, and since it maps as sequence to one of the same length, the overall attention matrix is square.
Here the compression introduces a second sequence length: the \VAEenc maps the $N$ points of a shower onto $d_s$
latent vectors and the \VAEdec maps them back, thus we have cross-attention between sequences of different lengths.
Figure~\ref{fig:attn} illustrates the three masks
that result, for one shower and one latent size chosen to make their structure
legible rather than to match any configuration used in production.
The two outer ones are rectangular: latent
vector $q$ is centred on point $qN/d_s$ of the cloud.
Only the latent self-attention, in the centre, is square.

Every mask follows the same rule: two elements attend to each other if and only if they lie within $w$ places of each other in the layer-ordered sequence.
The rule is written in positions rather than in layer indices because the latent vectors have
no layer index: each one summarises a stretch of consecutive points, which may cover part of a layer or several layers.

\paragraph{\textsc{VAE} masks.}
The \VAEenc summarises the cloud by cross-attention from the $d_s$ latent vectors
onto the $N$ points (figure~\ref{fig:attn}, left), and the \VAEdec rebuilds it by
the reverse cross-attention, from the points onto the latent vectors (right).
The default dimension of the latent vector is \(128\).
In both the \VAEenc and the \VAEdec latent vector $q$ is placed at point $qN/d_s$ and linked to the points within
$w = 8\sqrt{N/d_s}$ places of it.
Consecutive latent vectors are $N/d_s$ points apart, so each one reads a window wider than its own stretch, and neighbouring
latent vectors will read some of the same input points.
The \VAEdec then refines the rebuilt cloud with self-attention at full resolution, where $w$ is the mean occupancy of the two most populated layers.
Because $w$ counts points rather than layers, its reach in depth
varies along the shower: about two layers on either side around the shower maximum,
as in \as, and about five on average in the sparser layers away from it.

Only the parts of the attention matrix that are required by this scheme are computed: about a
tenth of it at full resolution, almost independently of energy, and a few percent in the cross-attention.
FlexAttention works in tiles of $128 \times 128$ entries and
needs to know, for every tile, whether the mask leaves it empty, fills it
completely, or cuts through it; only in the last case is the mask applied entry by
entry. Evaluating the mask on every entry to find this took about \SI{40}{\percent} of the
decompression time at \SI{90}{\giga\electronvolt}, so the tiles are instead
classified directly from the band edges, at a cost proportional to their number.

\paragraph{\LMF mask.}
The \LMF works on the latent sequence alone, with self-attention between the $d_s$
latent vectors (figure~\ref{fig:attn}, centre); the latent layers of the
\textsc{VAE} use the same mask. Here $w$ takes the full-resolution value but is
counted in latent vectors. Since one vector stands for $N/d_s$ points, it reaches
much further into the shower: at $d_s = 128$ the band of required attention computation covers about a third of the
latent sequence at \SI{10}{\giga\electronvolt} and nearly all of it at
\SI{90}{\giga\electronvolt}, so this is the least local of the masks. A tighter band
is left for future work; since $d_s \ll N$, the gain in speed would be modest.

\subsection{Shower assembly and calibration}\label{sec:assembly}
Once the \LMF has produced a latent, the \VAEdec expands it back to full
resolution --- one point for each of the hits \layerFM asked for --- and returns
the four features it generates: $x$, $z$, energy and time. The fifth, the layer
index, \layerFM has already assigned to every point, so the cloud is complete at
$(x, y, z, E, t)$.
Each of the three stages is therefore a single network evaluation --- \layerFM
distilled to a single step, the \LMF a single step by construction, and one pass of
the \VAEdec.
Two physical bounds are then imposed on the times: no hit may arrive before the earliest one
\textsc{Geant4} records in its layer, and none may fall outside the \SI{200}{\nano\second} readout window.
The total energy is calibrated by scaling the incident energy given as
condition at inference by a factor $f(E)$ that depends on the incident energy,
fitted per model to \textsc{Geant4} (appendix~\ref{sec:appcalib}); it refines the
single factor of \as~\cite{Buss:2026yrf} and changes the other distributions only
slightly.
Each shower is then projected onto the regular grid of readout cells
(figure~\ref{fig:granularity}), with the same post-processing as
\cc~\cite{Buss:2025cyw}: the grid is shifted by a random offset, drawn uniformly
within one cell for every shower, the energies falling in a cell are summed, its time
is the earliest among them, and cells below \SI{10}{\kilo\electronvolt} are
discarded.

The generated times also receive one correction after the projection.
A small fraction
of the generated points arrive earlier than a signal travelling at the speed of light from the
particle gun could reach them.
No \textsc{Geant4} point does so, and we call such
points \textit{unphysical hits}. The time of
every generated cell is therefore raised to a floor: the earliest time
\textsc{Geant4} records in that cell over the whole reference sample or, in cells
without a \textsc{Geant4} hit, the causal limit $d/c_\mathrm{eff}$, with $d$ the
distance from the gun to the cell and
\begin{equation}\label{ceff_eq}
    c_\mathrm{eff} = c\left(1 + \frac{\Delta}{2\,c\,t_\mathrm{min}}\right),
\end{equation}
where $\Delta$ is the cell diagonal and $t_\mathrm{min} \approx \SI{22}{\pico\second}$
the earliest arrival time in the sample. The limiting velocity $c_\mathrm{eff}$ exceeds
$c$ because $d$ is measured to the cell centre, while the earliest deposit can be up
to $\Delta/2$ closer to the gun; it is \SI{460}{\milli\meter\per\nano\second} for an
ECal cell and \SI{1260}{\milli\meter\per\nano\second} for an HCal one. This per-cell floor acts on
the cells after the projection, on top of the per-layer floor applied to the points
before it, and \textsc{Geant4} cells are left untouched. 

\section{Results}\label{sec:results}

Five \textsc{VAE}s are trained, with $d_s = 64$, 128, 256, 512 and 1024, so that the
length of the latent sequence spans more than an order of magnitude. On each of them a
standard \LMF is trained, and on those with $d_s = 128$, 256 and 512 also a large \LMF,
with twice the model dimension and two layers instead of one (table~\ref{tab:lmf}).

\begin{figure}[htbp]  
    \centering
    \includegraphics[width=\ifdefined\prdlayout\linewidth\else0.85\linewidth\fi]{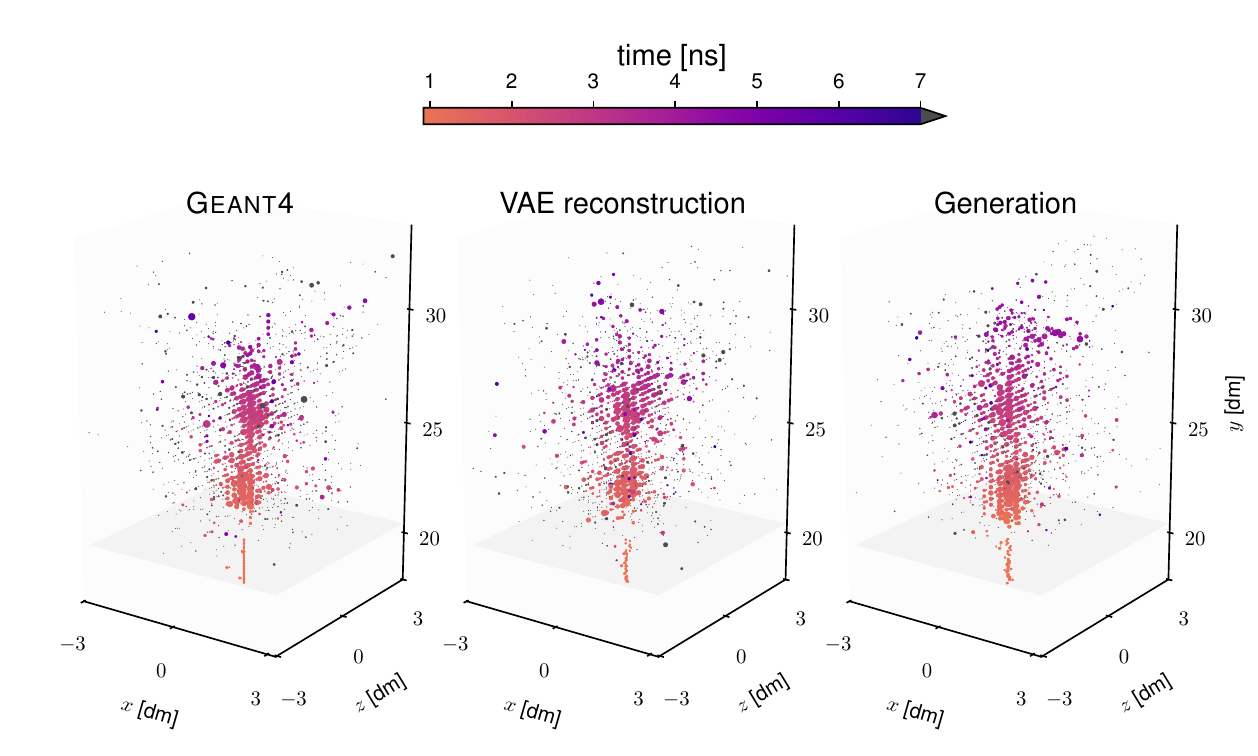}
    \caption{A $\pi^{+}$ shower at $E = \SI{50.2}{\giga\electronvolt}$: \textsc{Geant4}
    (left), its reconstruction by the \textsc{VAE}, encoded and decoded once (centre), and
    a shower generated by the large \LMF at $d_s = 256$ (right) with the incident energy
    and the per-layer numbers of points and earliest times of the \textsc{Geant4} shower
    as conditioning. The axes $x$ and $z$ are transverse and the depth $y$ vertical; the size of a cell
    grows with the square root of its energy and its colour gives the time of its earliest hit, from the start of the shower (red-orange)
    to \SI{7}{\nano\second} (dark indigo); later cells are grey. The
    grey plane is the boundary between the ECal and the HCal.}
    \label{fig:event_display}
\end{figure}

Figure~\ref{fig:event_display} shows one \textsc{Geant4} shower at
\SI{50}{\giga\electronvolt}, its reconstruction by the \textsc{VAE}, and a shower
generated by the large \LMF at $d_s = 256$. All the conditioning, i.e. incident energy, layer index, number of points and earliest time per layer, are taken directly from Geant4, bypassing \layerFM, to allow a consistent comparison. In all three, the pion leaves a narrow track through the ECal, and in the HCal the
energy is concentrated in two clusters, one right behind the ECal and one about
\SI{40}{\centi\metre} deeper, with a halo of late, low-energy deposits around them.
A visible difference is at the back of the HCal, where a narrow secondary track is more diffuse in the generated shower compared to the \textsc{Geant4} one. Five more showers are shown in
appendix~\ref{sec:appevents}.

All comparisons in this section are made at the physical readout granularity, after
the projection of section~\ref{sec:assembly}, for generated and simulated showers
alike.

\begin{widefigure}[!ht]
    \centering
    \includegraphics[width=0.95\linewidth]{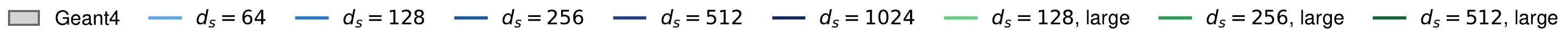}\\[2pt]
    \includegraphics[width=0.245\linewidth]{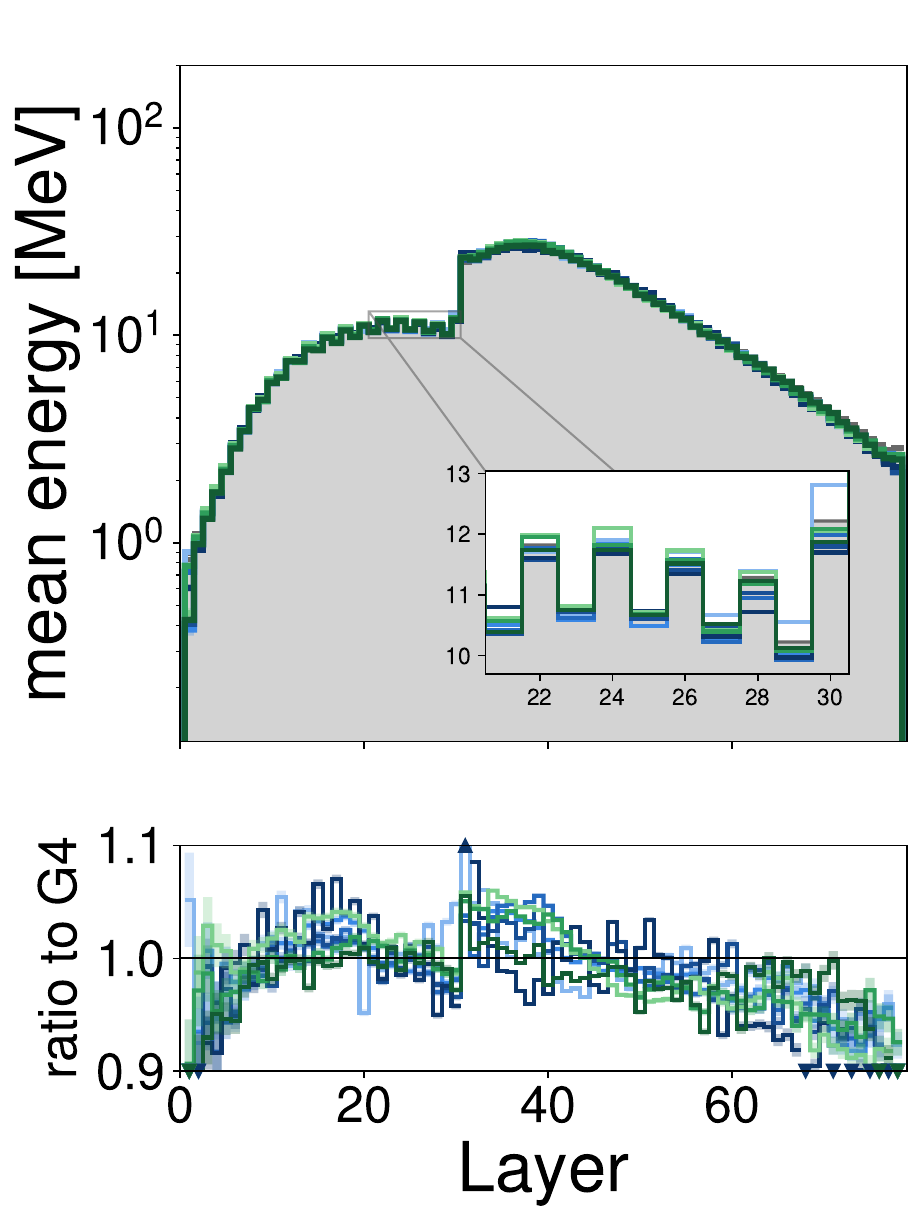}\hfill
    \includegraphics[width=0.245\linewidth]{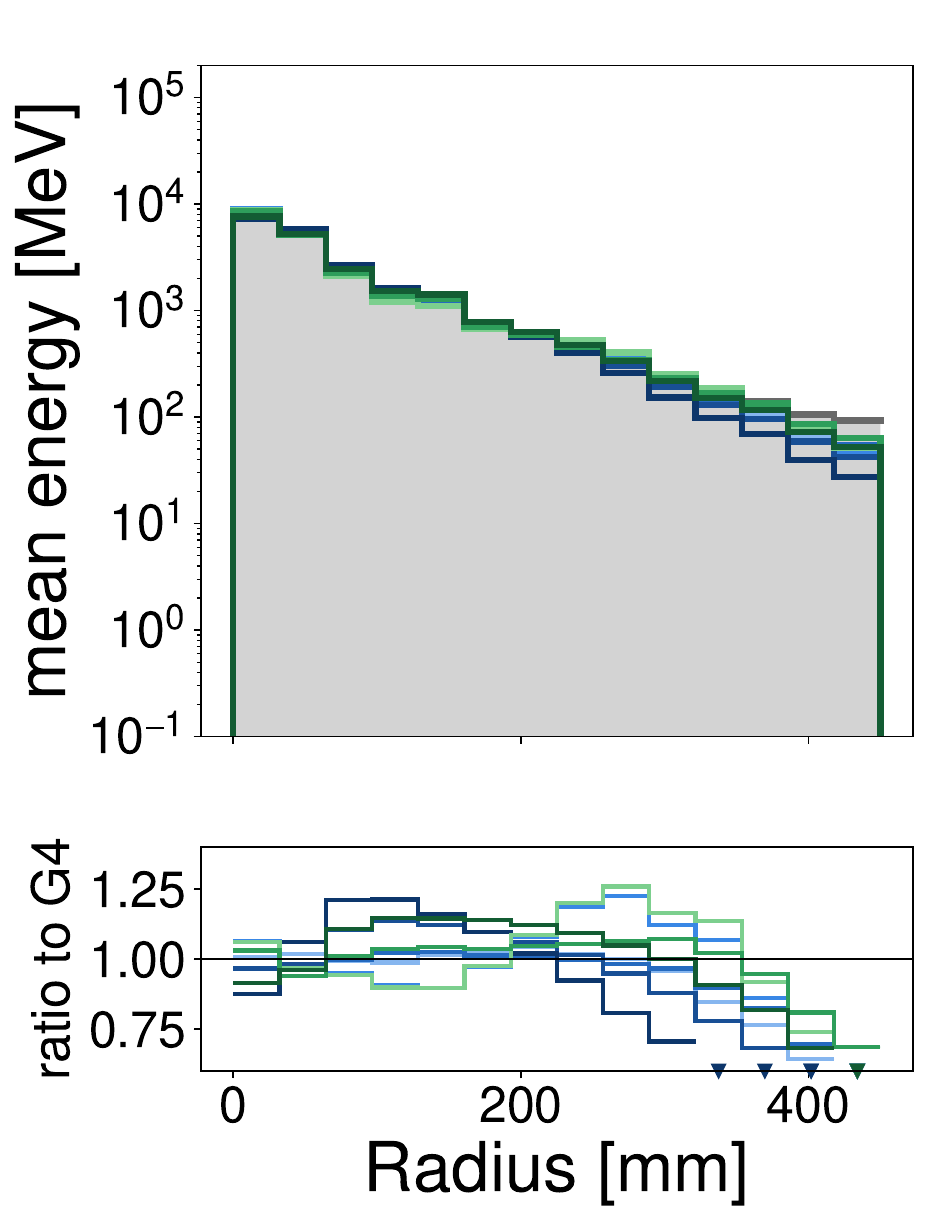}\hfill
    \includegraphics[width=0.245\linewidth]{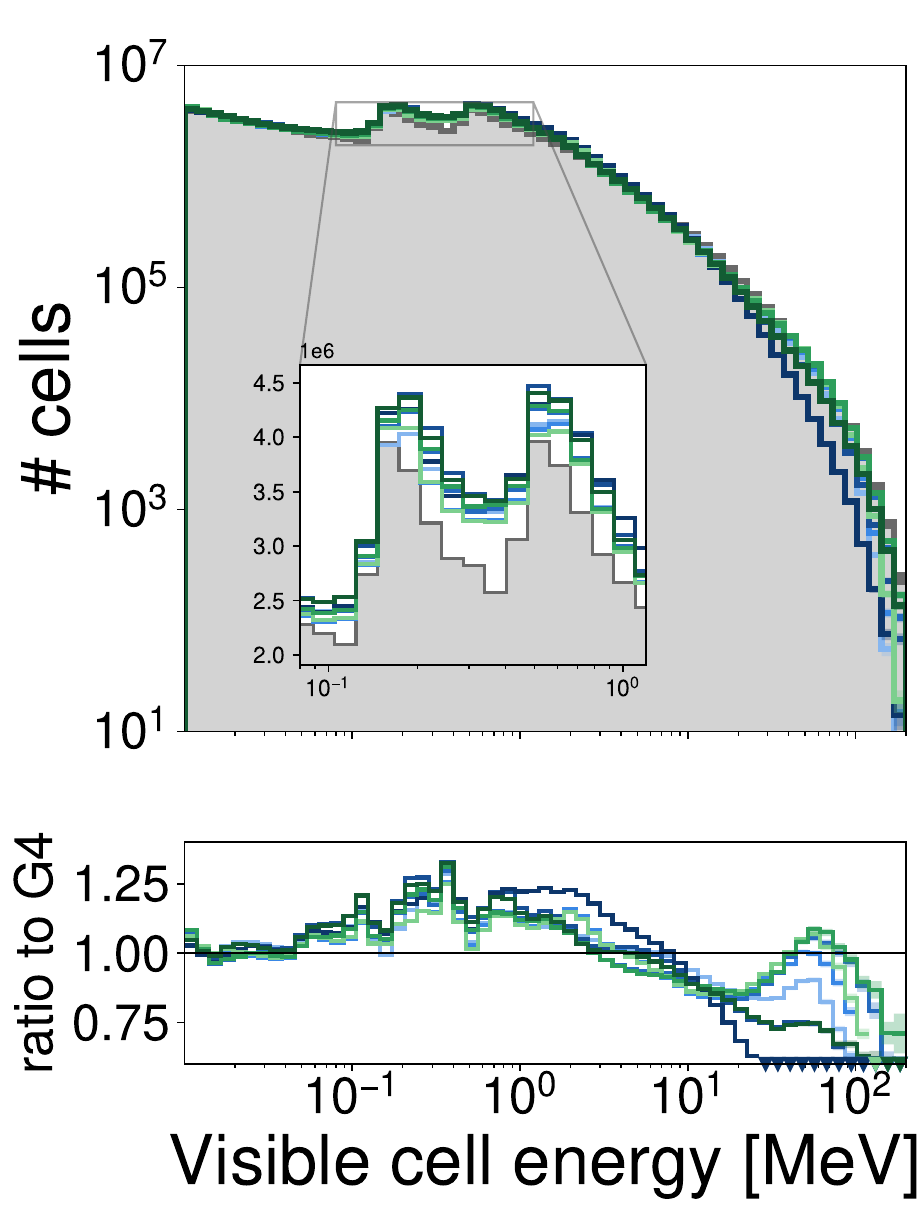}\hfill
    \includegraphics[width=0.245\linewidth]{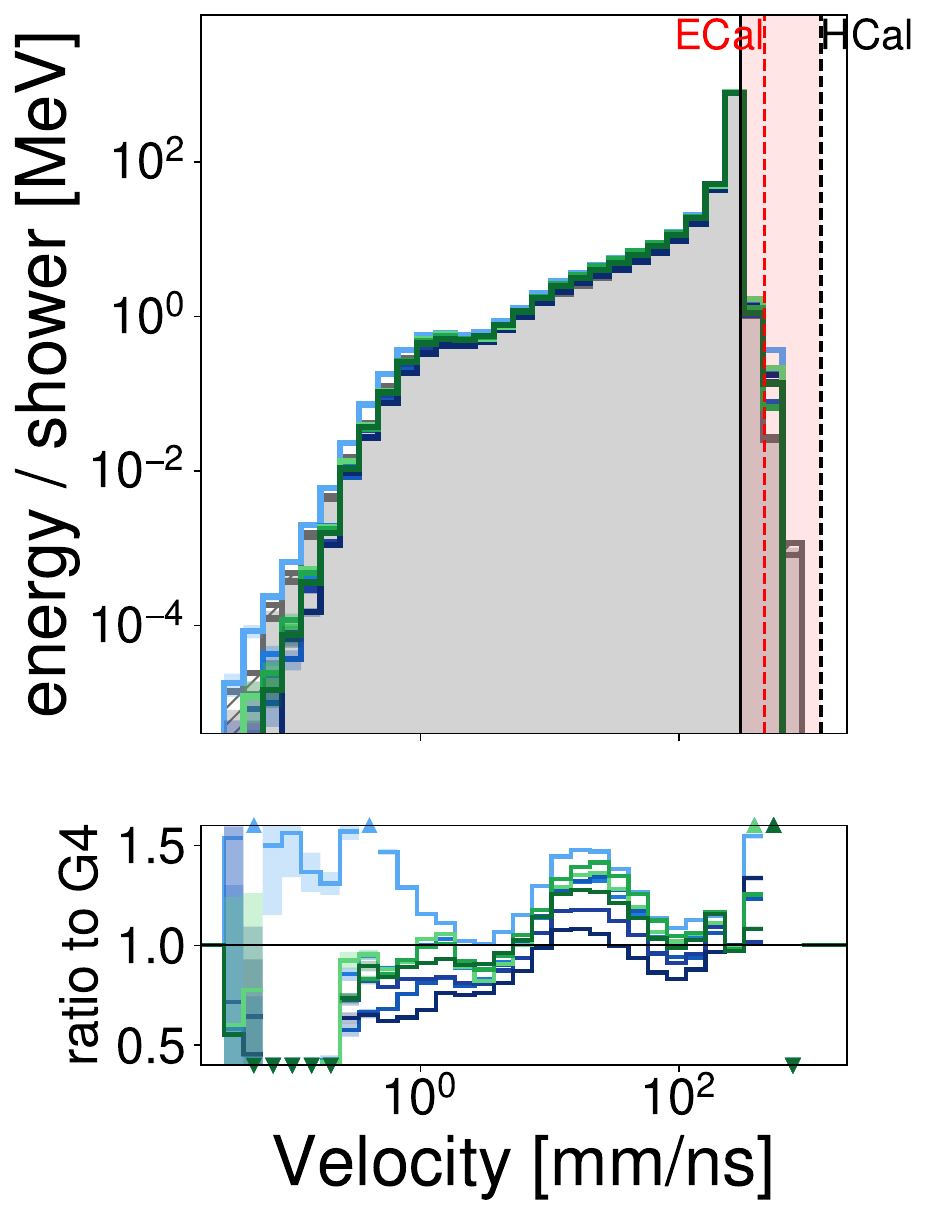}
    \caption{Shower observables at the readout granularity for \textsc{Geant4}
    (grey), the five latent sizes (blue) and the large \LMF at $d_s = 128$, 256 and
    512 (green), \num{100000} showers each between
    \SI{10}{\giga\electronvolt} and \SI{90}{\giga\electronvolt}. From left to
    right: longitudinal profile, radial profile, cell-energy spectrum and
    apparent-velocity profile (distance from the gun over arrival time). The dashed lines in the last panel
    are the largest apparent velocities $c_\mathrm{eff}$ of an ECal and an HCal
    cell, the solid line the speed of light. The insets show the last ECal layers of the
    longitudinal profile and the two MIP peaks of the cell-energy spectrum on a linear scale.
    The bands are the statistical uncertainty, and the lower panels the ratio to \textsc{Geant4}.}
    \label{fig:obs_ds}
\end{widefigure}

\subsection{Shower observables}\label{sec:observables}

Each model generates \num{100000} showers with incident energies drawn uniformly
between \SI{10}{\giga\electronvolt} and \SI{90}{\giga\electronvolt}, as in the
training data, and each stage is sampled in a single network evaluation. They are
compared with as many \textsc{Geant4} showers. Every generated sample, here and in the
rest of the paper, is corrected by the energy-dependent calibration factor $f(E)$ of
section~\ref{sec:model}, fitted for each model as described in
appendix~\ref{sec:appcalib}.

Figure~\ref{fig:obs_ds} shows four observables. The longitudinal profile is the
best reproduced, from the third layer to deep in the HCal, including the step at
the boundary between the two sections. Only in the last HCal layers, where little energy is left,
the models fall slightly below \textsc{Geant4}. This performance comes from \layerFM which determines the number of points in
each layer.
The \VAEdec then selects the energy of the points, and their lateral distribution,
but the multiplicity of the layers is already determined, which significantly improves the longitudinal performance.

Of the metrics we evaluate on, the radial profile differs the most between the latent sizes used. All of them follow
\textsc{Geant4} closely in the core of the shower, but between \SI{10}{\centi\meter}
and \SI{30}{\centi\meter} from the axis they scatter around it, with no ordering in
$d_s$. Beyond about \SI{32}{\centi\meter} from the axis, all latent sizes deposit too little
energy, $d_s = 1024$ most of all.
The discrepancy comes from the
\VAEdec, which most of the times places points close to the shower axis: the reconstructed showers of
appendix~\ref{sec:appreco} (figure~\ref{fig:obs_reco}) already show it.

The cell-energy spectrum it is well modeled by all latent size models. All of them get both MIP peaks of ECal and HCal. The plot shows the one systematic effect common to all latent sizes:
too many cells between about \SI{0.1}{\mega\electronvolt} and \SI{1}{\mega\electronvolt}
and too few above \SI{10}{\mega\electronvolt}. This effect is caused by \calochill model, which clusters the points less strongly than
\textsc{Geant4}. Fewer points therefore fall into the same readout cell, and the energy of a
shower is shared among more cells of lower energy. The reconstructed showers of appendix~\ref{sec:appreco}
(figure~\ref{fig:obs_reco}) show the same effect, including the excess at the two MIP
peaks near \SI{0.2}{\mega\electronvolt} and \SI{0.5}{\mega\electronvolt}, so it
originates in the \VAEdec and the generative stage adds a little error to it.

The rightmost plot shows the energy-weighted distribution of the apparent velocity of the shower cells, defined as the distance from the particle gun divided by the arrival time. In \textsc{Geant4} almost all the energy arrives
with the prompt electromagnetic front, between \SI{100}{\milli\meter\per\nano\second}
and the speed of light, and every model reproduces this closely. In the last panel of figure~\ref{fig:obs_ds}, the dashed vertical lines show $c_\mathrm{eff}$ (eq.~\ref{ceff_eq}) for the ECal and the HCal. Because of the time floor introduced in section~\ref{sec:assembly}, no generated cell can exceed this value. The models differ mostly in the
slower components, which carry little energy. Between \SI{10}{\milli\meter\per\nano\second}
and \SI{100}{\milli\meter\per\nano\second} most models overestimate the deposited
energy, and below \SI{1}{\milli\meter\per\nano\second}, in the slow neutron component,
the deviation from \textsc{Geant4} changes sign with the latent size, from an excess at
$d_s = 64$ to a deficit at $d_s = 1024$.

\begin{table}[htbp]  
    \centering
    \caption{Geometric mean over the observables of the normalised Wasserstein distance (WD)
    and of the Jensen--Shannon divergence (JSD) between each latent size and \textsc{Geant4},
    on the full-range samples of \num{100000} showers, from table~\ref{tab:ds_metrics}.
    The best model is in bold; the error is the standard deviation of the mean over 1000
    resamplings of every entry within its uncertainty, and the last row is the
    \textsc{Geant4} floor.}
    \label{tab:ds_metrics_avg}
    \footnotesize
    \setlength{\tabcolsep}{3pt}
\begin{tabular}{lcc}
\toprule
 & $\mathrm{WD}/\sigma_{\textsc{Geant4}}$ [$\times 10^{-2}$] & JSD [$\times 10^{-4}$] \\
\midrule
$d_s = 64$ & $3.98 \pm 0.18$ & $12.77 \pm 0.32$ \\
$d_s = 128$ & $3.41 \pm 0.18$ & $10.22 \pm 0.27$ \\
$d_s = 256$ & $3.65 \pm 0.11$ & $10.66 \pm 0.59$ \\
$d_s = 512$ & $3.94 \pm 0.25$ & $10.87 \pm 1.10$ \\
$d_s = 1024$ & $6.56 \pm 0.22$ & $23.68 \pm 0.42$ \\
$d_s = 128$, large \LMF & $3.88 \pm 0.51$ & $10.65 \pm 0.73$ \\
$d_s = 256$, large \LMF & $\mathbf{2.72 \pm 0.36}$ & $\mathbf{7.20 \pm 0.41}$ \\
$d_s = 512$, large \LMF & $3.61 \pm 0.19$ & $8.77 \pm 0.35$ \\
\textsc{Geant4} floor & $0.31 \pm 0.09$ & $0.17 \pm 0.04$ \\
\bottomrule
\end{tabular}

\end{table}

\begin{widefigure}[!ht]
    \centering
    \includegraphics[width=\modelslegendwidth]{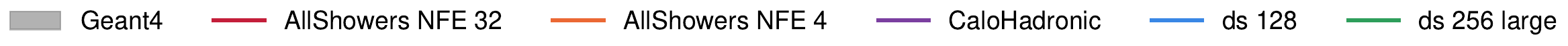}\\[2pt]
    \includegraphics[width=\modelspanelwidth]{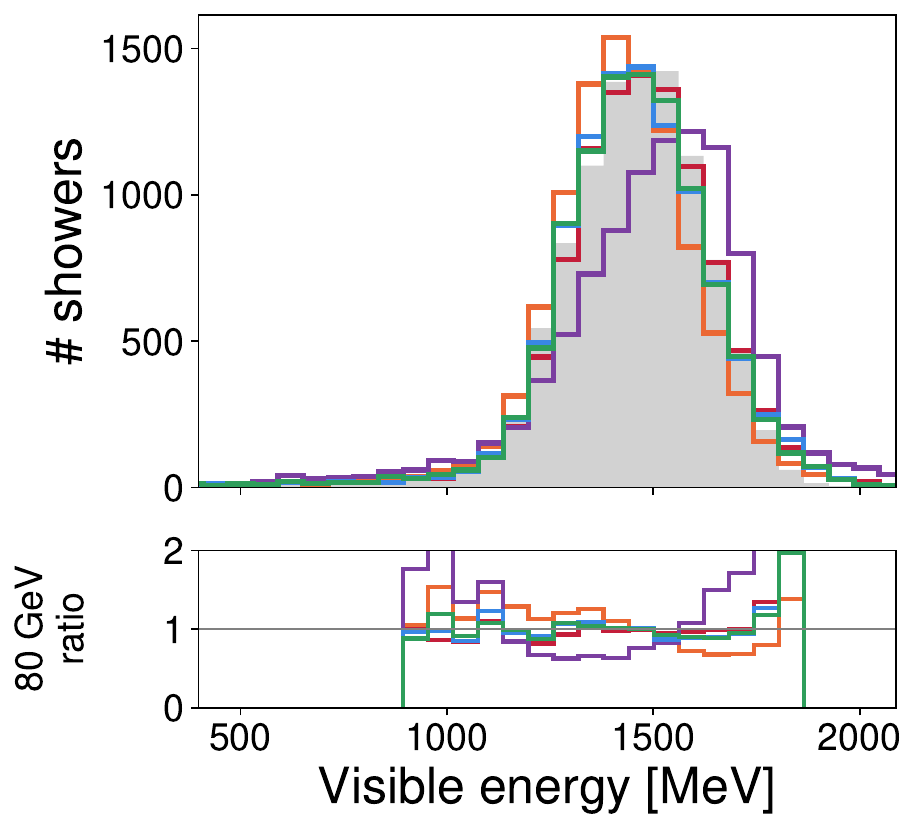}\hspace{0.02\linewidth}
    \includegraphics[width=\modelspanelwidth]{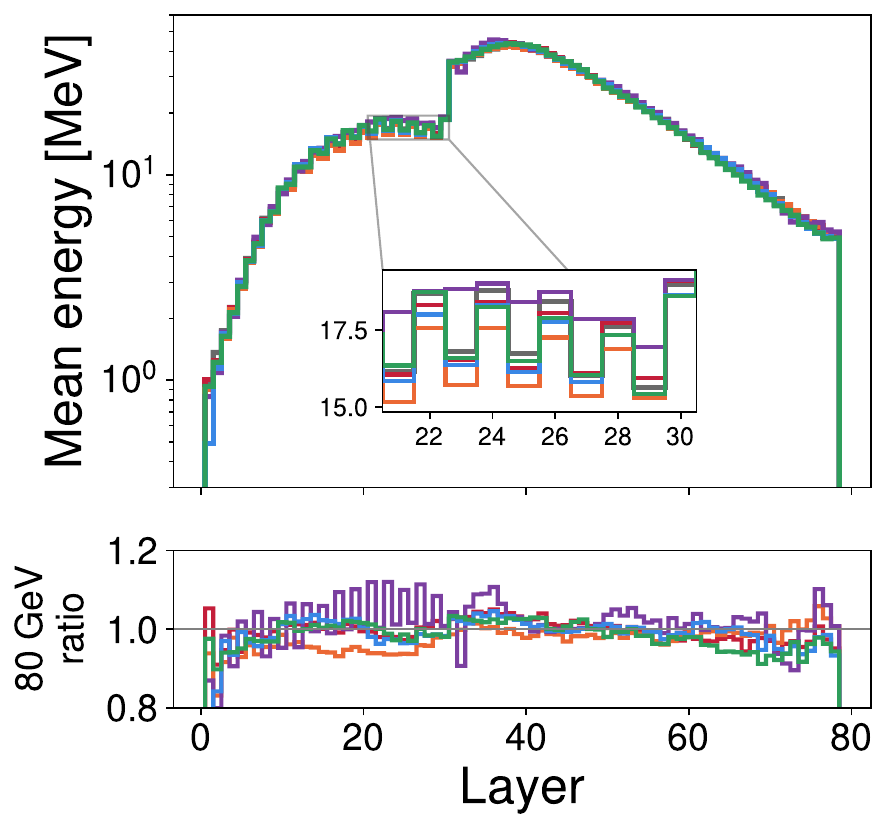}\hspace{0.02\linewidth}
    \includegraphics[width=\modelspanelwidth]{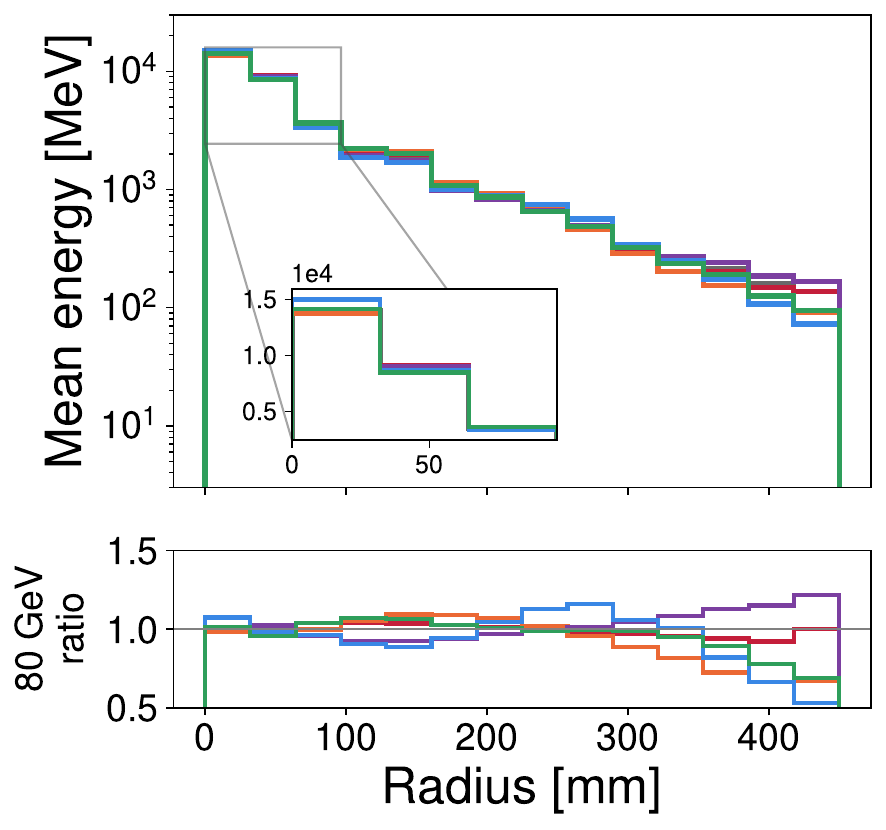}
    \caption{Energy sum (left), mean energy per layer (centre) and mean energy per
    shower in bins of the distance from the shower axis (right) at
    \SI{80}{\giga\electronvolt} for \textsc{Geant4} (grey), \as with
    32 and 4 network evaluations (NFE~32, NFE~4), \cc, $d_s = 128$ and the large \LMF
    at $d_s = 256$; the insets show the last ECal layers (centre) and the first three
    radial bins (right) on a linear scale. The lower panels give the ratio to \textsc{Geant4}. The same comparison at
    \SI{20}{\giga\electronvolt} and \SI{50}{\giga\electronvolt} is in
    figure~\ref{fig:obs_models_2050}.}
    \label{fig:obs_models}
\end{widefigure}

\medskip
To rank the latent sizes, two distances between the generated and the \textsc{Geant4}
distributions are used, the normalised Wasserstein distance (WD) and the
Jensen--Shannon divergence (JSD), both defined in appendix~\ref{sec:appdist}.
Table~\ref{tab:ds_metrics_avg} gives their geometric mean over the observables, and
table~\ref{tab:ds_metrics} in appendix~\ref{sec:appds} their value for each observable.
Per-shower observables are compared shower by shower, while profiles are energy-weighted cell distributions, accumulated over all showers. The radial profile is shown separately for the ECal and the HCal, which have cell sizes of 5 mm and 30 mm, respectively. The velocity is also split into a longitudinal component,
$v_L = (y - y_\mathrm{gun})/t$, and a radial one, $v_R = \rho/t$ with $\rho$ the distance
of the cell from the beam axis, both energy-weighted. Each value is the mean
over five disjoint subsets of both samples and its uncertainty their spread.

In table~\ref{tab:ds_metrics_avg} all latent sizes stay at least an order of magnitude
above the \textsc{Geant4} floor, the distance between two disjoint \textsc{Geant4}
subsets of the same size. For the standard \LMF, $d_s = 128$ gives the smallest WD, followed by
$d_s = 256$ and 512. In the JSD, the values for $d_s = 128$, 256
and 512 agree within their uncertainties.
$d_s = 64$ is further from \textsc{Geant4} in both distances, and $d_s = 1024$ is clearly
the furthest. This shows how a shorter latent sequence could be not only cheaper but, down to $d_s = 128$, even more accurate; only at $d_s = 64$ the accuracy falls again. The full observable by observable table is reported in appendix~\ref{sec:appds} table~\ref{tab:ds_metrics}, which shows that no latent size is best
everywhere.

The large \LMF at $d_s = 256$ has the smallest geometric mean of table~\ref{tab:ds_metrics_avg} in both
distances. Comparing the two \LMF{}s at $d_s = 512$, the large one has a smaller JSD than the standard one and a compatible
WD. Comparing the two \LMF{}s at $d_s = 128$, they agree in both distances: for this latent size the standard \LMF
is already large enough, and a larger network brings no further gain. On the other hand, for the
longer latent sequences at $d_s = 256$ and 512 the larger network improves the
agreement with \textsc{Geant4}. The analysis of the latent space in appendix~\ref{sec:applda} supports this
conclusion. For every latent size, the large \LMF retains more information about the incident energy than the standard one. This suggests that a longer latent sequence only pays off when the network has enough capacity to model it.

\subsection{Comparison with \as and \cc}\label{sec:cmp_models}

\begin{table}[htbp]  
    \centering
    \caption{Geometric mean over the observables of the WD and of the JSD to
    \textsc{Geant4} at 20 / 50 /
    \SI{80}{\giga\electronvolt}, and their mean over the three energies, from
    table~\ref{tab:three_energies}; the best model is in bold, the column $\pm$ is the
    largest uncertainty of the three energies, and the last row is the \textsc{Geant4} floor.}
    \label{tab:three_energies_avg}
    \footnotesize
    \setlength{\tabcolsep}{1.6pt}  
\begin{tabular}{lccccc}
\toprule
 & 20\,GeV & 50\,GeV & 80\,GeV & $\pm$ & Mean \\
\midrule
\multicolumn{6}{l}{$\mathrm{WD}/\sigma_{\textsc{Geant4}}$ [$\times 10^{-2}$]} \\
\midrule
\cc & 7.23 & 5.55 & 5.19 & 0.70 & 5.99 $\pm$ 0.34 \\
\as NFE~32 & \textbf{5.16} & \textbf{4.06} & \textbf{3.36} & 0.90 & \textbf{4.19} $\pm$ 0.40 \\
\as NFE~4 & 6.70 & 6.28 & 6.40 & 0.97 & 6.46 $\pm$ 0.43 \\
$d_s = 128$ & 6.87 & 5.52 & 5.24 & 0.61 & 5.88 $\pm$ 0.27 \\
$d_s = 256$, large \LMF & 5.84 & 4.97 & 5.00 & 0.41 & 5.27 $\pm$ 0.21 \\
\textsc{Geant4} floor & 1.82 & 1.64 & 1.61 & 0.42 & 1.69 $\pm$ 0.22 \\
\midrule
\multicolumn{6}{l}{JSD [$\times 10^{-4}$]} \\
\midrule
\cc & 51.07 & 37.33 & 33.88 & 2.90 & 40.76 $\pm$ 1.39 \\
\as NFE~32 & \textbf{18.73} & \textbf{13.29} & \textbf{13.68} & 1.45 & \textbf{15.23} $\pm$ 0.77 \\
\as NFE~4 & 29.21 & 25.34 & 26.58 & 2.97 & 27.04 $\pm$ 1.33 \\
$d_s = 128$ & 34.48 & 25.59 & 23.38 & 3.55 & 27.82 $\pm$ 1.46 \\
$d_s = 256$, large \LMF & 29.15 & 19.76 & 18.82 & 2.18 & 22.57 $\pm$ 0.91 \\
\textsc{Geant4} floor & 6.47 & 5.25 & 4.51 & 0.85 & 5.41 $\pm$ 0.44 \\
\bottomrule
\end{tabular}

\end{table}

The latent size $d_s = 128$ and the large \LMF at $d_s = 256$ of
table~\ref{tab:ds_metrics} are compared with the two generators that precede this
one on the same detector, \as~\cite{Buss:2026yrf} and \cc~\cite{Buss:2025cyw}, at
incident energies of \SI{20}{\giga\electronvolt}, \SI{50}{\giga\electronvolt} and
\SI{80}{\giga\electronvolt}. \as enters twice: with 32 network evaluations, the setting of
ref.~\cite{Buss:2026yrf}, and with 4 (2 midpoint steps), the fewest that still give
reasonable showers. The reference at each energy is
\num{10000} \textsc{Geant4} showers within $\pm\SI{0.5}{\giga\electronvolt}$; every
model contributes \num{10000} showers, all projected onto the readout cells in the
same way. Figure~\ref{fig:obs_models} shows the energy sum and the longitudinal and
radial profiles at \SI{80}{\giga\electronvolt}, and figure~\ref{fig:obs_models_2050} in
appendix~\ref{sec:app3E} the same at \SI{20}{\giga\electronvolt} and \SI{50}{\giga\electronvolt}.
The highest energy is shown here because its showers contain the most points, making it the most interesting case for the architecture presented in this work.
Table~\ref{tab:three_energies_avg} gives the geometric mean of the
distances over the observables, and table~\ref{tab:three_energies} in
appendix~\ref{sec:app3E} the distances for each observable. The velocity is not compared, since neither \as nor
\cc generates the arrival time. The training conditions of \as differ from those of the other models:
it is trained on a wider range of incident energies and on a larger and more varied
set of showers, covering several particle types and directions of incidence, which can
shift its accuracy at a given energy in either direction.

In the energy sum (figure~\ref{fig:obs_models}, left), all models except \cc reproduce the position and width of the peak at every energy. The \cc distribution is instead shifted towards higher visible energy and broadened (table~\ref{tab:three_energies}). A likely cause is that it is the only one of the five models generated without an energy calibration. In the longitudinal profile, \cc overestimates the mean energy
per layer in the ECal and \as at 4 evaluations underestimates it, while the other three
models reproduce it. \cc is also the only model that
does not reproduce the odd--even modulation of the mean deposited energy across
consecutive ECal layers (inset of the centre panel). This comes from how points are assigned to layers: \as and \calochill fix each point to its predicted layer, whereas \cc uses the predicted number of points per layer only as conditioning for the point generation.
In the HCal
all five models follow \textsc{Geant4} over the bulk of the shower; in its last layers,
where the deposits are smallest, both \calochill models fall slightly below it. The radial profile shows the largest differences between the models. All five
models follow \textsc{Geant4} up to about \SI{300}{\milli\meter} from the axis, apart
from a small excess of $d_s = 128$ between about \SI{230}{\milli\meter} and
\SI{290}{\milli\meter}. At larger radii \as at 32 evaluations remains close to
\textsc{Geant4}, \cc overestimates the energy, and \as at 4 evaluations and both
\calochill models underestimate it, $d_s = 128$ the most.

In table~\ref{tab:three_energies_avg}, \as at 32 evaluations has the smallest geometric
mean at every energy in both distances. Of the faster models, the large \LMF at $d_s = 256$
is the closest in both distances. In the WD it is
followed by $d_s = 128$ and \cc, which agree within their uncertainties, and by \as at
4 evaluations; in the JSD by \as at 4 evaluations and $d_s = 128$, which agree within
theirs, while \cc is clearly the furthest. No model reaches the
\textsc{Geant4} floor.
The distances for each observable and energy are given in table~\ref{tab:three_energies}
in appendix~\ref{sec:app3E}.

\subsection{Generation time}\label{sec:timing}
\begin{figure}[!ht]
    \centering
    \providecommand{\timingfigwidth}{\linewidth}
    \makebox[\linewidth][c]{\includegraphics[width=\timingfigwidth]{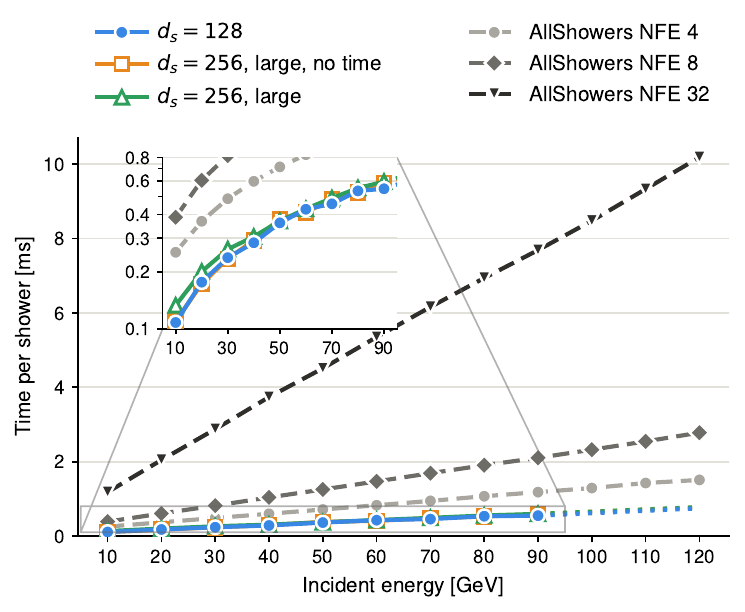}}
    \caption{Generation time per shower against incident energy on an NVIDIA A100
    (\SI{80}{GB}, PCIe) GPU at batch size 1400, for the full chain at $d_s = 128$, the large
    \LMF at $d_s = 256$ on a \textsc{VAE} trained without the hit time (orange) and with
    it (green, nearly on top of the orange line), compared with \as at 4, 8 and 32 network evaluations.
    The dotted
    line extends a straight-line fit between \SI{10}{\giga\electronvolt} and
    \SI{90}{\giga\electronvolt} to \SI{120}{\giga\electronvolt}, an upper estimate since the
    model is not trained above \SI{90}{\giga\electronvolt}.}
    \label{fig:timing_ds256}
\end{figure}

The main goal of this work is to reduce the time needed to generate hadronic showers at
the full granularity of the detector. Instead of sampling every point of the cloud with an
iterative sampler, \calochill generates a compressed representation, so the cost of the
sampling steps does not depend on the size of the shower and only the decompression
scales with it.
Figure~\ref{fig:timing_ds256} shows the generation time per shower of \calochill against
the incident energy at $d_s = 128$, with the large \LMF at $d_s = 256$, and with the same
large \LMF on a \textsc{VAE} trained without the hit time, next to \as at 4, 8 and 32 network evaluations, all on the
same NVIDIA A100 GPU at batch size 1400.

The configuration without the hit time is
included for a fair comparison of the generation time, since neither \as nor \cc
generates the time. \cc is not
shown, since even at its fastest batch size it is about 600 times slower than \calochill
(table~\ref{tab:timing_comparison} in appendix~\ref{sec:apptimingtable}).

The full
\calochill chain is timed, i.e.\ \layerFM distilled to a single step, the \LMF at a single network
evaluation and the \VAEdec. Over the whole energy range \calochill is faster than \as at every number of network
evaluations, and the gap widens with the energy, because the cost of \as grows with the
number of evaluations times the number of points, whereas \calochill runs the \VAEdec
over the points only once. In accuracy, the large
\LMF at $d_s = 256$ lies between \as at 32 and at 4 evaluations
(table~\ref{tab:three_energies_avg}) while being faster than both.

A straight-line fit in the incident energy $E$ is performed to the two \calochill models to extrapolate above
\SI{90}{\giga\electronvolt} (dotted line). The time depends linearly on the number of points in all models (appendix~\ref{sec:apptiming}). 
Most of this increase in \calochill comes from the \VAEdec, as shown in figure~\ref{fig:breakdown}. The intercept of the fit, 
at around \SI{0.08}{\milli\second}, is
the part of the time per shower that does not depend on the energy; its breakdown into
the stages of the chain is given in appendix~\ref{sec:appbreakdown}.

\subsection{Discussion}\label{sec:discussion}
\begin{figure}[htbp]
    \centering
    \includegraphics[width=\linewidth]{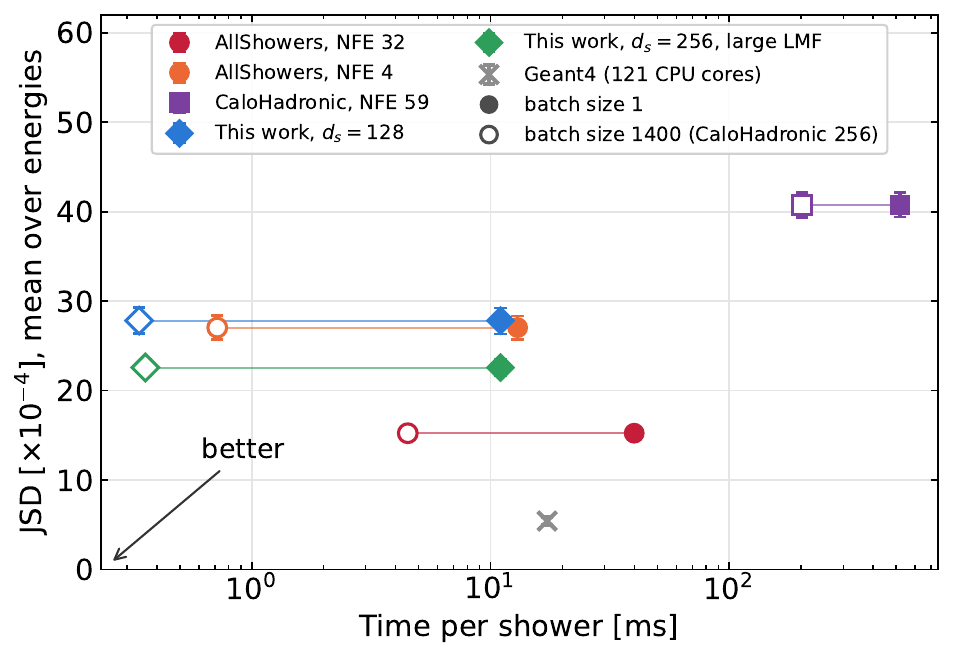}
    \caption{Accuracy against generation time. Vertical axis: JSD to
    \textsc{Geant4} of table~\ref{tab:three_energies_avg}, the geometric mean over the
    observables averaged over 20, 50 and \SI{80}{\giga\electronvolt}, with its error;
    \textsc{Geant4} against itself gives the statistical floor. Horizontal axis: time per
    shower on an NVIDIA A100 GPU from table~\ref{tab:timing_comparison}, at batch size 1
    (filled markers) and at batch size 1400 (open markers; 256 for \cc,
    the fastest batch size measured for it);
    for \textsc{Geant4} a cost-normalised CPU time is given (see appendix~\ref{sec:appcost}).}
    \label{fig:time_vs_quality}.
\end{figure}

Figure~\ref{fig:time_vs_quality} shows the accuracy of each generator, given by the
JSD of table~\ref{tab:three_energies_avg}
(geometric mean over the observables, averaged over the three energies), against its
time per shower on the GPU of table~\ref{tab:timing_comparison}, at batch size 1
(filled markers) and at batch size 1400 (open markers; 256 for \cc). 
\textsc{Geant4} against itself, 
at its CPU time divided by the 121 CPU cores that cost as much as one GPU (appendix~\ref{sec:appcost}), 
marks the statistical floor.
In both batch sizes, \cc at 59 evaluations is the slowest and the least accurate and
\as at 32 evaluations is the most accurate, but at batch size 1 it takes 3.6 times as long as the large \LMF.
At batch size 1,
$d_s = 128$ and the large \LMF at $d_s = 256$ take \SI{11}{\milli\second} per shower
and \as at 4 evaluations \SI{13}{\milli\second}. At batch size 1400 the
order in time is the same, but the difference to \as is larger (appendix~\ref{sec:apptimingtable}).
At equal hardware cost \textsc{Geant4} takes \SI{17}{\milli\second} per shower, so
$d_s = 128$ and the large \LMF at batch size 1400 are about 50 times faster.
\calochill with the large \LMF at $d_s = 256$ is both faster and more accurate than
\as at 4 evaluations (\SI{11}{\milli\second} and $22.6\times10^{-4}$ against
\SI{13}{\milli\second} and $27.0\times10^{-4}$), so it lies on the Pareto front of
figure~\ref{fig:time_vs_quality} together with \as at 32 evaluations.

\section{Conclusion}\label{sec:concl}

In this work we present \calochill, a generative model for hadronic showers in a highly granular calorimeter that no longer samples the hits of a shower one by one.
It generates a compressed summary of the shower in a learned latent space and expands it to the full point cloud only at the end, in three stages, \layerFM, the \LMF and the \VAEdec, each a single network evaluation.
Every hit in the ECal and the HCal comes with its position, energy and arrival time, which makes \calochill the first generative model to simulate the time structure of hadronic showers in both sections of a highly granular calorimeter.

The large \LMF at $d_s = 256$ is both faster and more accurate than \as at 4 evaluations, so for fast generation \calochill lies beyond the Pareto frontier of \as.
It also scales better with the shower size, since \as runs over all hits at every network evaluation and \calochill only once, in the \VAEdec, so the gap in time grows with the energy.
The longitudinal profile is the best reproduced observable, since \layerFM fixes the number of hits in each layer, and the time structure of the showers is reproduced as well.

A longer latent sequence improves the showers only when the \LMF is large enough to model it.
Passing \textsc{Geant4} showers through the \textsc{VAE} alone (appendix~\ref{sec:appreco}) shows that the deficit far from the shower axis and the excess of low-energy cells come from the compression.

Future work could scale $d_s$ with the number of hits of each shower and improve the sparse attention of the \textsc{VAE}~\cite{deepseekai2025deepseekv32pushingfrontieropen}.
Finally, the training should be extended to electromagnetic showers and to angular conditioning, as needed for a full analysis workflow.

\begin{acknowledgments}
We thank Thorsten Buss for fruitful discussions throughout the early stages of this work, especially on the sparse attention masks with FlexAttention and on the sliced Wasserstein distance as the reconstruction loss of the \textsc{VAE}. We also thank the members of the generative group at Universität Hamburg and DESY for valuable comments during the course of this research.
This research was supported in
part by the Maxwell computational resources operated at Deutsches Elektronen-Synchrotron DESY,
Hamburg, Germany. This project has received funding from the European Union’s Horizon 2020
Research and Innovation program under Grant Agreement No 101004761. We acknowledge
support by the Deutsche Forschungsgemeinschaft under Germany’s Excellence Strategy – EXC
2121 Quantum Universe – 390833306. 
\end{acknowledgments}

\bibliography{biblio}

\appendix
\appendix
\section{Code and model availability}\label{sec:appcode}
The code is available in the following Git repositories:
\begin{itemize}
    \item \layerFM: \url{https://github.com/FLC-QU-hep/LayerFM}
    \item \calochill: \url{https://github.com/FLC-QU-hep/CaloCHILL} 
\end{itemize}
The weights of the trained models are available at:
\begin{itemize}
    \item \layerFM: \url{https://huggingface.co/FLC-QU-hep/LayerFM}
    \item \calochill: \url{https://huggingface.co/FLC-QU-hep/CaloCHILL}
\end{itemize}

The full datasets, namely the one of \cc, the one used in this work and the same showers at the
physical cell granularity, are available at \url{https://doi.org/10.25592/uhhfdm.22077}.

\section{Distance measures}\label{sec:appdist}
The comparisons with \textsc{Geant4} use two distances between the distribution $q$ of an
observable in a generated sample and its distribution $p$ in \textsc{Geant4}. Both are
computed in one dimension: on the values per shower for per-shower observables, such as
the energy sum or the number of hits, and on the cells weighted by their energy for the
profiles and the velocities.

The Wasserstein distance is the area between the two cumulative distribution functions
$F_p$ and $F_q$,
\begin{equation}
    W(p, q) = \int \left| F_p(x) - F_q(x) \right| \mathrm{d}x ,
\end{equation}
i.e.\ the mean distance by which probability has to be moved to turn one distribution
into the other. Since it has the units of the observable, it is divided by the standard
deviation $\sigma_p$ of the full \textsc{Geant4} sample, $\mathrm{WD} = W(p, q)/\sigma_p$,
so that observables in different units can be compared. For example, a WD of 0.1
corresponds to a shift of the generated distribution by a tenth of the \textsc{Geant4}
width. The WD is mostly sensitive to differences in the bulk of the distribution.

The Jensen--Shannon divergence compares the binned distributions,
\begin{equation}
    \mathrm{JSD}(p, q) = \tfrac{1}{2}\,\mathrm{KL}(p\,\|\,m) + \tfrac{1}{2}\,\mathrm{KL}(q\,\|\,m),
\end{equation}
with $m = \tfrac{1}{2}(p + q)$ and the Kullback--Leibler divergence summed over the same
bins $i$ as $p$ and $q$, described below,
\begin{equation}
    \mathrm{KL}(p\,\|\,m) = \sum_i p_i \ln\frac{p_i}{m_i} .
\end{equation}
The JSD compares both samples with their mixture $m$; it is symmetric, finite and bounded, $0 \le \mathrm{JSD} \le \ln 2$. The
bins are given by 25 quantiles of the full \textsc{Geant4} sample, energy-weighted for the
profiles, and the outer edges are extended to $\pm\infty$ so that no entry is lost. Each
bin then contains the same fraction of the \textsc{Geant4} sample, so tails and core are
weighted equally and the JSD is mostly sensitive to differences in shape. This binning
does not work for the longitudinal velocity, because its energy-weighted \textsc{Geant4}
distribution is a sharp peak at the speed of light. All quantile bins would fall inside
the peak, and small shifts of the generated times would give large divergences. It is
therefore binned in 25 equal intervals of $\log_{10} v$, over the range that contains
nearly all the \textsc{Geant4} energy. The JSD is preferred to the Kullback--Leibler divergence
$\mathrm{KL}(p\,\|\,q)$, which is not symmetric and diverges as soon as a bin is empty in
the generated sample but not in \textsc{Geant4}.

We use both distances since the WD is mainly sensitive to a shift of the bulk of a
distribution and the JSD to differences in its shape, including the tails.

\section{Model details}\label{sec:appA}

The three stages of section~\ref{sec:model} are trained independently. Their
trainable parameter counts are collected in table~\ref{tab:params}; each stage is
then described in turn, with its hyper-parameters in tables~\ref{tab:layerfm},
\ref{tab:vae} and~\ref{tab:lmf}. The counts for the two point-level stages are given for the
baseline \textsc{VAE} with $d_s = 256$. For the other four latent sizes only the number of
learnable query vectors changes.

\begin{table}[htbp]
\centering
\caption{Number of trainable parameters of each stage, for the baseline
$d_s = 256$ configuration. The layer-level stage is \layerFM; the point-level
stages are the \textsc{VAE}, \VAEenc and \VAEdec together, and the \LMF.}
\label{tab:params}
\begin{tabular}{llr}
\toprule
Level & Stage & Parameters \\
\midrule
Layer-level & \layerFM (single-step student) & \num{754332} \\
\midrule
\multirow{2}{*}{Point-level} & \textsc{VAE} & \num{1553496} \\
                             & \LMF        & \num{1327967} \\
\midrule
\multicolumn{2}{l}{Total} & \num{3635795} \\
\bottomrule
\end{tabular}
\end{table}

\subsection{\layerFM}\label{sec:applayerfm}

\layerFM generates the per-layer observables of a shower in one shot. Its target
is the concatenation of two vectors of length 78, the number of points and the
earliest hit time in each sampling layer, so 156 numbers in total, and its only
conditioning input is the incident energy. The same network supports the
deposited energy per layer as a third block of 78, and conditioning on the
particle type and on the direction of incidence; neither is used for the single
$\pi^{+}$ sample studied here.

\begin{table}[htbp]
\centering
\caption{Hyper-parameters of \layerFM. Each transformation acts on each of the
78 layers independently; std.: standardised.}
\label{tab:layerfm}
\footnotesize
\setlength{\tabcolsep}{4pt}
\begin{tabular}{ll}
\toprule
Parameter & Value \\
\midrule
\multicolumn{2}{l}{\textit{Preprocessing}} \\
  points & dequantised, $\log(n+0.5)$, std. \\
  earliest hit time & quantile transform (1000) \\
  incident energy & $\log$, std. \\
  layers & 78 \\
\midrule
\multicolumn{2}{l}{\textit{Model}} \\
  architecture & fully connected \\
  embedding & 128 \\
  hidden widths & 256, 512 \\
  flow-time embedding & 6 \\
  conditioning & incident energy \\
\midrule
\multicolumn{2}{l}{\textit{Training}} \\
  objective & flow matching \\
  optimiser & Adam \\
  peak learning rate & $7\times10^{-4}$, one-cycle \\
  batch size & 1024 \\
  epochs & 1500 \\
  train / val.\ showers & \num{1806445} / \num{50000} \\
\midrule
\multicolumn{2}{l}{\textit{Distillation}} \\
  teacher & Heun, 200 steps \\
  student loss & squared error to teacher \\
  optimiser & Adam, $10^{-4}$, 1500 epochs \\
  final val.\ error & 0.018 \\
\midrule
\multicolumn{2}{l}{\textit{Sampling}} \\
  NFE & 1 \\
\bottomrule
\end{tabular}
\end{table}

\paragraph{Preprocessing.}
The three quantities are on very different scales and are transformed
independently, each layer being treated as its own feature. Since the point
counts are integers, they are dequantised by adding uniform noise, so that the flow
learns a continuous density. They are then passed through $\log(n + \alpha)$ with
$\alpha = 0.5$ and standardised. The earliest hit times are mapped onto a normal distribution by a
quantile transform with 1000 quantiles. We use a quantile transform instead of a
logarithm because a layer without energy deposits has no hit time, so in the late
layers a large fraction of the entries sits at a single value. The incident energy is likewise passed through
a logarithm and standardised.

\paragraph{Architecture and training.}
The network is a fully connected residual network with a 128-dimensional
per-layer embedding, two hidden blocks of width 256 and 512, and a
six-dimensional embedding of the flow time, \num{754332} parameters in all. The
point-count models of \cc and \as instead use a five-layer perceptron with widths
128, 256, 512, 256 and 128 and about \num{350000} parameters, which predicts only the
78 point counts. The network used here is shallower and has twice as many outputs. Training is as a continuous normalising flow with the conditional
flow-matching objective, Adam at a peak learning rate of $7\times10^{-4}$ on a
one-cycle schedule, batch size 1024, and 1500 epochs over \num{1806445} showers
with \num{50000} held out. The loss is a sum over the individual
observables, and the point counts and times are predicted by the same network. The main
difference to \as~\cite{Buss:2026yrf} is the sampling, where \as uses a Heun solver
with 100 evaluations and the distilled student used here a single one. The trained flow is integrated once, with a Heun solver at 200
steps, for every one of the \num{1856445} conditioning values of the sample, and
the noise that produced each output is stored alongside it. The student, which
has the same architecture as the teacher, is then trained to regress that
noise-to-sample map directly. Training the student takes a further 1500 epochs
with Adam at $10^{-4}$. The student is what the pipeline of
section~\ref{sec:model} uses, at a single network evaluation.

\subsection{The VAE}\label{sec:appvae}

The \textsc{VAE} turns a shower of up to 5376 points into $d_s$ latent vectors and
back. Both halves are transformers over the points of the cloud, and both use the
sparse attention of section~\ref{sec:attn}; the hyper-parameters are collected in
table~\ref{tab:vae}.

\begin{table}[htbp]
\centering
\caption{Hyper-parameters of the \textsc{VAE}. The latent size $d_s$ is the one
quantity scanned; all five \textsc{VAE}s share every other setting. std.: standardised.}
\label{tab:vae}
\footnotesize
\setlength{\tabcolsep}{4pt}
\begin{tabular}{ll}
\toprule
Parameter & Value \\
\midrule
\multicolumn{2}{l}{\textit{Preprocessing}} \\
  point features & $x$, $z$, energy, time \\
  $x$, $z$ & scaled to $[-1, 1]$ \\
  energy, time & $\log$, std. \\
\midrule
\multicolumn{2}{l}{\textit{Model}} \\
  latent points $d_s$ & 64, 128, 256, 512, 1024 \\
  latent dimension & 128 \\
  hidden dimension & 128 \\
  downsampling & cross-attention, learnable queries \\
  attention heads & 4 \\
  refinement blocks & 2 \\
  layer conditioning & 8 \\
  Fourier features & 4 frequencies, maximum 8 \\
  activation & GELU \\
  normalisation, dropout & layer norm, 0.1 \\
\midrule
\multicolumn{2}{l}{\textit{Training}} \\
  reconstruction loss & sliced Wasserstein, per layer \\
  random projections & 128 \\
  KL weight & $10^{-2}$ \\
  optimiser & Ranger (Lookahead + RAdam) \\
  peak learning rate & $10^{-3}$, 3500 warm-up steps \\
  gradient clipping & norm 2 \\
  batch size per device & 64 \\
\bottomrule
\end{tabular}
\end{table}

\paragraph{\VAEenc.}
Every point enters as its four generated features, $x$, $z$, energy and time,
lifted to the 128-dimensional working width by a small multi-layer perceptron
acting on a Fourier expansion of the input, as in \cc~\cite{Buss:2025cyw}. Two embeddings are then added to each
point, one of its layer index, as in \as~\cite{Buss:2026yrf},
and one of the 78 per-layer minimum times; the incident energy and the total
multiplicity are carried by the queries of the compression step instead. The cloud is
compressed in a single cross-attention step. A set of $d_s$ query vectors, free
parameters initialised from a normal distribution scaled by $d^{-1/2}$, attends
to the points of the shower and returns one token per query; a self-attention
block then mixes those $d_s$ tokens. Each query also carries the conditioning and
a fixed, non-trainable sinusoidal positional encoding. Because of this encoding the
latent is an ordered sequence, and the network of the generative stage needs its own
positional embedding to reproduce this order (appendix~\ref{sec:applmf}).

\paragraph{Latent.}
Two linear heads read the \VAEenc tokens and return the mean and log-variance of
a diagonal Gaussian posterior for each of them, of dimension 128. The mean is
passed through $2\tanh(\mu/2)$ and the log-variance is clamped to $[-2, 2]$, so
that the mean and the width of the posterior of each token are bounded. This keeps the
Kullback--Leibler term finite also for tokens that correspond to an empty region of the
shower. A latent is then a
$d_s \times 128$ array, drawn by the usual reparametrisation
$\mathbf{z} = \bm{\mu} + \bm{\sigma}\odot\bm{\epsilon}$.

\paragraph{\VAEdec.}
The latent is read back by a block that re-applies the conditioning, and is then
expanded by the mirror image of the compression step: a set of learnable queries,
one per output point, cross-attends to the $d_s$ latent tokens, with a scale and
shift derived from the pooled latent applied to the queries. These queries are
initialised like those of the \VAEenc, but without a positional encoding. Their
position in the shower is given by the layer embedding and the index within the layer,
both determined by \layerFM, which also sets the number of queries. Two further self-attention blocks refine the result at
full resolution, and a final projection returns the four per-point features.

\paragraph{Objective.}
The output of the \VAEdec is an unordered set of points, so there is no
correspondence between the $i$-th output and the $i$-th input point, and the
reconstruction loss has to compare two distributions of points. The exact Wassersten distance 
is too expensive for clouds of several thousand points at every
training step, and the Chamfer distance, a common cheaper alternative, tends to smooth
the result. We therefore use the sliced Wasserstein distance, which relies on the fact
that in one dimension the Wasserstein distance between two sets of equal size is given
by the difference of their sorted values. Both clouds are projected onto 128 random unit
directions, sorted along each direction and compared. The cost scales as
$\mathcal{O}(PN\log N)$ instead of $\mathcal{O}(N^3)$, and the estimate is unbiased
and differentiable.

The comparison is done separately in each calorimeter layer. This is possible
because the layer of every generated point is fixed by the conditioning and not
predicted, and it prevents the loss from being reduced by moving points between layers.
The objective is completed by a Kullback--Leibler term with weight $10^{-2}$, which
pulls the posterior of each token towards a unit Gaussian. Its effect on the information
stored in the latent is discussed in appendix~\ref{sec:applda}.

\subsection{The \latentMF}\label{sec:applmf}

The generative stage is a transformer acting on the $d_s$ latent vectors, with
the sparse attention of section~\ref{sec:attn}; its hyper-parameters are listed
in table~\ref{tab:lmf}.

\begin{table}[htbp]
\centering
\caption{Hyper-parameters of the standard and the large \LMF; a value written across both
columns holds for both. The network is a transformer acting on the $d_s$ latent vectors, with
the sparse attention mask of section~\ref{sec:attn}. EMA: exponential moving average.}
\label{tab:lmf}
\footnotesize
\setlength{\tabcolsep}{4pt}
\begin{tabular}{lcc}
\toprule
Parameter & Standard & Large \\
\midrule
\multicolumn{3}{l}{\textit{Preprocessing}} \\
  target & \multicolumn{2}{c}{posterior mean $\mathbf{z}_{\mu}$, rescaled} \\
  scale, noise level & \multicolumn{2}{c}{from the training set} \\
\midrule
\multicolumn{3}{l}{\textit{Model}} \\
  architecture & \multicolumn{2}{c}{transformer} \\
  model dimension & 256 & 512 \\
  attention heads & \multicolumn{2}{c}{8} \\
  layers & 1 & 2 \\
  dropout & \multicolumn{2}{c}{0.1} \\
  time embedding & \multicolumn{2}{c}{32} \\
  positional embedding & \multicolumn{2}{c}{learned, zero-initialised} \\
  parameters ($d_s = 256$) & $1.3\times10^{6}$ & $7.7\times10^{6}$ \\
\midrule
\multicolumn{3}{l}{\textit{Objective}} \\
  $(r, t)$ & \multicolumn{2}{c}{uniform} \\
  fraction with $r \neq t$ & \multicolumn{2}{c}{0.5} \\
  adaptive loss weight & \multicolumn{2}{c}{$p = 1$, $c = 10^{-3}$} \\
  Gaussian skip & \multicolumn{2}{c}{per latent dimension} \\
  output head & \multicolumn{2}{c}{zero-initialised} \\
  $\mathrm{d}u/\mathrm{d}t$ & \multicolumn{2}{c}{finite difference, step $10^{-2}$} \\
\midrule
\multicolumn{3}{l}{\textit{Training}} \\
  optimiser & \multicolumn{2}{c}{Ranger (Lookahead + RAdam)} \\
  peak learning rate & \multicolumn{2}{c}{$3\times10^{-5}$, one-cycle} \\
  weight decay & \multicolumn{2}{c}{0} \\
  gradient clipping & \multicolumn{2}{c}{norm 2} \\
  EMA & \multicolumn{2}{c}{inverse, power $2/3$, maximum 0.9999} \\
\midrule
\multicolumn{3}{l}{\textit{Sampling}} \\
  NFE & \multicolumn{2}{c}{1} \\
\bottomrule
\end{tabular}
\end{table}

\paragraph{Training target.}
The \LMF is trained on the posterior mean $\bm{\mu}$ instead of a sample from the
posterior, rescaled to unit variance by the spread of $\bm{\mu}$ across showers. Before decompression, a generated latent is rescaled back and
Gaussian noise of the posterior width $\bm{\sigma}$ added, so that the \VAEdec
receives inputs distributed as in training. We train on the mean because, at a
Kullback--Leibler weight of $10^{-2}$, a single posterior draw is dominated by noise, and
a model trained on such draws learns mostly that noise (appendix~\ref{sec:applda}).

Since the tokens of the latent come from ordered queries with a positional encoding
in the \VAEenc, and the \VAEdec uses this order, the network has a learnable,
zero-initialised positional embedding over the token index. Without it, the generated
tokens have the correct marginal distribution but no ordering, and the \VAEdec cannot
decode them correctly.

\paragraph{Parametrisation.}
Along the path $\mathbf{z}_t = (1-t)\,\mathbf{x} + t\,\bm{\epsilon}$ the model
predicts the average velocity over $[r, t]$,
$\mathbf{u}(\mathbf{z}_t, r, t) = \frac{1}{t-r}\int_r^t \mathbf{v}(\mathbf{z}_\tau, \tau)\,\mathrm{d}\tau$,
so that a single evaluation at $r = 0$, $t = 1$ carries a sample the whole way,
$\mathbf{x} = \bm{\epsilon} - \mathbf{u}_\theta(\bm{\epsilon}, 0, 1)$. Instead of the
integral, training uses the mean-flow identity
$\mathbf{u} = \mathbf{v}(\mathbf{z}_t, t) - (t - r)\,\mathrm{d}\mathbf{u}/\mathrm{d}t$,
where $\mathbf{v}(\mathbf{z}_t, t)$ is the instantaneous velocity at time $t$; the
network output is regressed onto this target with the gradient stopped through it.
Two further choices help single-step sampling. The
average velocity of the corresponding Gaussian path, which is known in closed form, is
added to the output, so that the network only predicts the non-Gaussian residual. In
addition, the output head is zero-initialised, so that training starts from the
Gaussian solution.

Half of each batch has $r = t$, where the term in $\mathrm{d}\mathbf{u}/\mathrm{d}t$
vanishes and the objective reduces to ordinary flow matching onto the known velocity
$\mathbf{v}$. For the other half $r < t$, drawn uniformly, and the target, which is built from the
network's own predictions, extends the training to larger gaps. Since large gaps are rare,
the case $r = 0$, $t = 1$ used for sampling is learned mostly indirectly. The derivative along the path is
taken by finite differences with a step of $10^{-2}$, and samples are weighted by
$w = (\lVert\Delta\rVert^2 + c)^{-p}$ with $p = 1$ and $c = 10^{-3}$, detached
from the gradient.

\paragraph{Training.}
The network is an evolution of the one used for the ECal diffusion of
\cc~\cite{Buss:2025cyw}, acting on the $d_s$ latent vectors rather than on the
points of a shower. It is optimised with Ranger at a peak learning rate of
$3\times10^{-5}$ on a one-cycle schedule, without weight decay, with the gradient
norm clipped at 2, dropout 0.1 and an exponential moving average of the weights kept for
evaluation. Sampling uses a single network evaluation. The standard \LMF is sampled from the
checkpoint at the end of training. For the large \LMF this is not possible, since late in
training its output at $r = 0$, $t = 1$, the flow-time pair used for sampling, grows and the
generated showers deviate from \textsc{Geant4} in the energy sum, the number of hits and
the transverse centre of gravity. At each latent size, \num{10000} showers at 20, 50 and
\SI{80}{\giga\electronvolt} are therefore generated from the checkpoints at 566k, 850k,
1.2M iterations and the end of training, and the one closest to \textsc{Geant4} in the
geometric mean of both distances over the observables is used: 566k iterations at
$d_s = 128$, 1.2M at $d_s = 256$ and 850k at $d_s = 512$.

\section{Distances to \textsc{Geant4} for every observable}\label{sec:appds}
Table~\ref{tab:ds_metrics} gives, for every observable, the distances to \textsc{Geant4}
whose geometric means are summarised in table~\ref{tab:ds_metrics_avg} and discussed in
section~\ref{sec:observables}.

All observables are computed after the projection onto the readout cells, from the cells
with a deposit; $x$ and $z$ are the transverse coordinates and $y$ the depth.
\begin{itemize}
    \item[--] \textit{Energy sum}: total energy deposited in the calorimeter per shower.
    \item[--] \textit{ECal and HCal energy sum}: energy deposited per shower in the ECal and in the HCal.
    \item[--] \textit{Number of hits}: number of cells with a deposit per shower.
    \item[--] \textit{Centre of gravity $x$, $y$, $z$}: energy-weighted mean position of the cells
          of a shower along each coordinate; $y$ measures the shower depth, $x$ and $z$ its
          transverse position.
    \item[--] \textit{Cell energy}: spectrum of the cell energies, on a logarithmic scale.
    \item[--] \textit{Longitudinal profile}: energy-weighted distribution of the layer index of the cells.
    \item[--] \textit{Radial profile ECal and HCal}: energy-weighted distribution of the transverse
          distance of the cells from the shower axis, separately in the ECal and the HCal,
          whose cells have different sizes.
    \item[--] \textit{Velocity}: distribution of the apparent velocity of the cells, the distance
          from the particle gun to the cell divided by its arrival time.
    \item[--] \textit{Longitudinal velocity}: energy-weighted distribution of the depth of the cell
          layer, measured from the gun, divided by the arrival time.
    \item[--] \textit{Radial velocity}: energy-weighted distribution of the transverse distance of
          the cell from the shower axis divided by the arrival time.
\end{itemize}

\begin{widetable}[t]
    \centering
    \caption{Distance between each latent size and \textsc{Geant4} for every
    observable, on the full-range samples of \num{100000} showers: the Wasserstein
    distance normalised to the \textsc{Geant4} standard deviation (top) and the
    Jensen--Shannon divergence (bottom), mean and spread over five disjoint
    subsets. The best model per observable is in bold,
    and each block ends with the geometric mean over the observables of the entries
    of every column, in the units of the block; its error is the standard deviation
    of that mean over 1000 resamplings of every entry within its uncertainty. The last
    column is the \textsc{Geant4} floor.}
    \label{tab:ds_metrics}
    \small
    \setlength{\tabcolsep}{4pt}
    \renewcommand{\arraystretch}{0.85}  
    \resizebox{\linewidth}{!}{
\begin{tabular}{llccccccccc}
\toprule
& & & & & & & \multicolumn{3}{c}{large \LMF} & \\
\cmidrule(lr){8-10}
& & $d_s = 64$ & $d_s = 128$ & $d_s = 256$ & $d_s = 512$ & $d_s = 1024$ & $d_s = 128$ & $d_s = 256$ & $d_s = 512$ & \textsc{Geant4} \\
\midrule
\multirow{15}{*}{\rotatebox[origin=c]{90}{$\mathrm{WD}/\sigma_{\textsc{Geant4}}$ [$\times 10^{-2}$]}} & Energy sum & $1.2 \pm 0.2$ & $1.2 \pm 0.3$ & $1.3 \pm 0.2$ & $1.5 \pm 0.4$ & $2.5 \pm 0.7$ & $1.7 \pm 0.5$ & $\mathbf{1.1 \pm 0.3}$ & $2.6 \pm 0.5$ & $0.8 \pm 0.1$ \\
& ECal energy sum & $1.5 \pm 0.5$ & $1.3 \pm 0.3$ & $1.3 \pm 0.2$ & $1.5 \pm 0.5$ & $1.2 \pm 0.3$ & $2.1 \pm 0.3$ & $\mathbf{1.1 \pm 0.3}$ & $1.4 \pm 0.4$ & $1.0 \pm 0.1$ \\
& HCal energy sum & $2.3 \pm 0.3$ & $2.2 \pm 0.4$ & $1.9 \pm 0.2$ & $1.9 \pm 0.4$ & $7.5 \pm 0.2$ & $1.9 \pm 0.3$ & $\mathbf{1.0 \pm 0.1}$ & $2.5 \pm 0.7$ & $1.5 \pm 0.6$ \\
& Number of hits & $16.7 \pm 0.9$ & $13.7 \pm 0.8$ & $15.8 \pm 0.8$ & $23.4 \pm 0.7$ & $23.5 \pm 0.3$ & $\mathbf{13.6 \pm 0.6}$ & $16.4 \pm 0.5$ & $22.0 \pm 0.5$ & $1.2 \pm 0.3$ \\
& Centre of gravity $x$ & $20.7 \pm 0.3$ & $9.3 \pm 0.2$ & $14.2 \pm 1.1$ & $12.8 \pm 0.3$ & $19.9 \pm 0.6$ & $8.6 \pm 0.9$ & $\mathbf{6.2 \pm 1.3}$ & $6.5 \pm 0.6$ & $1.4 \pm 0.3$ \\
& Centre of gravity $y$ & $3.2 \pm 0.7$ & $1.9 \pm 0.7$ & $2.0 \pm 0.6$ & $1.8 \pm 0.5$ & $4.9 \pm 0.3$ & $2.6 \pm 1.2$ & $\mathbf{1.6 \pm 0.8}$ & $1.7 \pm 0.5$ & $1.2 \pm 0.3$ \\
& Centre of gravity $z$ & $15.6 \pm 0.8$ & $15.3 \pm 0.4$ & $12.8 \pm 0.3$ & $15.0 \pm 0.4$ & $22.5 \pm 0.5$ & $14.4 \pm 0.5$ & $\mathbf{3.5 \pm 0.7}$ & $7.0 \pm 0.7$ & $1.3 \pm 0.3$ \\
& Cell energy & $\mathbf{2.74 \pm 0.03}$ & $3.44 \pm 0.01$ & $3.96 \pm 0.02$ & $4.80 \pm 0.01$ & $5.97 \pm 0.02$ & $2.903 \pm 0.005$ & $3.97 \pm 0.02$ & $4.36 \pm 0.01$ & $0.05 \pm 0.03$ \\
& Longitudinal profile & $1.5 \pm 0.1$ & $1.6 \pm 0.1$ & $1.8 \pm 0.1$ & $1.36 \pm 0.02$ & $2.32 \pm 0.04$ & $2.5 \pm 0.1$ & $1.77 \pm 0.01$ & $\mathbf{1.10 \pm 0.05}$ & $0.1 \pm 0.1$ \\
& Radial profile ECal & $6.1 \pm 0.1$ & $6.9 \pm 0.1$ & $9.6 \pm 0.1$ & $7.7 \pm 0.1$ & $9.4 \pm 0.1$ & $6.9 \pm 0.1$ & $6.1 \pm 0.1$ & $\mathbf{4.5 \pm 0.1}$ & $0.2 \pm 0.2$ \\
& Radial profile HCal & $3.8 \pm 0.1$ & $3.7 \pm 0.1$ & $4.2 \pm 0.1$ & $5.45 \pm 0.03$ & $9.66 \pm 0.04$ & $3.55 \pm 0.02$ & $\mathbf{2.08 \pm 0.03}$ & $5.91 \pm 0.05$ & $0.1 \pm 0.1$ \\
& Velocity & $2.08 \pm 0.03$ & $1.27 \pm 0.01$ & $\mathbf{1.09 \pm 0.01}$ & $4.92 \pm 0.03$ & $9.64 \pm 0.04$ & $2.31 \pm 0.03$ & $1.21 \pm 0.01$ & $2.79 \pm 0.01$ & $0.03 \pm 0.01$ \\
& Longitudinal velocity & $5.20 \pm 0.03$ & $2.30 \pm 0.01$ & $1.96 \pm 0.04$ & $\mathbf{1.55 \pm 0.01}$ & $3.22 \pm 0.03$ & $2.8 \pm 0.1$ & $4.12 \pm 0.05$ & $3.48 \pm 0.05$ & $0.05 \pm 0.02$ \\
& Radial velocity & $3.56 \pm 0.03$ & $5.47 \pm 0.05$ & $6.66 \pm 0.04$ & $\mathbf{2.9 \pm 0.1}$ & $6.24 \pm 0.04$ & $5.1 \pm 0.1$ & $3.8 \pm 0.1$ & $3.42 \pm 0.04$ & $0.10 \pm 0.03$ \\
\cmidrule(l){3-11}
& Geometric mean & $3.98 \pm 0.18$ & $3.41 \pm 0.18$ & $3.65 \pm 0.11$ & $3.94 \pm 0.25$ & $6.56 \pm 0.22$ & $3.88 \pm 0.51$ & $\mathbf{2.72 \pm 0.36}$ & $3.61 \pm 0.19$ & $0.31 \pm 0.09$ \\
\midrule
\multirow{15}{*}{\rotatebox[origin=c]{90}{JSD [$\times 10^{-4}$]}} & Energy sum & $3.6 \pm 0.7$ & $4.2 \pm 0.8$ & $3.8 \pm 0.7$ & $4.9 \pm 2.0$ & $10.4 \pm 1.4$ & $3.8 \pm 0.6$ & $\mathbf{2.9 \pm 0.8}$ & $4.4 \pm 0.9$ & $2.6 \pm 0.5$ \\
& ECal energy sum & $\mathbf{12.8 \pm 1.7}$ & $56.4 \pm 4.1$ & $21.8 \pm 2.4$ & $19.2 \pm 1.5$ & $18.0 \pm 2.4$ & $51.3 \pm 5.4$ & $16.8 \pm 1.9$ & $19.9 \pm 2.3$ & $2.6 \pm 0.5$ \\
& HCal energy sum & $5.5 \pm 0.9$ & $5.0 \pm 0.2$ & $6.5 \pm 0.7$ & $6.0 \pm 0.5$ & $63.8 \pm 3.0$ & $5.0 \pm 0.7$ & $\mathbf{3.5 \pm 0.9}$ & $8.3 \pm 2.7$ & $3.3 \pm 0.8$ \\
& Number of hits & $40.0 \pm 2.8$ & $\mathbf{27.6 \pm 3.4}$ & $33.4 \pm 3.4$ & $69.1 \pm 4.8$ & $71.4 \pm 2.5$ & $\mathbf{27.6 \pm 3.8}$ & $36.2 \pm 3.4$ & $61.9 \pm 4.5$ & $3.0 \pm 0.8$ \\
& Centre of gravity $x$ & $189.3 \pm 5.0$ & $27.2 \pm 1.8$ & $52.5 \pm 6.6$ & $71.6 \pm 3.7$ & $202.5 \pm 11.9$ & $22.8 \pm 2.6$ & $\mathbf{8.5 \pm 2.6}$ & $12.0 \pm 1.7$ & $2.9 \pm 0.5$ \\
& Centre of gravity $y$ & $5.2 \pm 0.6$ & $\mathbf{3.6 \pm 0.8}$ & $3.8 \pm 1.4$ & $3.8 \pm 1.2$ & $17.1 \pm 2.0$ & $4.8 \pm 1.8$ & $3.9 \pm 0.8$ & $\mathbf{3.6 \pm 0.5}$ & $2.7 \pm 0.4$ \\
& Centre of gravity $z$ & $89.0 \pm 3.5$ & $61.4 \pm 3.9$ & $54.7 \pm 2.9$ & $79.3 \pm 3.4$ & $175.1 \pm 5.2$ & $51.0 \pm 4.1$ & $\mathbf{9.5 \pm 3.1}$ & $15.0 \pm 2.6$ & $3.8 \pm 1.4$ \\
& Cell energy & $4.5 \pm 0.1$ & $5.06 \pm 0.04$ & $7.4 \pm 0.1$ & $9.10 \pm 0.05$ & $8.9 \pm 0.1$ & $\mathbf{4.35 \pm 0.01}$ & $7.8 \pm 0.1$ & $8.20 \pm 0.04$ & $0.004 \pm 0.001$ \\
& Longitudinal profile & $1.0 \pm 0.1$ & $0.9 \pm 0.1$ & $1.3 \pm 0.1$ & $0.75 \pm 0.02$ & $1.8 \pm 0.1$ & $1.6 \pm 0.1$ & $1.07 \pm 0.04$ & $\mathbf{0.43 \pm 0.03}$ & $0.04 \pm 0.01$ \\
& Radial profile ECal & $55.0 \pm 0.6$ & $44.3 \pm 0.6$ & $34.1 \pm 0.4$ & $21.2 \pm 0.4$ & $25.3 \pm 0.3$ & $45.7 \pm 0.4$ & $20.3 \pm 0.5$ & $\mathbf{14.7 \pm 0.3}$ & $0.02 \pm 0.01$ \\
& Radial profile HCal & $4.0 \pm 0.1$ & $7.2 \pm 0.2$ & $5.5 \pm 0.2$ & $11.4 \pm 0.1$ & $41.2 \pm 0.3$ & $6.9 \pm 0.1$ & $\mathbf{2.9 \pm 0.1}$ & $13.2 \pm 0.2$ & $0.01 \pm 0.01$ \\
& Velocity & $91.7 \pm 0.2$ & $63.8 \pm 0.2$ & $\mathbf{55.8 \pm 0.2}$ & $66.4 \pm 0.1$ & $75.3 \pm 0.1$ & $66.8 \pm 0.2$ & $56.9 \pm 0.2$ & $63.0 \pm 0.1$ & $0.0017 \pm 0.0004$ \\
& Longitudinal velocity & $3.47 \pm 0.03$ & $0.96 \pm 0.01$ & $1.34 \pm 0.03$ & $\mathbf{0.77 \pm 0.02}$ & $1.94 \pm 0.04$ & $1.4 \pm 0.1$ & $3.3 \pm 0.1$ & $2.7 \pm 0.1$ & $0.005 \pm 0.002$ \\
& Radial velocity & $11.0 \pm 0.2$ & $8.2 \pm 0.1$ & $9.6 \pm 0.1$ & $\mathbf{4.5 \pm 0.1}$ & $14.0 \pm 0.2$ & $7.9 \pm 0.2$ & $5.5 \pm 0.1$ & $4.9 \pm 0.1$ & $0.04 \pm 0.01$ \\
\cmidrule(l){3-11}
& Geometric mean & $12.77 \pm 0.32$ & $10.22 \pm 0.27$ & $10.66 \pm 0.59$ & $10.87 \pm 1.10$ & $23.68 \pm 0.42$ & $10.65 \pm 0.73$ & $\mathbf{7.20 \pm 0.41}$ & $8.77 \pm 0.35$ & $0.17 \pm 0.04$ \\
\bottomrule
\end{tabular}
}
\end{widetable}

\section{Information content of the latent space}\label{sec:applda}
\begin{figure}[htbp]
\centering
\includegraphics[width=\ldafigwidth]{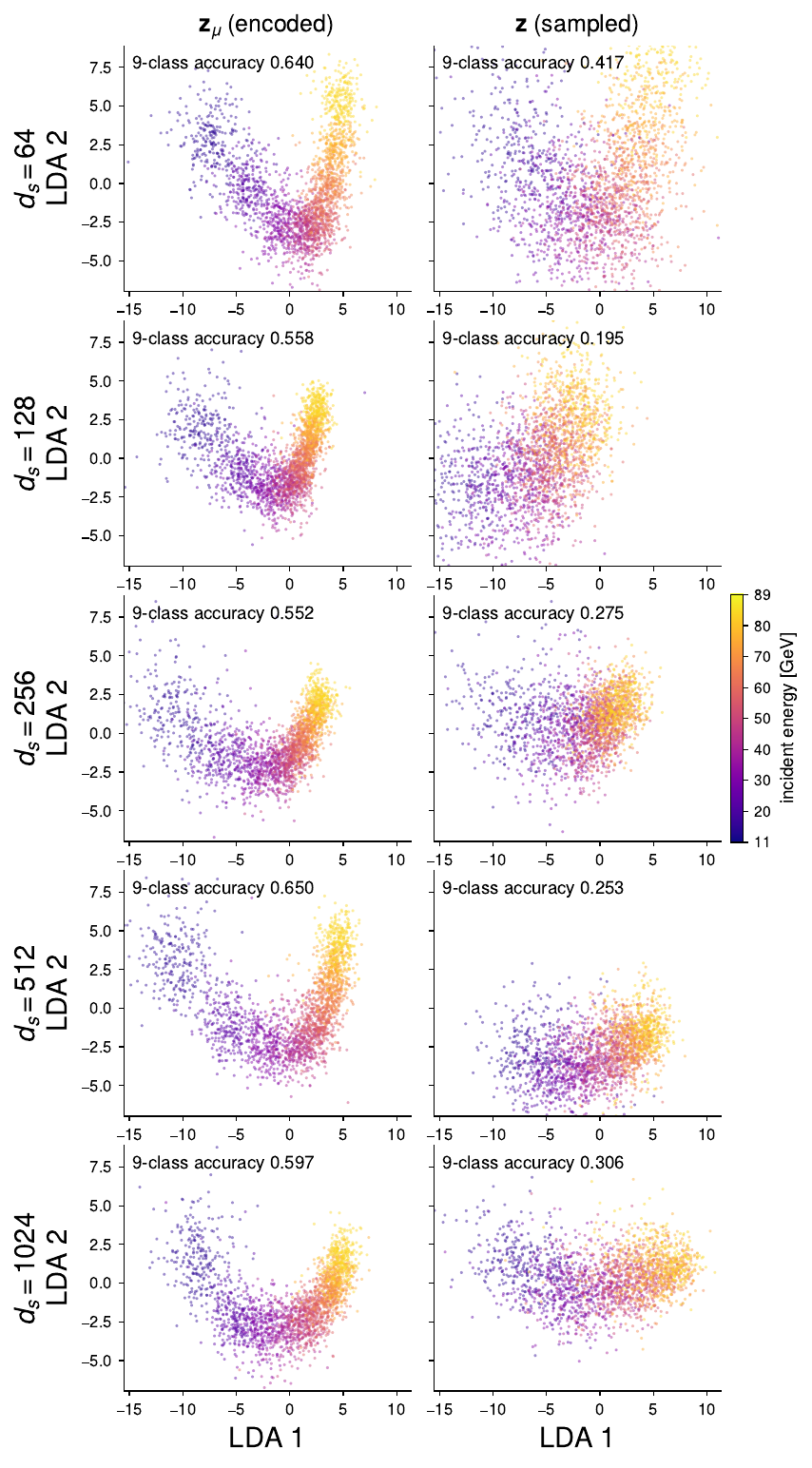}
\caption{Linear discriminant analysis of the latent representation, coloured by
incident energy, for the posterior mean $\mathbf{z}_{\mu}$ (left) and the latent
$\mathbf{z}$ generated by the \LMF (right), at $d_s = 64$, 128, 256, 512 and
1024 (top to bottom). The annotation in each panel is the nine-class accuracy of the
corresponding classifier, against $1/9 \approx 0.11$ for chance.}
\label{fig:lda_energy}
\end{figure}

The \textsc{VAE} is trained with a Kullback--Leibler term that pulls the
approximate posterior towards the prior, so a single draw from that posterior
carries appreciably less information about the shower than the posterior mean
does, which is why the \LMF is trained on the posterior mean
$\mathbf{z}_{\mu}$ (appendix~\ref{sec:applmf}). The question is then how much of
the information in $\mathbf{z}_{\mu}$ the \LMF reproduces, and how this depends on the
latent size, which is studied in figure~\ref{fig:lda_energy}. For each shower the latent representation is projected onto
the two most discriminant directions found by a linear discriminant analysis (LDA)
fitted to separate nine classes in incident energy. The left column shows the posterior
mean $\mathbf{z}_{\mu}$, the target of the \LMF; the right column a latent
$\mathbf{z}$ generated by the \LMF for the same conditioning; the rows are
the latent sizes $d_s = 64$, 128, 256, 512 and 1024. At every latent size the showers are split at random into two
sets of equal size. The LDA is fitted on the posterior means of the first set only,
and the same projection, without refitting, is applied to the second set: to its
posterior means in the left column and to the latents generated for the same showers in
the right column. The number quoted in
each panel is the nine-class accuracy of the LDA on that set, to be compared with
the $1/9 \approx 0.11$ expected from chance. The sample is a balanced grid of 4500
distinct showers, 500 for each of the nine energy labels, the same showers for
every latent size. The same analysis with nine classes in the number of points per
shower instead of the incident energy gives similar results, since the number of
points grows with the incident energy.

We use a supervised projection on purpose. Principal component analysis orders
directions by variance alone, and in a latent dominated by the per-shower noise
of the posterior the directions of largest variance are not those along which
the incident energy varies, so the classes overlap in the leading principal
components. A $t$-SNE embedding preserves local neighbourhoods but not the
distances between groups, so the separation between energies is hard to interpret and
depends on the chosen settings. LDA directly measures how much of the class information
is linearly accessible, and quantifies it with an accuracy. Since the projection is
fitted using the labels, the separation it shows is an upper bound on what an
unsupervised method would find, so only the comparison between the panels is
meaningful, not the absolute separation in a single panel.

At every latent size the posterior mean separates the classes well, with energy
accuracies of $0.640$, $0.558$, $0.552$, $0.650$ and $0.597$ for $d_s = 64$, 128,
256, 512 and 1024; its projection traces a single curved trajectory along which
the energy increases monotonically. The generated latents keep that ordering,
with low energies on one side and high ones on the other, but spread each class
over a wider region, and the accuracies drop to $0.417$, $0.195$, $0.275$, $0.253$
and $0.306$. The \LMF reproduces the largest fraction of the
information at $d_s = 64$, with $65\%$ of the posterior-mean accuracy, while for the
larger sizes the fraction lies between $35\%$ and $51\%$ without a clear trend.

\begin{figure}[!ht]
\centering
\includegraphics[width=\ldabigfigwidth]{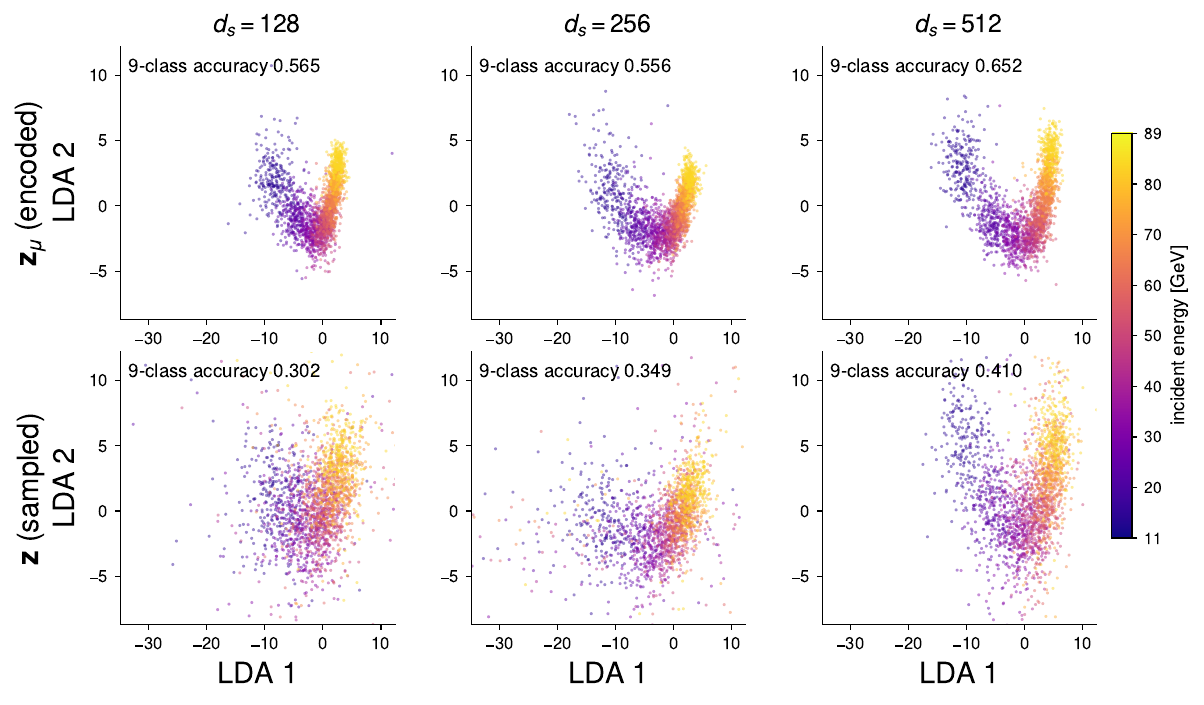}
\caption{As figure~\ref{fig:lda_energy}, for the large \LMF at
$d_s = 128$, 256 and 512 (left to right).}
\label{fig:lda_biglmf}
\end{figure}

Figure~\ref{fig:lda_biglmf} repeats the analysis for the large \LMF at $d_s = 128$,
256 and 512, on
the same balanced grid of showers.
The posterior mean comes from the same \VAEenc, and its accuracies of
$0.565$, $0.556$ and $0.652$ agree with figure~\ref{fig:lda_energy} to within
$0.02$. The generated
latents reach $0.302$, $0.349$ and $0.410$, between $53\%$ and $63\%$ of the
posterior-mean accuracy, against $35\%$ to $50\%$ for the \LMF of
figure~\ref{fig:lda_energy} at the same latent sizes; with the number of points
instead of the incident energy the accuracies are again similar.
At every latent size the large \LMF retains more of the energy information than the
standard one; the retained fraction saturates from $d_s = 256$ on.

\section{Reconstruction by the \textsc{VAE}}\label{sec:appreco}

A generated shower passes through two learned stages, the \LMF and the \textsc{VAE},
and its distance to \textsc{Geant4} in section~\ref{sec:observables} mixes the
errors of both. To isolate the contribution of the \textsc{VAE}, \num{20000} of the
\textsc{Geant4} showers of figure~\ref{fig:obs_ds} are passed through the
\textsc{VAE} of each latent size: every shower is compressed once by the \VAEenc,
a single latent is sampled from the resulting posterior, and this latent is
decompressed.
Neither \layerFM nor the \LMF takes part. The reconstructed showers are not
calibrated in energy: unlike the generated ones, whose incident energy is scaled
by the factor $f(E)$ of section~\ref{sec:model}, they are conditioned on the
true incident energy.
The reconstructed showers are projected onto the readout cells and corrected for
the time floor exactly as the generated ones, and compared with the same
\num{20000} \textsc{Geant4} showers they were reconstructed from.
Figure~\ref{fig:obs_reco} shows the four observables of figure~\ref{fig:obs_ds},
and table~\ref{tab:ds_reco_metrics} the distances of table~\ref{tab:ds_metrics},
computed in the same way.

\begin{widefigure}[!ht]
    \centering
    \includegraphics[width=0.7\linewidth]{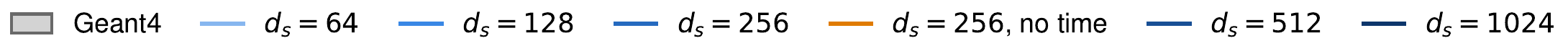}\\[2pt]
    \includegraphics[width=0.245\linewidth]{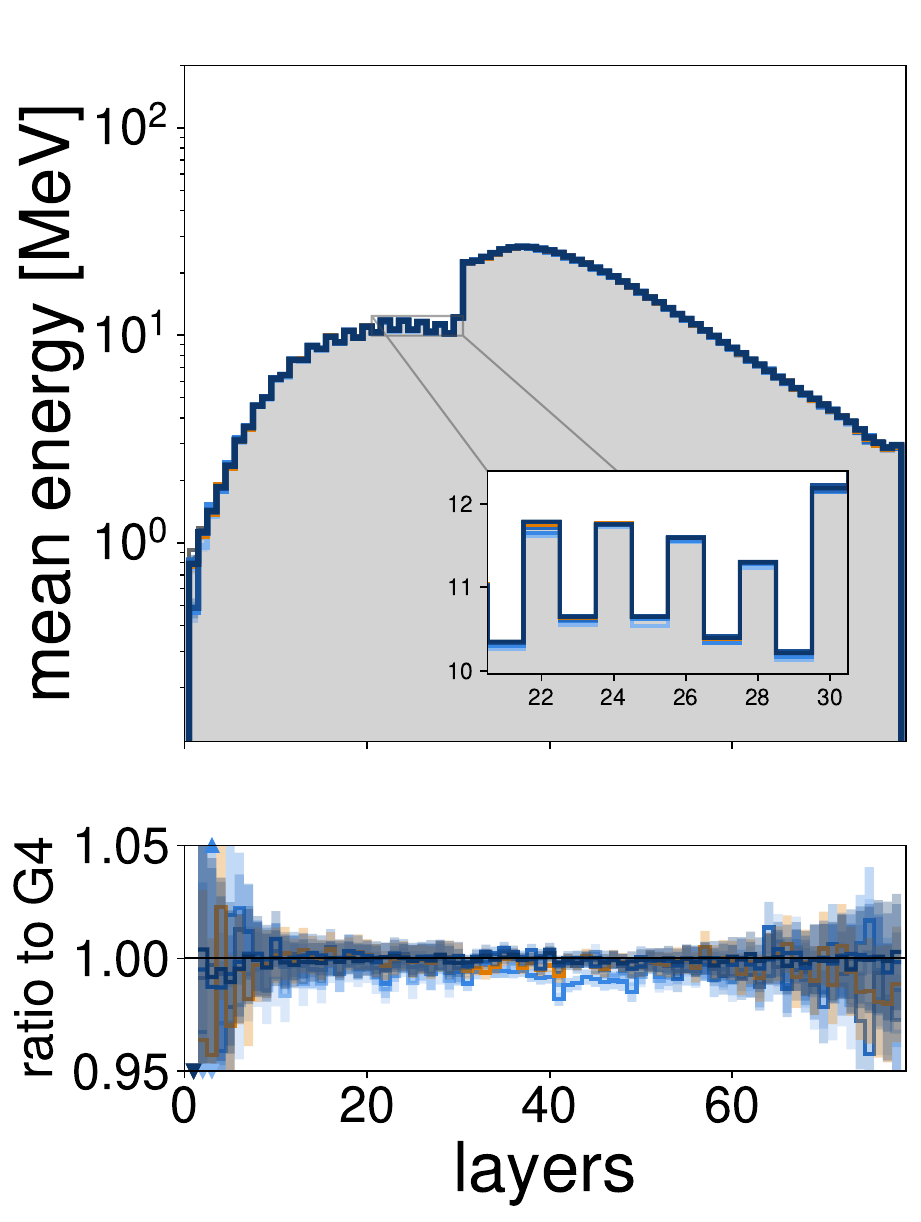}\hfill
    \includegraphics[width=0.245\linewidth]{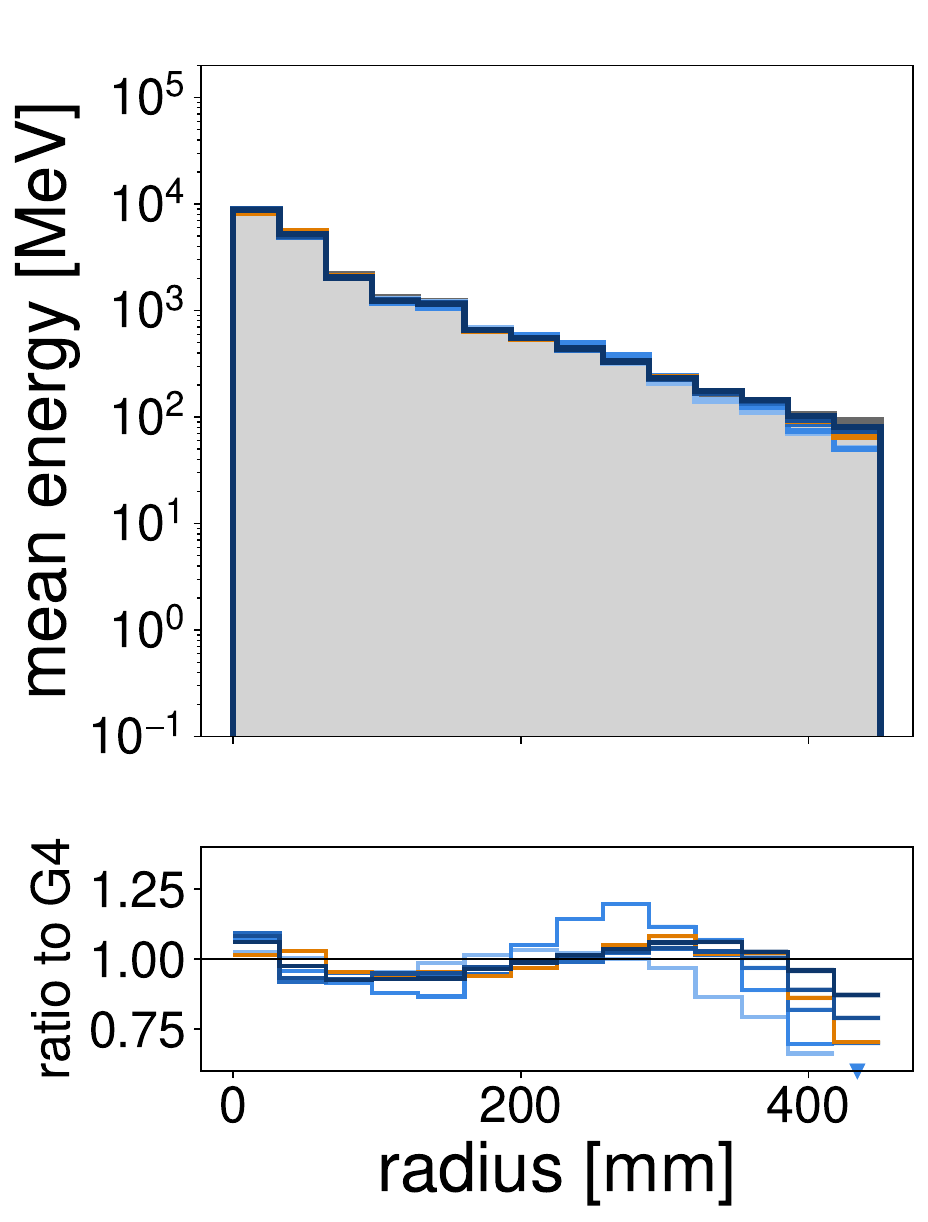}\hfill
    \includegraphics[width=0.245\linewidth]{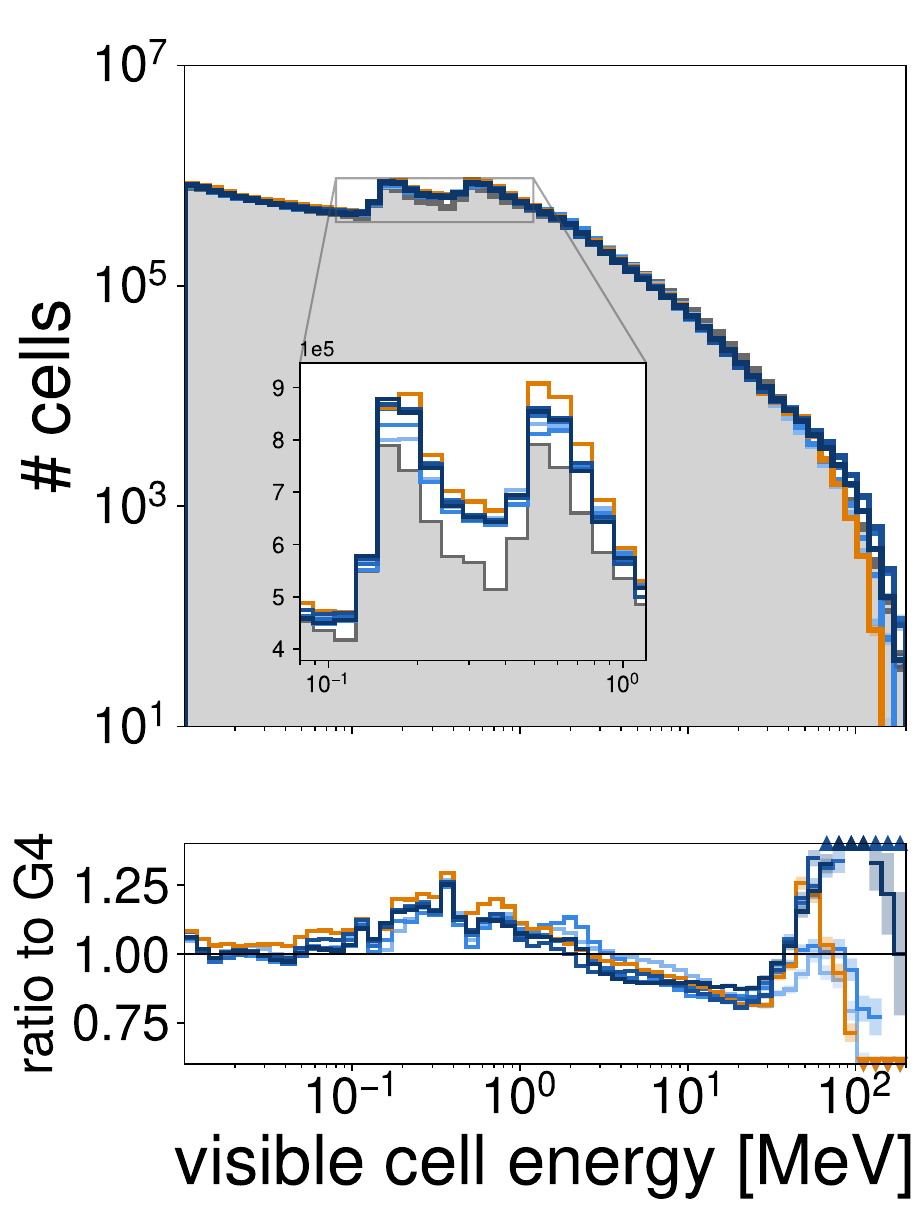}\hfill
    \includegraphics[width=0.245\linewidth]{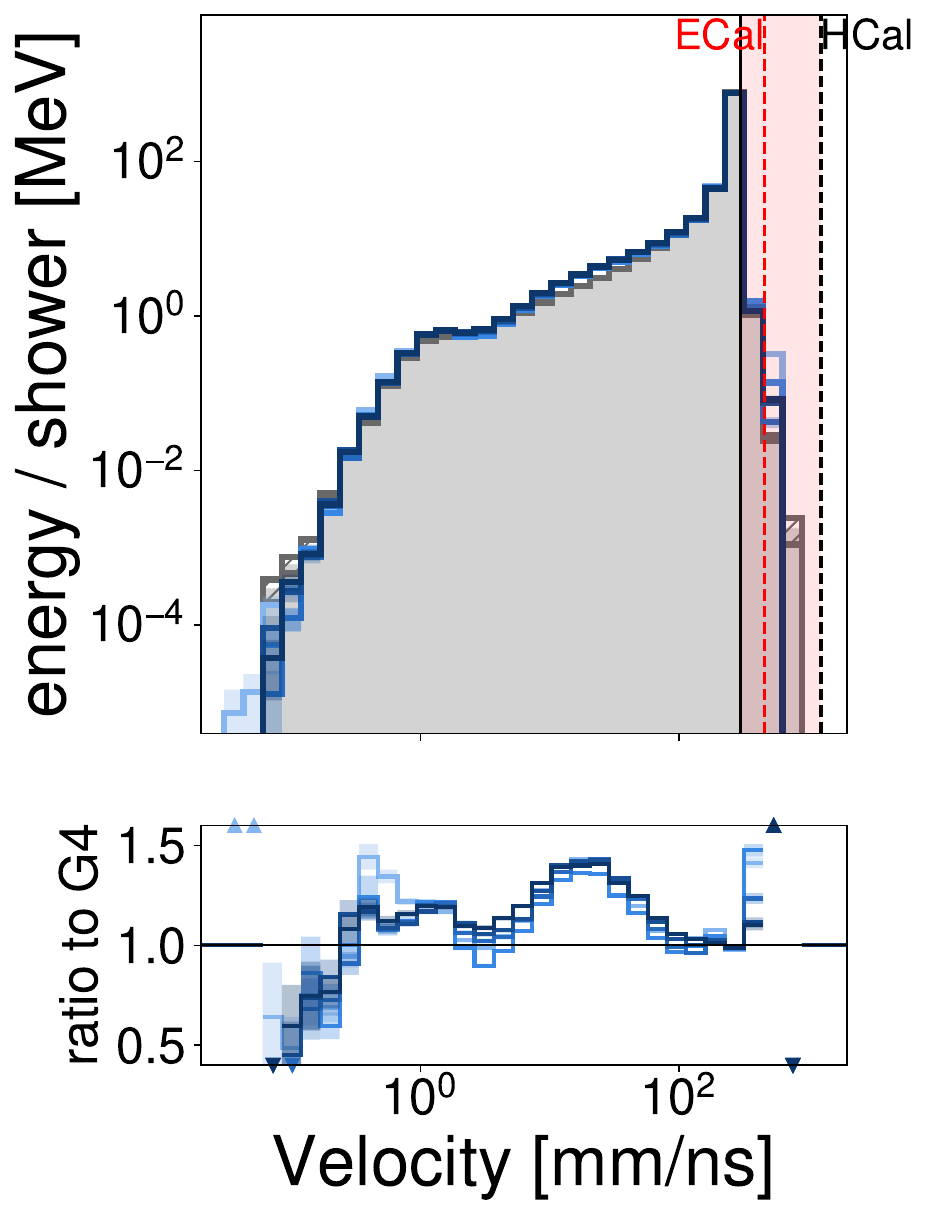}
    \caption{As figure~\ref{fig:obs_ds}, for \num{20000} \textsc{Geant4} showers
    (grey) and the same showers compressed and decompressed by the \textsc{VAE} of each
    latent size, without \layerFM, the \LMF and energy calibration. The orange line is
    the \textsc{VAE} at $d_s = 256$ trained without the hit time, absent from the
    velocity panel. From left to
    right: mean energy per layer, mean energy per shower in bins of the distance
    from the shower axis, the energy spectrum of the cells, and the energy per
    shower in bins of the apparent velocity of the cells.}
    \label{fig:obs_reco}
\end{widefigure}

The longitudinal profile is reproduced at every latent size, with distances at
the \textsc{Geant4} floor, so given the number of points in each layer the
\textsc{VAE} distributes the energy along the shower correctly. The energy sum, too,
is at the floor for all five \textsc{VAE}s in the Wasserstein distance, and only
$d_s = 64$ lies slightly above it in the Jensen--Shannon divergence. The radial profile shows the same features as
after generation, but smaller, with a slight excess in the core, a slight deficit just
outside it and a deficit far from the axis. In the Wasserstein distance it is smallest for
$d_s = 1024$ in both sections; in the Jensen--Shannon divergence, which weighs the shape
of the whole profile rather than the shift of its mean, the ECal profile improves
steadily from $d_s = 64$ (55 [$\times 10^{-4}$]) to $d_s = 1024$ (19). The \textsc{VAE} trained without the hit time
reconstructs the radial profile best of all in both sections. As after generation, the cell-energy spectrum has
too many low-energy cells. The velocity distribution is again nearly the same
for all latent sizes in the Wasserstein distance, while in the Jensen--Shannon
divergence $d_s = 64$ is clearly the furthest from \textsc{Geant4}. Weighted by energy
(figure~\ref{fig:obs_reco}, right), the prompt front above \SI{100}{\milli\meter\per\nano\second},
\SI{95}{\percent} of the energy, is reproduced to within \SI{1}{\percent} at every size,
while between a few tenths of a \si{\milli\meter\per\nano\second} and
\SI{100}{\milli\meter\per\nano\second} the reconstruction deposits \SIrange{7}{21}{\percent}
too much energy, the same excess as the generated showers of figure~\ref{fig:obs_ds}. Part of these differences may come from the
projection onto the readout cells and not from the \textsc{VAE} itself. Points that lie
slightly apart are counted in separate cells, so a small
spread of the reconstructed points already changes the number of cells, their
energies and their times.

\begin{widetable}[!ht]
    \centering
    \caption{As table~\ref{tab:ds_metrics}, for \num{20000} \textsc{Geant4} showers
    compressed and decompressed by the \textsc{VAE} of each latent size and by the
    \textsc{VAE} at $d_s = 256$ trained without the hit time, without energy calibration,
    compared with the same \textsc{Geant4} showers. The \textsc{VAE} without time has no
    entry for the three velocities, and its geometric mean, marked $\dagger$, is over the
    other eleven observables and not compared with the others. The best
    latent size per observable is in bold, and each block ends with the geometric
    mean over the observables of the entries of every column, with its error the
    standard deviation over 1000 resamplings of the entries within their
    uncertainties. The last column is
    \textsc{Geant4} against itself on subsets of the same size.}
    \label{tab:ds_reco_metrics}
    \small
    \setlength{\tabcolsep}{4pt}
    \renewcommand{\arraystretch}{0.85}  
    \resizebox{\linewidth}{!}{
\begin{tabular}{llccccccc}
\toprule
& & $d_s = 64$ & $d_s = 128$ & $d_s = 256$ & $d_s = 512$ & $d_s = 1024$ & $d_s = 256$, no time & \textsc{Geant4} \\
\midrule
\multirow{15}{*}{\rotatebox[origin=c]{90}{$\mathrm{WD}/\sigma_{\textsc{Geant4}}$ [$\times 10^{-2}$]}} & Energy sum & $2.7 \pm 1.3$ & $2.8 \pm 1.2$ & $\mathbf{1.9 \pm 1.0}$ & $2.1 \pm 0.7$ & $2.1 \pm 0.7$ & $2.1 \pm 0.7$ & $2.4 \pm 0.9$ \\
& ECal energy sum & $2.4 \pm 1.2$ & $\mathbf{1.6 \pm 0.5}$ & $3.1 \pm 0.2$ & $2.9 \pm 1.3$ & $2.6 \pm 1.0$ & $2.1 \pm 0.7$ & $2.8 \pm 0.7$ \\
& HCal energy sum & $3.7 \pm 1.3$ & $3.1 \pm 1.2$ & $2.3 \pm 0.7$ & $3.1 \pm 1.2$ & $\mathbf{2.2 \pm 0.8}$ & $3.1 \pm 1.2$ & $2.6 \pm 0.9$ \\
& Number of hits & $13.2 \pm 1.9$ & $\mathbf{11.5 \pm 1.2}$ & $11.9 \pm 0.7$ & $12.9 \pm 2.1$ & $12.4 \pm 4.0$ & $20.5 \pm 1.8$ & $2.2 \pm 0.7$ \\
& Centre of gravity $x$ & $20.0 \pm 0.7$ & $3.6 \pm 0.7$ & $3.6 \pm 0.9$ & $3.4 \pm 1.4$ & $\mathbf{3.0 \pm 0.7}$ & $4.7 \pm 2.0$ & $3.9 \pm 0.7$ \\
& Centre of gravity $y$ & $3.3 \pm 1.4$ & $2.8 \pm 0.3$ & $\mathbf{2.1 \pm 0.3}$ & $\mathbf{2.1 \pm 0.7}$ & $3.0 \pm 0.7$ & $2.8 \pm 1.4$ & $2.7 \pm 1.1$ \\
& Centre of gravity $z$ & $12.2 \pm 0.8$ & $6.1 \pm 2.1$ & $3.8 \pm 0.7$ & $3.3 \pm 1.0$ & $3.3 \pm 0.9$ & $\mathbf{3.2 \pm 0.9}$ & $3.6 \pm 0.4$ \\
& Cell energy & $\mathbf{3.14 \pm 0.01}$ & $3.24 \pm 0.05$ & $3.4 \pm 0.1$ & $3.45 \pm 0.02$ & $3.36 \pm 0.03$ & $3.47 \pm 0.04$ & $0.1 \pm 0.1$ \\
& Longitudinal profile & $\mathbf{0.34 \pm 0.02}$ & $\mathbf{0.3 \pm 0.1}$ & $\mathbf{0.3 \pm 0.1}$ & $\mathbf{0.3 \pm 0.1}$ & $\mathbf{0.3 \pm 0.1}$ & $\mathbf{0.3 \pm 0.1}$ & $0.2 \pm 0.1$ \\
& Radial profile ECal & $5.9 \pm 0.1$ & $7.4 \pm 0.1$ & $5.9 \pm 0.2$ & $4.8 \pm 0.4$ & $\mathbf{3.9 \pm 0.3}$ & $4.1 \pm 0.1$ & $0.3 \pm 0.1$ \\
& Radial profile HCal & $4.3 \pm 0.3$ & $4.2 \pm 0.1$ & $4.5 \pm 0.2$ & $3.3 \pm 0.2$ & $2.7 \pm 0.2$ & $\mathbf{2.6 \pm 0.2}$ & $0.4 \pm 0.3$ \\
& Velocity & $\mathbf{2.28 \pm 0.04}$ & $2.7 \pm 0.1$ & $2.7 \pm 0.1$ & $2.5 \pm 0.1$ & $2.4 \pm 0.1$ & -- & $0.05 \pm 0.01$ \\
& Longitudinal velocity & $4.0 \pm 0.1$ & $\mathbf{2.7 \pm 0.1}$ & $3.3 \pm 0.2$ & $3.8 \pm 0.2$ & $4.1 \pm 0.1$ & -- & $0.13 \pm 0.04$ \\
& Radial velocity & $\mathbf{3.9 \pm 0.1}$ & $6.9 \pm 0.2$ & $9.4 \pm 0.2$ & $9.1 \pm 0.2$ & $8.2 \pm 0.1$ & -- & $0.21 \pm 0.05$ \\
\cmidrule(l){3-9}
& Geometric mean & $3.99 \pm 0.81$ & $3.22 \pm 0.46$ & $3.16 \pm 0.50$ & $3.11 \pm 0.47$ & $\mathbf{2.93 \pm 0.32}$ & $2.87 \pm 0.61^\dagger$ & $0.68 \pm 0.19$ \\
\midrule
\multirow{15}{*}{\rotatebox[origin=c]{90}{JSD [$\times 10^{-4}$]}} & Energy sum & $18.8 \pm 2.6$ & $14.9 \pm 3.8$ & $15.4 \pm 4.3$ & $12.6 \pm 4.6$ & $\mathbf{11.7 \pm 2.7}$ & $13.9 \pm 4.3$ & $14.1 \pm 3.0$ \\
& ECal energy sum & $90.3 \pm 8.1$ & $155.8 \pm 13.2$ & $99.5 \pm 11.0$ & $92.5 \pm 13.7$ & $\mathbf{58.2 \pm 7.6}$ & $94.4 \pm 8.7$ & $15.1 \pm 5.3$ \\
& HCal energy sum & $19.5 \pm 5.1$ & $15.4 \pm 1.3$ & $14.4 \pm 3.7$ & $14.2 \pm 4.1$ & $\mathbf{13.4 \pm 4.1}$ & $15.1 \pm 5.3$ & $16.3 \pm 2.7$ \\
& Number of hits & $36.9 \pm 10.1$ & $\mathbf{31.6 \pm 6.8}$ & $32.3 \pm 4.6$ & $35.7 \pm 9.0$ & $34.7 \pm 13.7$ & $65.7 \pm 11.3$ & $16.8 \pm 6.2$ \\
& Centre of gravity $x$ & $175.4 \pm 18.4$ & $18.7 \pm 2.3$ & $19.8 \pm 6.2$ & $15.7 \pm 3.1$ & $\mathbf{12.8 \pm 2.3}$ & $16.6 \pm 6.2$ & $15.0 \pm 4.6$ \\
& Centre of gravity $y$ & $13.4 \pm 4.5$ & $15.1 \pm 3.1$ & $\mathbf{10.3 \pm 3.4}$ & $12.0 \pm 3.5$ & $12.8 \pm 2.7$ & $13.9 \pm 3.9$ & $14.4 \pm 4.6$ \\
& Centre of gravity $z$ & $62.1 \pm 7.4$ & $24.1 \pm 9.7$ & $16.0 \pm 6.4$ & $12.2 \pm 2.6$ & $\mathbf{11.3 \pm 3.1}$ & $18.4 \pm 5.5$ & $15.8 \pm 2.2$ \\
& Cell energy & $\mathbf{4.59 \pm 0.04}$ & $5.1 \pm 0.1$ & $6.3 \pm 0.1$ & $6.43 \pm 0.04$ & $6.0 \pm 0.1$ & $6.5 \pm 0.1$ & $0.013 \pm 0.005$ \\
& Longitudinal profile & $0.18 \pm 0.02$ & $\mathbf{0.14 \pm 0.04}$ & $\mathbf{0.13 \pm 0.02}$ & $0.19 \pm 0.02$ & $\mathbf{0.14 \pm 0.04}$ & $\mathbf{0.13 \pm 0.03}$ & $0.21 \pm 0.03$ \\
& Radial profile ECal & $55.1 \pm 1.0$ & $41.1 \pm 2.0$ & $30.0 \pm 0.6$ & $23.3 \pm 1.2$ & $19.2 \pm 0.8$ & $\mathbf{17.0 \pm 0.9}$ & $0.10 \pm 0.04$ \\
& Radial profile HCal & $4.2 \pm 0.3$ & $9.0 \pm 0.3$ & $8.5 \pm 0.9$ & $6.4 \pm 0.8$ & $7.7 \pm 0.7$ & $\mathbf{2.1 \pm 0.2}$ & $0.1 \pm 0.1$ \\
& Velocity & $89.4 \pm 0.3$ & $64.8 \pm 0.2$ & $55.2 \pm 0.2$ & $57.9 \pm 0.2$ & $\mathbf{54.5 \pm 0.1}$ & -- & $0.010 \pm 0.001$ \\
& Longitudinal velocity & $2.0 \pm 0.1$ & $\mathbf{1.2 \pm 0.1}$ & $2.1 \pm 0.1$ & $2.4 \pm 0.1$ & $3.0 \pm 0.1$ & -- & $0.04 \pm 0.02$ \\
& Radial velocity & $\mathbf{10.2 \pm 0.3}$ & $11.0 \pm 0.3$ & $14.5 \pm 0.3$ & $14.4 \pm 0.4$ & $12.1 \pm 0.2$ & -- & $0.2 \pm 0.1$ \\
\cmidrule(l){3-9}
& Geometric mean & $15.81 \pm 0.73$ & $12.28 \pm 1.06$ & $11.62 \pm 1.00$ & $11.21 \pm 0.92$ & $\mathbf{10.24 \pm 1.04}$ & $10.50 \pm 1.41^\dagger$ & $0.94 \pm 0.25$ \\
\bottomrule
\end{tabular}
}
\end{widetable}

The latent sizes are much closer to one another than after generation. From
$d_s = 256$ on, the centre of gravity lies close to the \textsc{Geant4} floor, and at
$d_s = 128$ only its $z$ coordinate is about twice the floor. After generation it is
an order of magnitude above the floor, as is the energy sum at $d_s = 512$ and 1024, so
these errors come from the \LMF. At $d_s = 64$, instead, the centre of gravity is
already as far off after reconstruction as after generation. Here the error comes from
the \textsc{VAE}, while the \LMF learns the latent well, as also seen in
appendix~\ref{sec:applda}. In the geometric mean over the observables, $d_s = 64$ is
the worst. The \textsc{VAE} at $d_s = 1024$ performs best, opposite to what is seen
after generation, since its latent is not learned well enough by the \LMF
(appendix~\ref{sec:applda}).

\section{Time of first hit in a cell}\label{sec:apptime}
\begin{figure}[htbp]
    \centering
    \includegraphics[width=\timefirsthitfigwidth]{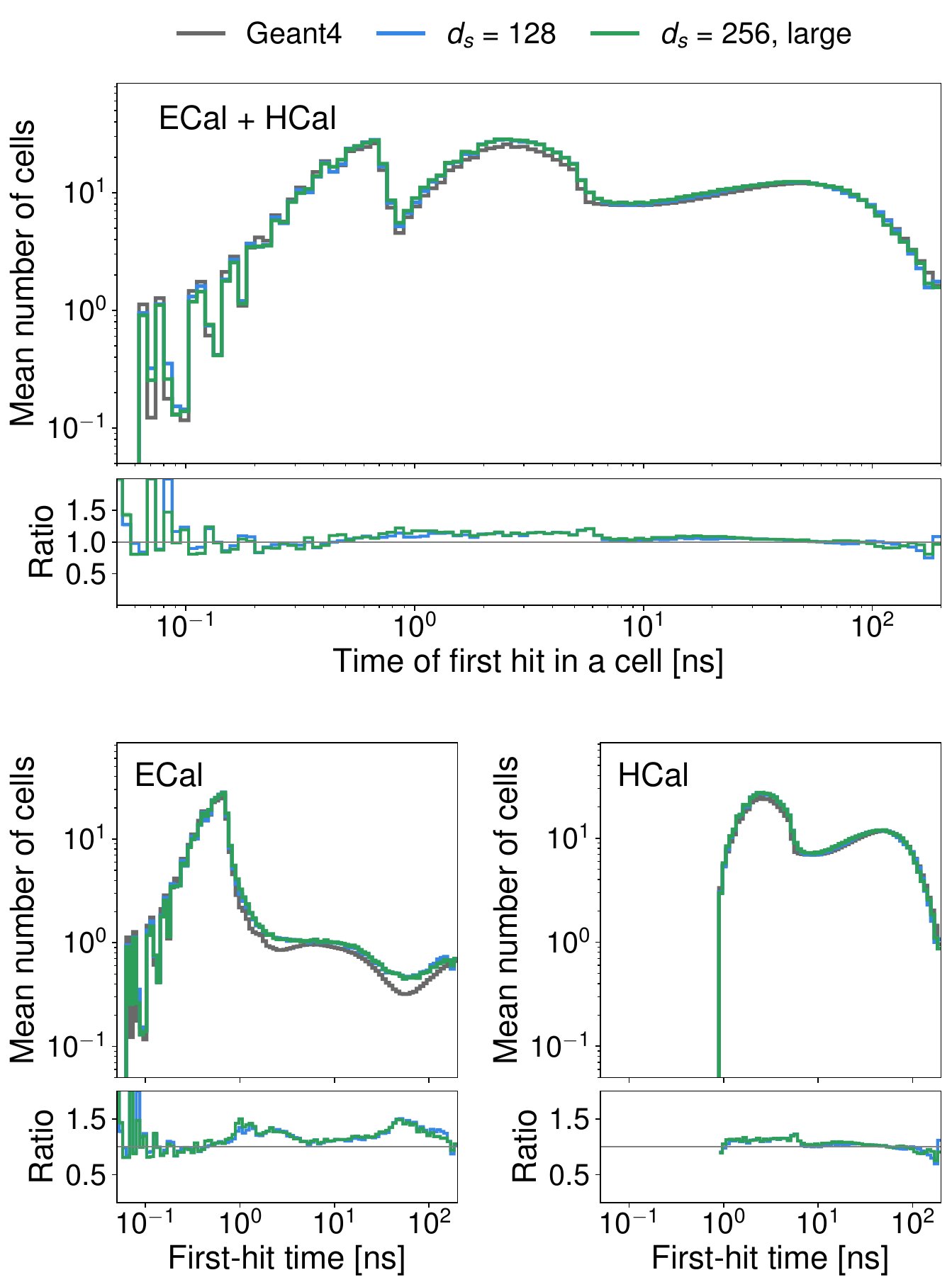}
    \caption{Mean number of cells per shower against the time of the first hit in the
    cell, in logarithmic time bins, for \num{100000} \textsc{Geant4} showers (grey),
    $d_s = 128$ (blue) and the large \LMF at $d_s = 256$ (green), with incident
    energies between 10 and \SI{90}{\giga\electronvolt}. Top: ECal and HCal together;
    bottom: ECal (left) and HCal (right). Below each panel, the ratio to \textsc{Geant4}.}
    \label{fig:time_first_hit}
\end{figure}

Figure~\ref{fig:time_first_hit} shows the distribution of the cell time, the time of
the first hit in each readout cell, for the two models of section~\ref{sec:cmp_models},
$d_s = 128$ and the large \LMF at $d_s = 256$, on their full-range samples of
figure~\ref{fig:obs_ds}, in the ECal and the HCal together and separately. The
generated times include the floor of section~\ref{sec:assembly}. \as and \cc are not
shown, since neither generates the arrival time.

In \textsc{Geant4}, most ECal cells are hit within the first nanosecond by the
prompt electromagnetic front. The shower reaches the HCal only after about
\SI{1}{\nano\second}, and its prompt component there peaks at a few nanoseconds. The
late hits, in both sections, form a slow component that extends from about
\SI{10}{\nano\second} to the end of the \SI{200}{\nano\second} window. Both models reproduce
all three, and the shape of the slow component over more than two orders of magnitude
in time. Both also have more cells than \textsc{Geant4}, in the ECal and in the HCal, which
corresponds to the excess in the number of hits seen in section~\ref{sec:observables}. The prompt
ECal front agrees with \textsc{Geant4}. In the ECal the excess sits in the delayed cells
and is largest around the minimum near \SI{50}{\nano\second}; in the HCal it sits in the
prompt component, while the later HCal cells agree with \textsc{Geant4} and the latest,
beyond \SI{100}{\nano\second}, fall slightly below it.

\section{Energy calibration}\label{sec:appcalib}

\begin{figure}[tbp]  
    \centering
    \includegraphics[width=\calibfigwidth,height=0.9\textheight,keepaspectratio]{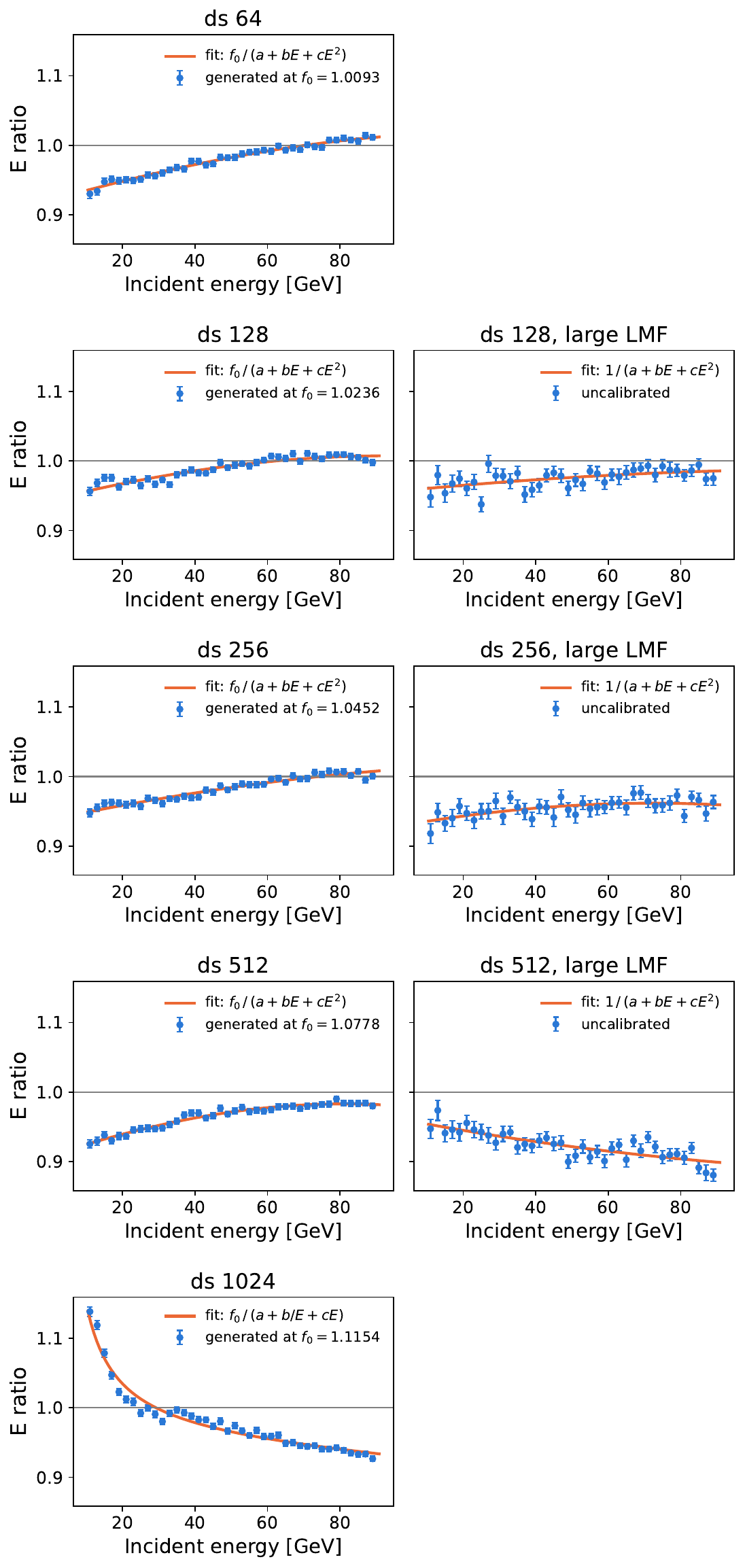}
    \caption{Response, the mean visible energy of the generated showers divided by
    that of \textsc{Geant4}, in bins of \SI{2}{\giga\electronvolt} of incident energy.
    One row per latent size. Left: the \LMF, rescaled by a constant factor (points), and the response
    implied by the fitted factor $f(E)$ (line). Right: the
    large \LMF at $d_s = 128$, 256 and 512,
    uncalibrated (points), and $1/f(E)$ of the factor applied to it (line).}
    \label{fig:calib_fit}
\end{figure}

The factor $f(E)$ of section~\ref{sec:model} is fitted for each model on showers
generated without calibration over the full range of incident energies. Their
response $r(E)$, the mean visible energy of the generated showers divided by that of
\textsc{Geant4} in bins of \SI{2}{\giga\electronvolt}, depends on the incident energy
(figure~\ref{fig:calib_fit}), and the factor applied is its inverse,
$f(E) = 1/r(E)$. Among the forms tried, the Akaike information criterion selects
$a + bE + cE^2$ for $d_s = 64$ to 512 and $a + b/E + cE$ for $d_s = 1024$, which
reproduce the required factor to within \SIrange{0.7}{1.5}{\percent} and
\SI{2.3}{\percent}. Every generated sample compared with \textsc{Geant4} in this paper
uses $f(E)$.

The response at $d_s = 1024$ has a completely different shape from that of the other
latent sizes: instead of rising slowly with the energy, it starts well above one at
the lowest energies and falls throughout the range, which is why only this size needs
a $1/E$ term. A possible cause is that the \LMF does not learn the latent of this size
well (appendix~\ref{sec:appreco}), and the calibration has to compensate for this,
but the response alone cannot confirm it.

The large \LMF is calibrated in the same way, with $a + bE + cE^2$ (right column of
figure~\ref{fig:calib_fit}). Its response varies less with the energy and lies closer
to one than that of the standard \LMF, so it needs a smaller correction, although its
points scatter more about the fit, which reproduces them to within
\SI{3.3}{\percent} in every bin. At $d_s = 512$ the response falls with the energy,
whereas at $d_s = 128$ and 256 it rises.

\section{Comparison with \as and \cc}\label{sec:appcmp}
This appendix gives the details of the comparison with \as and \cc in
sections~\ref{sec:cmp_models} and~\ref{sec:timing} and extends it with two further tests.
Appendix~\ref{sec:app3E} gives the distances to \textsc{Geant4} for every observable at 20,
50 and \SI{80}{\giga\electronvolt}, whose geometric means are summarised in
table~\ref{tab:three_energies_avg}, and appendix~\ref{sec:apptimingtable} the timing
measurements on which section~\ref{sec:timing} is based. Appendices~\ref{sec:appclassifier}
and~\ref{sec:applinearity} add a classifier test and the linearity and resolution of the
reconstructed energy, which are not discussed in the main text.

\subsection{Distances to \textsc{Geant4} at 20, 50 and 80~GeV}\label{sec:app3E}

Figure~\ref{fig:obs_models_2050} repeats the comparison of figure~\ref{fig:obs_models} at
\SI{20}{\giga\electronvolt} and \SI{50}{\giga\electronvolt} for \cc, \as with 32 and 4
network evaluations, $d_s = 128$ and the large \LMF at $d_s = 256$, each against
\num{10000} \textsc{Geant4} showers within $\pm\SI{0.5}{\giga\electronvolt}$ of the
incident energy. As at \SI{80}{\giga\electronvolt}, the energy sum of \cc is shifted to
higher values and is broader than that of \textsc{Geant4}, whereas the other models follow
\textsc{Geant4} closely. All models reproduce the mean energy per layer to within about
\SI{10}{\percent}, except in the last HCal layers, where the deposits are smallest: there \as,
with both 32 and 4 evaluations, lies above \textsc{Geant4} and both models of this work lie
below it. As at \SI{80}{\giga\electronvolt}, \cc does not reproduce the odd--even modulation
of the mean deposited energy across consecutive ECal layers (inset of the centre panels),
whereas the other models do, for the reason given in section~\ref{sec:cmp_models}. In the radial profile all models agree with \textsc{Geant4} near the shower axis
and deviate from it beyond about \SI{200}{\milli\meter}. \as with 32 evaluations remains the
closest at both energies, whereas with 4 evaluations it falls below \textsc{Geant4} beyond
about \SI{250}{\milli\meter}. Both models of this work overestimate the energy between about
\SI{200}{\milli\meter} and \SI{320}{\milli\meter}, the large \LMF by less than
$d_s = 128$, and underestimate it at the largest radii, where \cc lies above
\textsc{Geant4} at \SI{50}{\giga\electronvolt}.

\begin{widefigure}[!ht]
    \centering
    \includegraphics[width=\modelslegendwidth]{figures/models_3E_large_do01/legend.pdf}\\[2pt]
    \includegraphics[width=\modelspanelwidth]{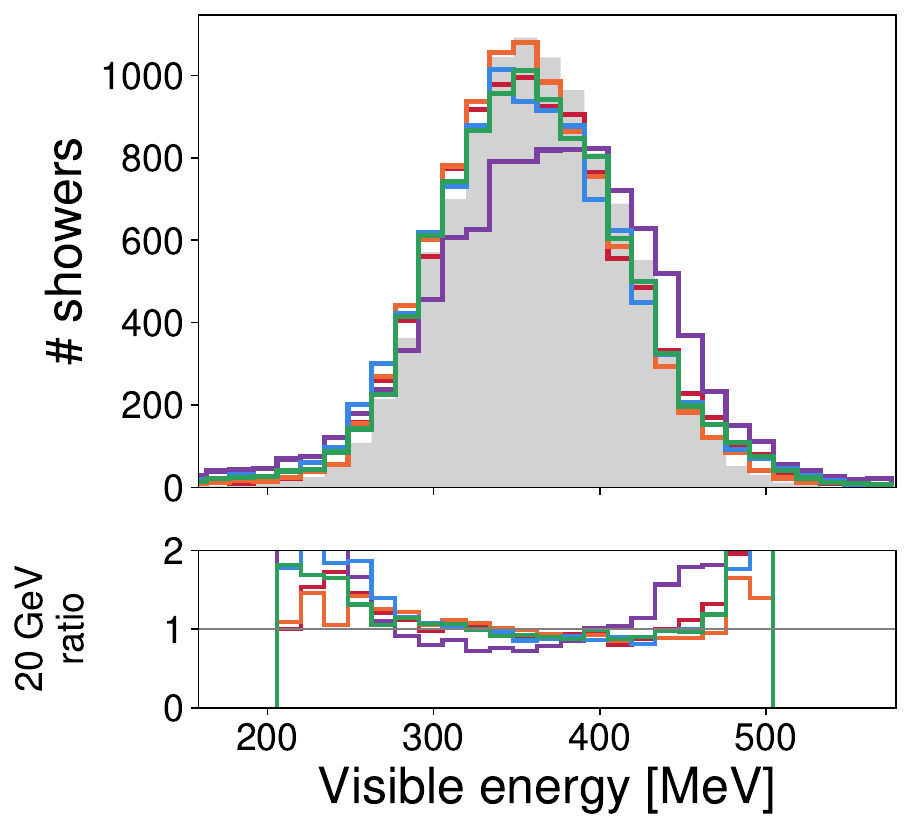}\hspace{0.02\linewidth}
    \includegraphics[width=\modelspanelwidth]{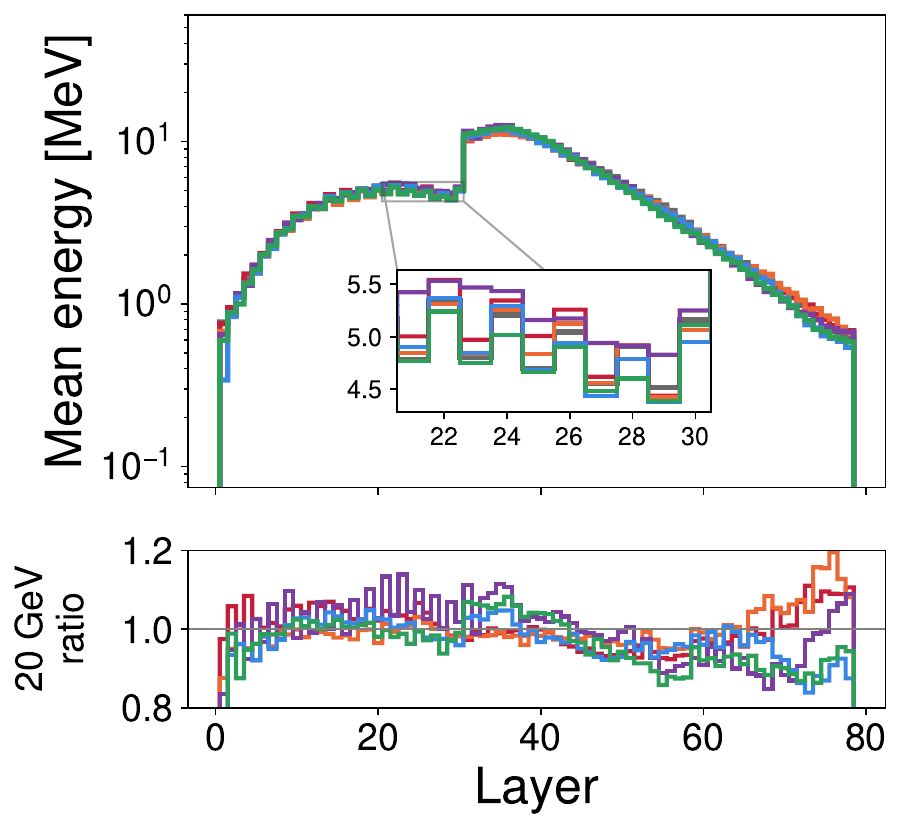}\hspace{0.02\linewidth}
    \includegraphics[width=\modelspanelwidth]{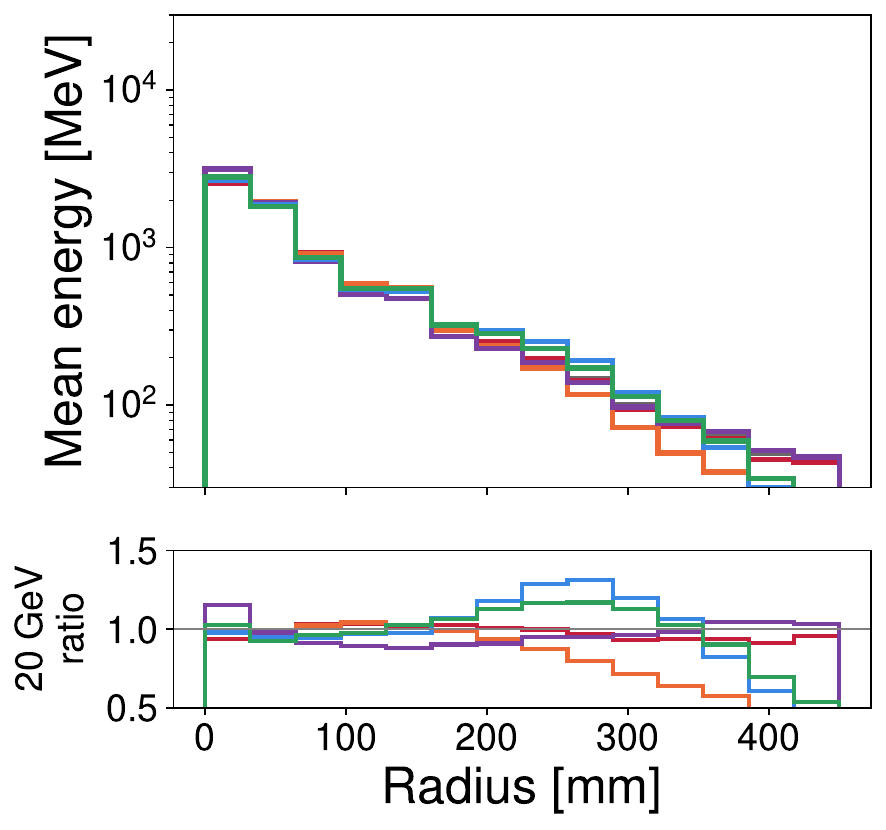}\\[4pt]
    \includegraphics[width=\modelspanelwidth]{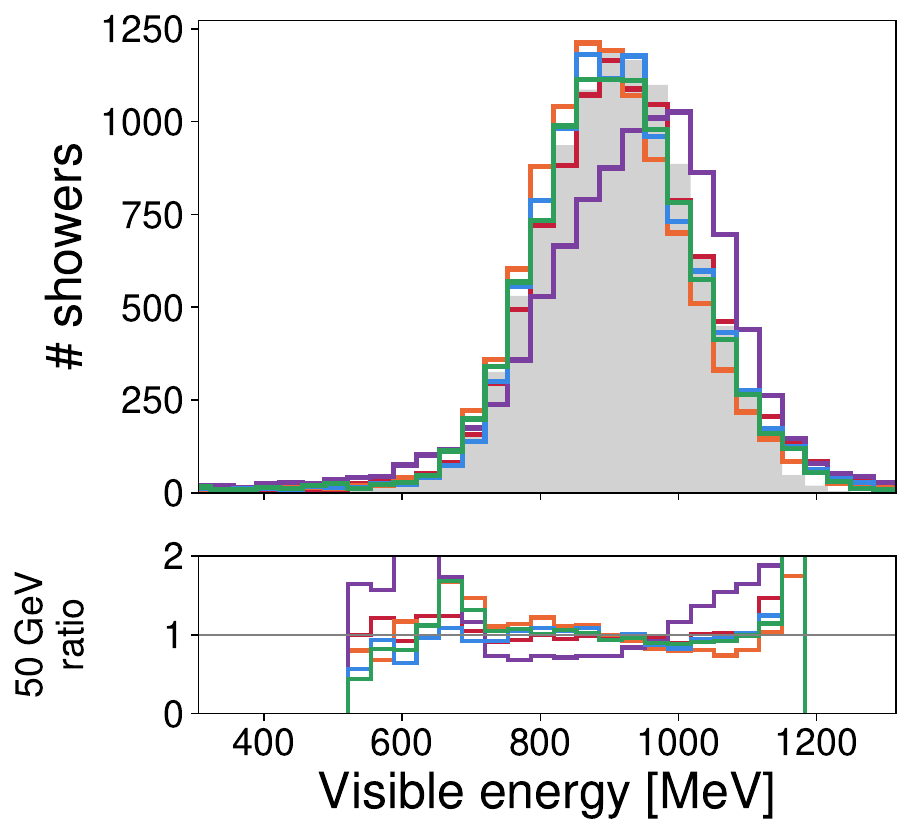}\hspace{0.02\linewidth}
    \includegraphics[width=\modelspanelwidth]{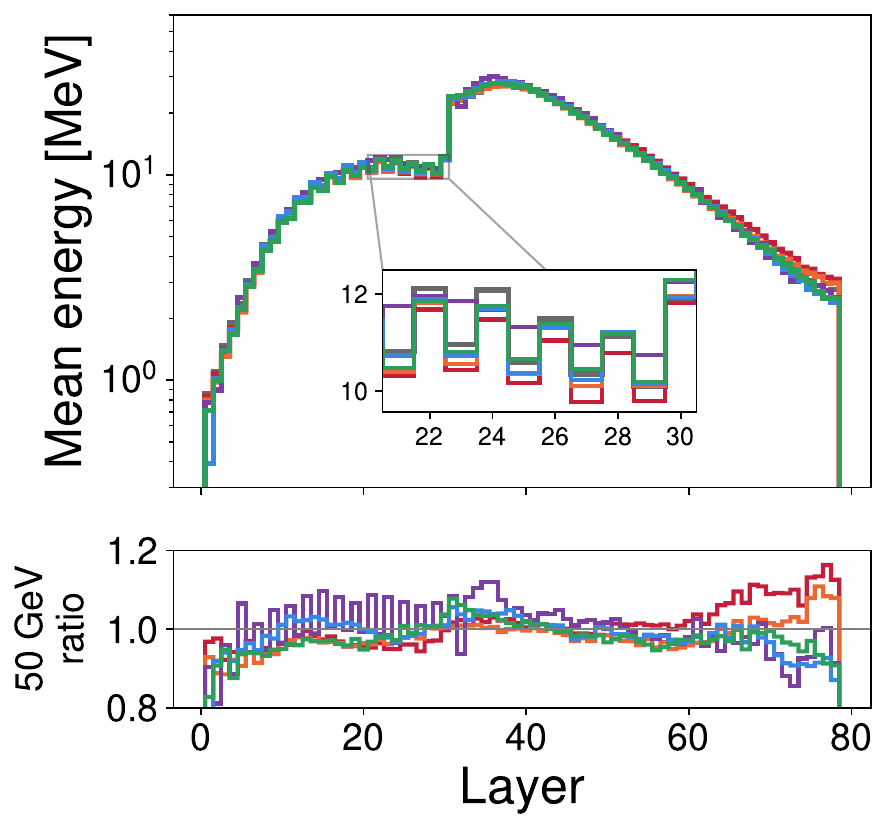}\hspace{0.02\linewidth}
    \includegraphics[width=\modelspanelwidth]{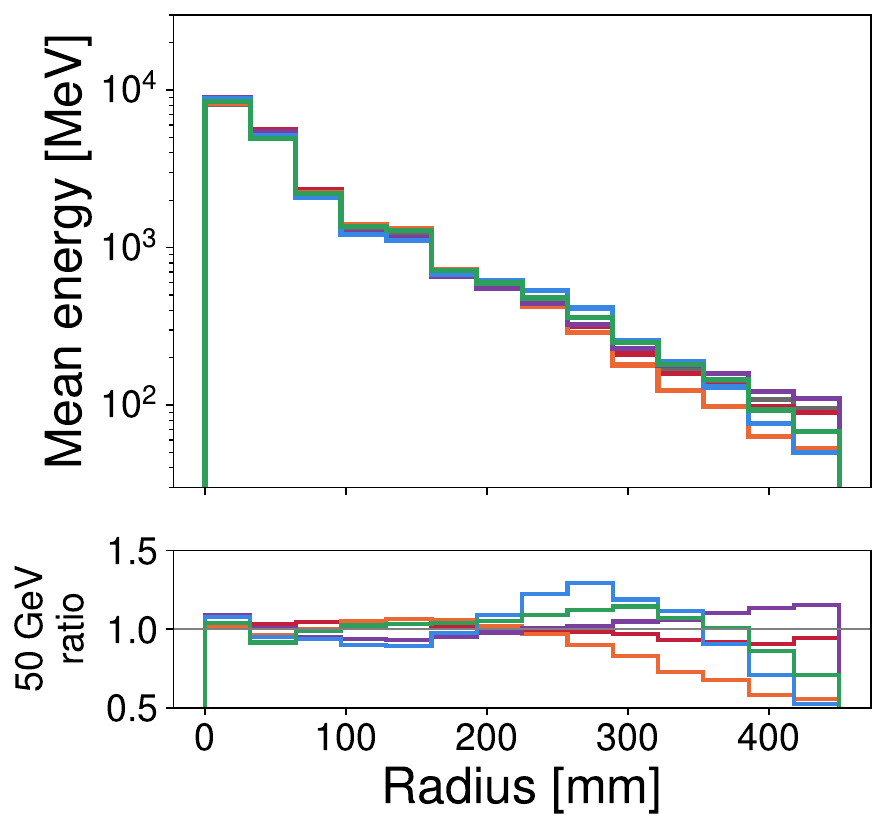}
    \caption{As figure~\ref{fig:obs_models}, at \SI{20}{\giga\electronvolt} (top) and
    \SI{50}{\giga\electronvolt} (bottom).}
    \label{fig:obs_models_2050}
\end{widefigure}

Table~\ref{tab:three_energies} quantifies these differences with the distances of
section~\ref{sec:observables}. \as with 32 network evaluations has the smallest geometric
mean of both distances at every energy, on average 4.2 in the normalised Wasserstein
distance and 15.2 in the Jensen--Shannon divergence, and is the closest model in the number
of hits and in the radial profile at every energy. Among the faster models, the large \LMF
has the smallest averages, 5.3 and 22.6, followed in the Wasserstein distance by
$d_s = 128$ (5.9), \cc (6.0) and \as with 4 evaluations (6.5), and in the divergence by
\as with 4 evaluations (27.0), $d_s = 128$ (27.8) and \cc (40.8); the order of the two
models of this work is the same as in table~\ref{tab:ds_metrics_avg}.

For the large \LMF, the number of hits deviates most from \textsc{Geant4}. \layerFM
reproduces the number of points per layer, but the \VAEdec clusters them less strongly
than \textsc{Geant4}, so they occupy more cells, and at 50 and \SI{80}{\giga\electronvolt}
the normalised Wasserstein distance of the large \LMF in the number of hits is the
largest of the five models. $d_s = 128$ deviates most from
\textsc{Geant4} in the centre of gravity in $z$ at every energy. The excess of cells is
already present when real showers are compressed and decompressed
(section~\ref{sec:observables}), the offset in the centre of gravity only in part
(appendix~\ref{sec:appreco}); the remainder originates in the generated latent. Compared
with $d_s = 128$, the large \LMF at $d_s = 256$ has a much smaller offset in the centre
of gravity in $z$ and a smaller one in $x$ at 20 and \SI{50}{\giga\electronvolt}, but its
number of hits is not closer to \textsc{Geant4}. \cc deviates most from \textsc{Geant4} in the energy sum and in the ECal energy sum, whose
divergence is 8 to 13 times that of any other model. Among
the other four models, the closest in the energy sum is \as with 4 evaluations at
\SI{20}{\giga\electronvolt}, $d_s = 128$ at \SI{50}{\giga\electronvolt} and the large
\LMF at \SI{80}{\giga\electronvolt}, in each case within the uncertainties of the next
closest; \as with 4 evaluations moves away from \textsc{Geant4} as the energy increases.

\begin{widetable}[!ht]
    \centering
    \caption{Normalised Wasserstein distance and Jensen--Shannon divergence to \textsc{Geant4} per
    observable. Each entry gives the values at 20 / 50 / \SI{80}{\giga\electronvolt} with the
    largest of their three statistical uncertainties, and the best model at each energy is in
    bold. Each block ends with the geometric mean over the observables, in the units of the
    block, with the largest of its three errors, each the standard deviation over 1000
    resamplings of the entries within their uncertainties, followed by the mean of the three
    geometric means with the error propagated from theirs. The last column is
    \textsc{Geant4} against itself on subsets of the same size.}
    \label{tab:three_energies}
    \footnotesize
    \setlength{\tabcolsep}{3pt}
    \resizebox{\linewidth}{!}{
\begin{tabular}{lcccccc}
\toprule
 & \cc & \as NFE~32 & \as NFE~4 & $d_s = 128$ & $d_s = 256$, large \LMF & \textsc{Geant4} \\
\midrule
\multicolumn{7}{l}{$\mathrm{WD}/\sigma_{\textsc{Geant4}}$ [$\times 10^{-2}$], at 20 / 50 / 80\,GeV} \\
\midrule
Energy sum & 30.3 / 30.7 / 38.8 $\pm$ 4.3 & 11.6 / 9.2 / 7.8 $\pm$ 2.3 & \textbf{10.8} / 12.8 / 15.3 $\pm$ 3.3 & 14.8 / \textbf{6.8} / 7.8 $\pm$ 2.2 & 11.7 / 8.5 / \textbf{6.7} $\pm$ 2.8 & 4.1 / 4.7 / 4.4 $\pm$ 2.4 \\
ECal energy sum & 7.0 / 6.2 / 7.3 $\pm$ 2.3 & 4.4 / 3.6 / \textbf{2.3} $\pm$ 1.9 & \textbf{4.0} / 4.2 / 5.0 $\pm$ 1.8 & 4.1 / \textbf{3.2} / 3.4 $\pm$ 1.2 & 4.7 / \textbf{3.2} / 2.7 $\pm$ 1.3 & 3.1 / 3.7 / 2.4 $\pm$ 1.1 \\
HCal energy sum & 9.6 / 8.4 / 8.2 $\pm$ 3.0 & 7.2 / 5.9 / 6.0 $\pm$ 1.4 & 7.0 / 6.4 / 8.4 $\pm$ 1.7 & 6.1 / 4.5 / \textbf{4.3} $\pm$ 1.9 & \textbf{5.7} / \textbf{4.3} / 6.2 $\pm$ 1.1 & 2.9 / 2.6 / 4.3 $\pm$ 1.9 \\
Number of hits & 29.6 / 19.8 / 15.1 $\pm$ 2.5 & \textbf{17.1} / \textbf{7.5} / \textbf{9.8} $\pm$ 2.7 & 28.4 / 18.6 / 17.2 $\pm$ 3.2 & 30.3 / 23.7 / 22.7 $\pm$ 3.0 & 29.8 / 29.7 / 27.3 $\pm$ 3.2 & 3.2 / 3.6 / 3.5 $\pm$ 0.9 \\
Centre of gravity $x$ & 8.2 / 4.9 / 4.7 $\pm$ 2.6 & 6.7 / \textbf{4.0} / \textbf{4.0} $\pm$ 2.6 & 12.6 / 13.8 / 9.0 $\pm$ 2.1 & 11.4 / 9.3 / 10.0 $\pm$ 3.7 & \textbf{4.2} / 6.8 / 9.6 $\pm$ 2.6 & 5.5 / 5.1 / 5.1 $\pm$ 2.7 \\
Centre of gravity $y$ & 4.5 / 3.8 / 4.6 $\pm$ 1.4 & 3.6 / 5.5 / 4.2 $\pm$ 2.3 & 5.1 / 4.4 / 5.2 $\pm$ 1.7 & \textbf{3.2} / \textbf{3.2} / 4.4 $\pm$ 1.2 & 3.7 / 3.7 / \textbf{3.5} $\pm$ 1.1 & 4.2 / 3.9 / 3.9 $\pm$ 1.7 \\
Centre of gravity $z$ & \textbf{4.7} / 4.8 / \textbf{3.8} $\pm$ 1.8 & \textbf{4.7} / \textbf{4.7} / 4.7 $\pm$ 1.8 & 9.2 / 8.2 / 7.9 $\pm$ 2.3 & 19.1 / 16.5 / 12.6 $\pm$ 2.5 & 5.1 / 6.1 / 5.6 $\pm$ 1.3 & 3.8 / 5.0 / 3.5 $\pm$ 1.5 \\
Cell energy & 3.34 / \textbf{1.86} / \textbf{1.81} $\pm$ 0.04 & 3.5 / 2.5 / 1.9 $\pm$ 0.1 & 6.3 / 5.5 / 5.2 $\pm$ 0.1 & \textbf{2.6} / 3.4 / 3.4 $\pm$ 0.1 & 4.1 / 4.0 / 3.6 $\pm$ 0.1 & 0.1 / 0.1 / 0.1 $\pm$ 0.1 \\
Longitudinal profile & 3.4 / 2.2 / 1.1 $\pm$ 0.2 & 2.8 / 2.6 / 1.3 $\pm$ 0.3 & \textbf{0.8} / \textbf{1.0} / 1.2 $\pm$ 0.2 & 2.3 / 1.6 / \textbf{0.9} $\pm$ 0.5 & 2.9 / 1.7 / 1.5 $\pm$ 0.2 & 0.6 / 0.3 / 0.4 $\pm$ 0.2 \\
Radial profile & 3.4 / 2.4 / 2.4 $\pm$ 0.2 & \textbf{1.9} / \textbf{1.2} / \textbf{0.9} $\pm$ 0.2 & 6.9 / 5.6 / 4.5 $\pm$ 0.2 & 5.0 / 4.3 / 3.5 $\pm$ 0.3 & 5.3 / 2.4 / 3.2 $\pm$ 0.2 & 0.5 / 0.3 / 0.3 $\pm$ 0.2 \\
\cmidrule(l){2-7}
Geometric mean & 7.23 / 5.55 / 5.19 $\pm$ 0.70 & \textbf{5.16} / \textbf{4.06} / \textbf{3.36} $\pm$ 0.90 & 6.70 / 6.28 / 6.40 $\pm$ 0.97 & 6.87 / 5.52 / 5.24 $\pm$ 0.61 & 5.84 / 4.97 / 5.00 $\pm$ 0.41 & 1.82 / 1.64 / 1.61 $\pm$ 0.42 \\
Mean over energies & 5.99 $\pm$ 0.34 & \textbf{4.19} $\pm$ 0.40 & 6.46 $\pm$ 0.43 & 5.88 $\pm$ 0.27 & 5.27 $\pm$ 0.21 & 1.69 $\pm$ 0.22 \\
\midrule
\multicolumn{7}{l}{JSD [$\times 10^{-4}$], at 20 / 50 / 80\,GeV} \\
\midrule
Energy sum & 194.6 / 212.7 / 302.9 $\pm$ 42.3 & 58.4 / 63.1 / 47.8 $\pm$ 15.4 & \textbf{46.9} / 67.3 / 86.9 $\pm$ 23.8 & 70.0 / 51.5 / 56.0 $\pm$ 10.7 & 54.2 / \textbf{48.2} / \textbf{46.1} $\pm$ 14.1 & 25.5 / 28.7 / 26.2 $\pm$ 7.2 \\
ECal energy sum & 938.7 / 792.2 / 700.0 $\pm$ 88.7 & \textbf{30.3} / 34.2 / 29.1 $\pm$ 12.4 & 43.7 / \textbf{32.5} / 36.1 $\pm$ 9.2 & 109.8 / 62.5 / 78.0 $\pm$ 20.7 & 82.8 / 46.0 / \textbf{27.9} $\pm$ 10.4 & 27.0 / 26.2 / 30.0 $\pm$ 12.4 \\
HCal energy sum & 142.2 / 104.5 / 91.7 $\pm$ 30.5 & 63.3 / 51.0 / 59.7 $\pm$ 19.5 & 51.5 / 52.1 / 101.7 $\pm$ 11.6 & 59.2 / 53.0 / \textbf{44.9} $\pm$ 16.6 & \textbf{51.2} / \textbf{39.0} / 54.0 $\pm$ 13.0 & 32.7 / 25.6 / 27.5 $\pm$ 7.0 \\
Number of hits & 223.2 / 114.2 / 77.3 $\pm$ 30.6 & \textbf{74.0} / \textbf{39.3} / \textbf{51.8} $\pm$ 13.9 & 147.1 / 77.3 / 70.3 $\pm$ 23.3 & 198.5 / 108.0 / 100.5 $\pm$ 44.4 & 169.9 / 154.8 / 128.0 $\pm$ 43.4 & 42.2 / 34.2 / 22.7 $\pm$ 12.5 \\
Centre of gravity $x$ & 38.6 / 33.7 / \textbf{38.2} $\pm$ 12.7 & 33.9 / 32.6 / 38.4 $\pm$ 11.9 & 81.3 / 89.2 / 58.5 $\pm$ 23.1 & 63.0 / 50.9 / 54.3 $\pm$ 11.3 & \textbf{30.6} / \textbf{32.1} / 44.5 $\pm$ 10.4 & 31.2 / 32.9 / 31.2 $\pm$ 7.1 \\
Centre of gravity $y$ & 44.9 / 39.4 / 30.8 $\pm$ 9.5 & \textbf{30.8} / \textbf{24.3} / 38.1 $\pm$ 9.4 & 35.8 / 25.4 / 34.4 $\pm$ 9.3 & 32.5 / 27.4 / 35.4 $\pm$ 18.0 & 32.2 / 27.9 / \textbf{26.0} $\pm$ 7.0 & 27.8 / 29.2 / 27.2 $\pm$ 9.1 \\
Centre of gravity $z$ & 31.3 / 32.5 / \textbf{31.5} $\pm$ 7.2 & \textbf{29.5} / 33.1 / 35.6 $\pm$ 10.3 & 49.2 / 56.9 / 53.1 $\pm$ 13.5 & 107.0 / 99.1 / 81.8 $\pm$ 24.4 & 35.9 / \textbf{31.5} / 38.4 $\pm$ 11.4 & 22.2 / 35.8 / 29.8 $\pm$ 11.8 \\
Cell energy & 6.3 / 3.5 / 3.2 $\pm$ 0.1 & \textbf{4.0} / \textbf{3.3} / \textbf{2.6} $\pm$ 0.2 & 17.7 / 14.9 / 15.5 $\pm$ 0.4 & 6.4 / 5.2 / 4.4 $\pm$ 0.2 & 11.1 / 8.2 / 6.2 $\pm$ 0.3 & 0.06 / 0.03 / 0.02 $\pm$ 0.01 \\
Longitudinal profile & 4.1 / 2.3 / 1.4 $\pm$ 0.3 & 2.1 / 1.7 / 1.0 $\pm$ 0.3 & \textbf{1.0} / \textbf{0.9} / 0.9 $\pm$ 0.3 & 2.2 / 1.2 / \textbf{0.8} $\pm$ 0.4 & 4.4 / 1.5 / 1.0 $\pm$ 0.2 & 0.6 / 0.3 / 0.3 $\pm$ 0.1 \\
Radial profile & 14.8 / 7.6 / 7.8 $\pm$ 1.1 & \textbf{2.4} / \textbf{0.3} / \textbf{0.4} $\pm$ 0.5 & 11.8 / 7.4 / 5.5 $\pm$ 0.4 & 8.6 / 7.3 / 4.5 $\pm$ 0.3 & 6.6 / 1.9 / 2.2 $\pm$ 0.3 & 0.2 / 0.1 / 0.1 $\pm$ 0.1 \\
\cmidrule(l){2-7}
Geometric mean & 51.07 / 37.33 / 33.88 $\pm$ 2.90 & \textbf{18.73} / \textbf{13.29} / \textbf{13.68} $\pm$ 1.45 & 29.21 / 25.34 / 26.58 $\pm$ 2.97 & 34.48 / 25.59 / 23.38 $\pm$ 3.55 & 29.15 / 19.76 / 18.82 $\pm$ 2.18 & 6.47 / 5.25 / 4.51 $\pm$ 0.85 \\
Mean over energies & 40.76 $\pm$ 1.39 & \textbf{15.23} $\pm$ 0.77 & 27.04 $\pm$ 1.33 & 27.82 $\pm$ 1.46 & 22.57 $\pm$ 0.91 & 5.41 $\pm$ 0.44 \\
\bottomrule
\end{tabular}
}
\end{widetable}

\subsection{Classifier test}\label{sec:appclassifier}
The distances of appendix~\ref{sec:app3E} compare one observable at a time. To test whether the observables
of each shower are also jointly compatible with \textsc{Geant4}, we train a classifier.
For every model a fully connected network with two hidden layers of 2048 units is
trained to separate its showers from \textsc{Geant4} showers, using
\num{10000} showers per energy from 10 to \SI{90}{\giga\electronvolt} on each side,
split into \SI{60}{\percent} for training, \SI{20}{\percent} for validation and
\SI{20}{\percent} for testing. Each shower is described by eight features: the centre of
gravity in $x$, $y$ and $z$, the visible energy divided by the incident energy, and
the energy-weighted mean and root mean square of the layer index and of the distance
from the shower axis. The number of hits is not included, since it depends on the
point-count stage. The model for this stage was first developed for \cc, its conditioning
was widened by \as and its output extended by \layerFM (section~\ref{sec:stage_layerfm}), so
these generators are likely to use very similar models for the number of points per
layer, and the classifier tests only the generation of the shower itself. The network is trained for 100 epochs and evaluated at the epoch with the lowest validation
loss.
Table~\ref{tab:classifier} gives the area under the ROC curve (AUC) on the test set,
averaged over three trainings with different random splits; an AUC of 0.5 means that the
generated showers cannot be told apart from \textsc{Geant4}. It also gives the
out-of-distribution (OOD) fraction, in percent: the classifier output is calibrated with an isotonic
regression fitted on the validation set, and a test shower counts as OOD when its
calibrated output is below 0.01 or above 0.99, i.e.\ when the classifier is more than
\SI{99}{\percent} sure of its origin. A high OOD fraction of \textsc{Geant4} showers means
that the generator misses regions of phase space (mode dropping), and a high OOD fraction
of generated showers means that it produces unphysical artefacts. For a perfect generator both fractions would be zero, so
lower values are better.

\begin{table}[htbp]
    \centering
    \caption{AUC of the classifier separating each model from \textsc{Geant4}, and the
    fraction of \textsc{Geant4} and of generated test showers classified with more than
    \SI{99}{\percent} confidence (OOD); mean and standard deviation over three trainings.
    Lower is better in every column; the lowest value is in bold.}
    \label{tab:classifier}
    \footnotesize
    \setlength{\tabcolsep}{2pt}
    \begin{tabular}{lccc}
    \toprule
    & & \multicolumn{2}{c}{OOD [\%]} \\
    \cmidrule(lr){3-4}
    Model & AUC & \textsc{Geant4} & generated \\
    \midrule
    \as NFE~32 & $\mathbf{0.614 \pm 0.002}$ & $0.13 \pm 0.09$ & $0.26 \pm 0.12$ \\
    $d_s = 256$, large \LMF & $0.702 \pm 0.002$ & $0.93 \pm 0.31$ & $0.32 \pm 0.15$ \\
    \as NFE~4 & $0.726 \pm 0.003$ & $\mathbf{0.06 \pm 0.04}$ & $0.20 \pm 0.10$ \\
    \cc & $0.735 \pm 0.001$ & $0.13 \pm 0.11$ & $2.93 \pm 1.05$ \\
    $d_s = 128$ & $0.755 \pm 0.001$ & $3.67 \pm 0.31$ & $\mathbf{0.10 \pm 0.03}$ \\
    \bottomrule
    \end{tabular}
\end{table}

\as with 32 network evaluations is the hardest to separate from \textsc{Geant4},
followed by the large \LMF at $d_s = 256$, which is ahead of \as with 4 evaluations,
\cc and $d_s = 128$. The OOD fractions show that the \LMF models leave more of the
\textsc{Geant4} phase space uncovered than \as and \cc, while \cc generates about
\SI{2.9}{\percent} of showers that do not occur in \textsc{Geant4}.

\subsection{Linearity and resolution of the reconstructed energy}\label{sec:applinearity}
\begin{figure}[htbp]
    \centering
    \includegraphics[width=\linearityfigwidth]{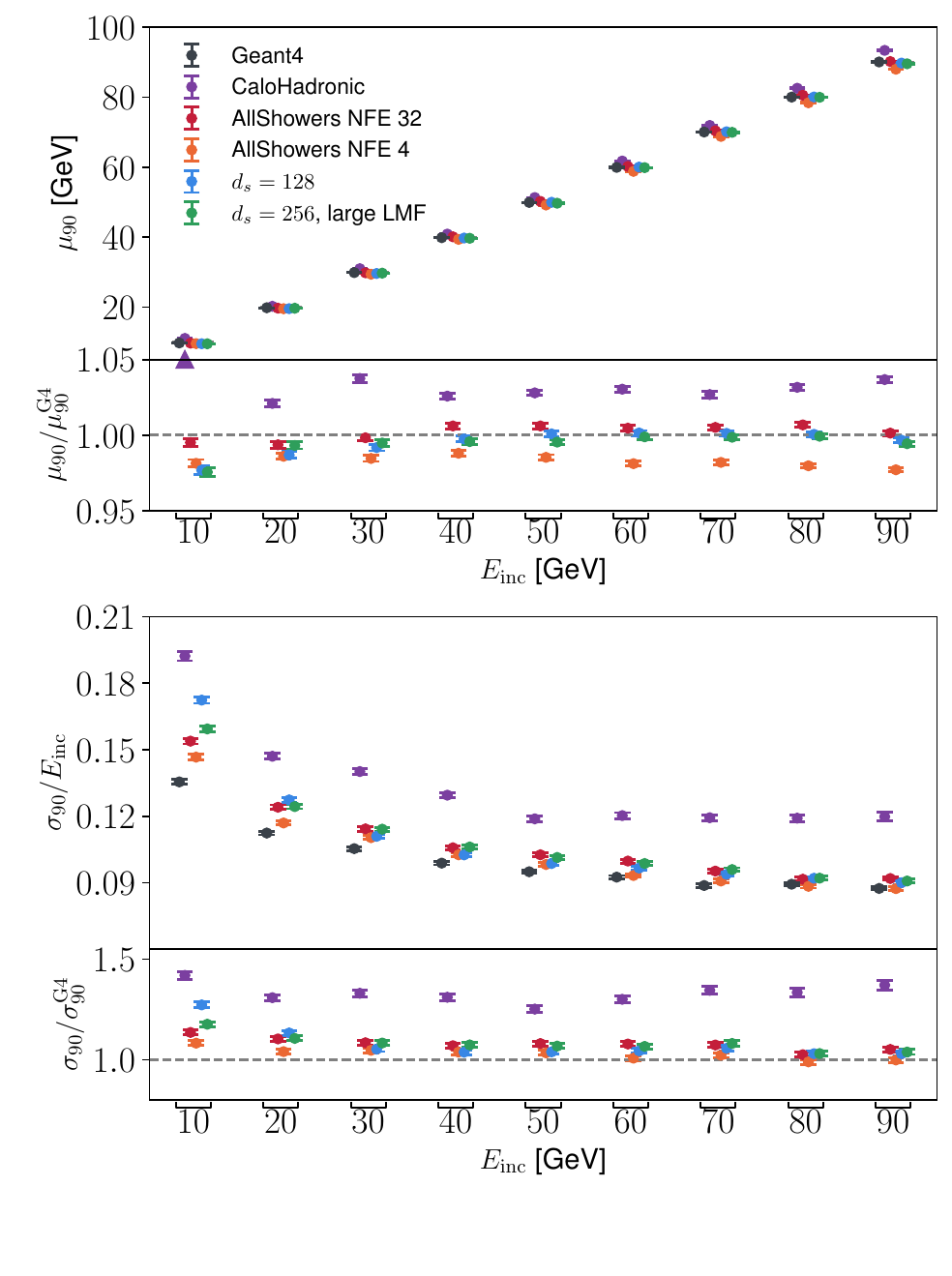}
    \caption{Linearity (top) and resolution (bottom) of the reconstructed energy of
    $\pi^{+}$ showers for \textsc{Geant4}, \cc, \as with 32 and 4 network evaluations
    (NFE~32, NFE~4), $d_s = 128$ and the large \LMF at $d_s = 256$, with \num{10000} showers
    per energy and generator. Below each, the ratio to \textsc{Geant4}; the point beyond the range of the
    upper ratio panel is marked by a triangle at its edge. At each energy the models are drawn side by side above a bracket.}
    \label{fig:linearity}
\end{figure}

Figure~\ref{fig:linearity} gives the linearity and the resolution of the reconstructed
energy of $\pi^{+}$ showers for all models, at incident
energies from 10 to \SI{90}{\giga\electronvolt} in steps of \SI{10}{\giga\electronvolt},
with \num{10000} showers per energy. At each energy the \SI{5}{\percent} of
the showers with the lowest and the \SI{5}{\percent} with the highest visible energy are
discarded, and the mean $\mu_{90}$ and standard deviation $\sigma_{90}$ of the remaining
\SI{90}{\percent} are computed. The visible energy is converted to GeV with a single factor,
fitted by least squares such that $\mu_{90}$ of \textsc{Geant4} matches the incident energy,
and applied unchanged to every generator. The uncertainties are the standard deviation over
200 bootstrap resamplings of the showers.

Both \LMF models are linear (figure~\ref{fig:linearity}, top): their $\mu_{90}$ agrees with \textsc{Geant4} to within
\SI{1}{\percent} from \SI{30}{\giga\electronvolt} upwards and lies slightly below it only at
the lowest energies. \as with 32 evaluations is as close at every energy, with 4
evaluations it is slightly low throughout, and \cc is high. Every generator yields a broader
energy distribution than \textsc{Geant4} (figure~\ref{fig:linearity}, bottom). At high energy the \LMF models are a few per cent
broader, similar to \as, and the excess grows towards low energy; \cc is the broadest at
every energy. At \SI{10}{\giga\electronvolt} \as is closer to \textsc{Geant4} than the \LMF
models in both linearity and resolution. A possible reason is the training range: \as is
trained on incident energies from 1 to \SI{130}{\giga\electronvolt}, well around
\SI{10}{\giga\electronvolt}, whereas for the \LMF models \SI{10}{\giga\electronvolt} is the
lower end of the training range.

\subsection{Generation time}\label{sec:apptimingtable}
\begin{table}[htbp]  
    \centering
    \caption{Generation time per shower and speed-up relative to \textsc{Geant4} on a single
    CPU core. The \textsc{Geant4} rows and the \cc rows at a single network evaluation and
    on the CPU are taken from ref.~\cite{Buss:2026yrf}, which takes them from
    ref.~\cite{Buss:2025cyw}; all other rows are measured here as described in the text.}
    \label{tab:timing_comparison}
    \ifdefined\prdlayout  
    {\footnotesize\setlength{\tabcolsep}{3pt}\resizebox{\linewidth}{!}{
\begin{tabular}{llrrrr}
\toprule
Hardware & Model & NFE & Batch & Time [s] & Speed-up \\
\midrule
CPU & \textsc{Geant4} & -- & 1 & 2.09 & $1.0\times$ \\
\cmidrule(l){2-6}
 & \cc & 1 & 1 & 0.59 & $3.5\times$ \\
 &  &  & 16 & 0.73 & $2.8\times$ \\
\cmidrule(l){3-6}
 &  & 59 & 1 & 34.8 & $<0.1\times$ \\
 &  &  & 16 & 43.3 & $<0.1\times$ \\
\cmidrule(l){2-6}
 & \as & 4 & 1 & 0.50 & $4.2\times$ \\
\cmidrule(l){3-6}
 &  & 32 & 1 & 3.8 & $0.6\times$ \\
\cmidrule(l){2-6}
 & This work, $d_s = 128$ & 1 & 1 & 0.14 & $14.9\times$ \\
\cmidrule(l){2-6}
 & \shortstack[l]{This work, $d_s = 256$,\\ large \LMF} & 1 & 1 & 0.22 & $9.5\times$ \\
\midrule
GPU & \cc & 1 & 1 & 0.0086 & $243.0\times$ \\
 &  &  & 16 & 0.0033 & $633.3\times$ \\
\cmidrule(l){3-6}
 &  & 59 & 1 & 0.52 & $4.0\times$ \\
 &  &  & 16 & 0.27 & $7.7\times$ \\
 &  &  & 256 & 0.20 & $10.3\times$ \\
\cmidrule(l){2-6}
 & \as & 4 & 1 & 0.013 & $161.0\times$ \\
 &  &  & 16 & 0.0013 & $1589.9\times$ \\
 &  &  & 1400 & 0.00072 & $2919.1\times$ \\
\cmidrule(l){3-6}
 &  & 32 & 1 & 0.040 & $52.3\times$ \\
 &  &  & 16 & 0.0052 & $404.4\times$ \\
 &  &  & 1400 & 0.0045 & $464.5\times$ \\
\cmidrule(l){2-6}
 & This work, $d_s = 128$ & 1 & 1 & 0.011 & $189.6\times$ \\
 &  &  & 16 & 0.0010 & $2087.5\times$ \\
 &  &  & 1400 & 0.00034 & $6203.1\times$ \\
\cmidrule(l){2-6}
 & \shortstack[l]{This work, $d_s = 256$,\\ large \LMF} & 1 & 1 & 0.011 & $189.3\times$ \\
 &  &  & 16 & 0.0011 & $1938.4\times$ \\
 &  &  & 1400 & 0.00036 & $5839.3\times$ \\
\cmidrule(l){2-6}
 & \shortstack[l]{This work, $d_s = 256$,\\ large \LMF, no time} & 1 & 1400 & 0.00034 & $6205.1\times$ \\
\bottomrule
\end{tabular}
}}
    \else
    \small
\begin{tabular}{llrrrr}
\toprule
Hardware & Model & NFE & Batch & Time [s] & Speed-up \\
\midrule
CPU & \textsc{Geant4} & -- & 1 & 2.09 & $1.0\times$ \\
\cmidrule(l){2-6}
 & \cc & 1 & 1 & 0.59 & $3.5\times$ \\
 &  &  & 16 & 0.73 & $2.8\times$ \\
\cmidrule(l){3-6}
 &  & 59 & 1 & 34.8 & $<0.1\times$ \\
 &  &  & 16 & 43.3 & $<0.1\times$ \\
\cmidrule(l){2-6}
 & \as & 4 & 1 & 0.50 & $4.2\times$ \\
\cmidrule(l){3-6}
 &  & 32 & 1 & 3.8 & $0.6\times$ \\
\cmidrule(l){2-6}
 & This work, $d_s = 128$ & 1 & 1 & 0.14 & $14.9\times$ \\
\cmidrule(l){2-6}
 & \shortstack[l]{This work, $d_s = 256$,\\ large \LMF} & 1 & 1 & 0.22 & $9.5\times$ \\
\midrule
GPU & \cc & 1 & 1 & 0.0086 & $243.0\times$ \\
 &  &  & 16 & 0.0033 & $633.3\times$ \\
\cmidrule(l){3-6}
 &  & 59 & 1 & 0.52 & $4.0\times$ \\
 &  &  & 16 & 0.27 & $7.7\times$ \\
 &  &  & 256 & 0.20 & $10.3\times$ \\
\cmidrule(l){2-6}
 & \as & 4 & 1 & 0.013 & $161.0\times$ \\
 &  &  & 16 & 0.0013 & $1589.9\times$ \\
 &  &  & 1400 & 0.00072 & $2919.1\times$ \\
\cmidrule(l){3-6}
 &  & 32 & 1 & 0.040 & $52.3\times$ \\
 &  &  & 16 & 0.0052 & $404.4\times$ \\
 &  &  & 1400 & 0.0045 & $464.5\times$ \\
\cmidrule(l){2-6}
 & This work, $d_s = 128$ & 1 & 1 & 0.011 & $189.6\times$ \\
 &  &  & 16 & 0.0010 & $2087.5\times$ \\
 &  &  & 1400 & 0.00034 & $6203.1\times$ \\
\cmidrule(l){2-6}
 & \shortstack[l]{This work, $d_s = 256$,\\ large \LMF} & 1 & 1 & 0.011 & $189.3\times$ \\
 &  &  & 16 & 0.0011 & $1938.4\times$ \\
 &  &  & 1400 & 0.00036 & $5839.3\times$ \\
\cmidrule(l){2-6}
 & \shortstack[l]{This work, $d_s = 256$,\\ large \LMF, no time} & 1 & 1400 & 0.00034 & $6205.1\times$ \\
\bottomrule
\end{tabular}

    \fi
\end{table}

Table~\ref{tab:timing_comparison} complements figure~\ref{fig:timing_ds256} with the time
per shower at several batch sizes and on a CPU. The full chain is timed, from the conditioning to
the points in physical units, on an NVIDIA A100 (\SI{80}{GB}, PCIe) GPU at batch sizes 1, 16
and 1400, and on a single core of an AMD EPYC 7402 CPU at batch size 1; each time is the
mean over 10 to \SI{90}{\giga\electronvolt}, with 20 warm-up and 30 timed batches per
energy. The largest batch size, 1400, is chosen close to the largest at which the \calochill
chain fits in the memory of the GPU at every incident energy. \as is timed
in the same way with its own code, weights and Euler solver, within \SI{8}{\percent} of
ref.~\cite{Buss:2026yrf} on the CPU. A possible reason for the difference is the software
environment, such as the PyTorch version and its compiled kernels, or the clock frequency
of the node. For \cc at 59 evaluations on the GPU, the full chain is
measured here. Its times are \SI{0.12}{\second} per shower longer than in the reference,
because they also include the conversion of the network output into showers, which \cc
performs on the CPU, one shower at a time. The
remaining \cc rows are taken from the reference and therefore exclude this step.

At small batch sizes the time of both \as and \calochill is dominated by a constant cost,
the point-count stage, the preparation of the inputs and the launch of many small GPU
operations: a linear fit of the time of \as against the number of evaluations gives
\SI{0.77}{\milli\second} for this cost at batch size 16 and \SI{0.14}{\milli\second} per
evaluation. At batch size 1400 the constant drops to \SI{0.17}{\milli\second} and the time of \as
is dominated by the network evaluations over all points. Since \calochill passes over
the points only once, it is about twice as fast as \as with 4 evaluations at this batch
size.
\cc is given at batch size 256, its fastest, since every batch is padded to its largest
shower and its dense attention slows down with the square of the padded length.

\calochill is the fastest generator in table~\ref{tab:timing_comparison} in every
configuration except batch size 1 on the GPU, where \cc with a single evaluation, timed
without the conversion into showers, is faster. At batch size 1400 \calochill generates
a shower in \SI{0.34}{\milli\second}, about 6200 times faster than \textsc{Geant4}. \as
with 32 evaluations, which is the most accurate, is 13 times slower on the GPU, and on a
single CPU core it is slower than \textsc{Geant4}.

\section{Cost-equivalent CPU time of \textsc{Geant4}}\label{sec:appcost}
\textsc{Geant4} runs on CPUs and the generators on a GPU, so their times per shower in
table~\ref{tab:timing_comparison} are measured on different hardware and cannot be
compared directly. To place \textsc{Geant4} on the time axis of
figure~\ref{fig:time_vs_quality}, the GPU is converted into the number of CPU cores that
can be bought for the same price. Showers are simulated independently of each other and its time
per shower at the cost of one GPU is its time on one core divided by $n$.

The NVIDIA A100 with \SI{80}{GB} (PCIe) used for the timing costs about \$12{,}000
(\$9{,}500--14{,}000~\cite{JarvisLabsA100Price}), including its memory.
A CPU core costs its share of an AMD EPYC 7402, the processor of the \textsc{Geant4}
timing, listed at \$1{,}783 for 24 cores~\cite{AMDEPYC7402}, plus its share of
the memory. A \textsc{Geant4} simulation of a full detector needs about \SI{2}{GB} per
job~\cite{ECCE:2022computing}, but memory is bought in modules. The smallest
standard configuration of the EPYC 7402 has one \SI{8}{GB} module in each of its eight
memory channels~\cite{AMDEPYC7402}, i.e.\ \SI{2.7}{GB} per core, at \$9.17 per GB of
DDR4 server memory~\cite{DatacenterDiskRAMIndex}. This gives about \$99 per core, so
one GPU costs as much as 121 cores. With \SI{2.09}{\second} per shower on one core
(table~\ref{tab:timing_comparison}), \textsc{Geant4} takes \SI{17}{\milli\second} per
shower at the cost of one GPU, and 15 to \SI{22}{\milli\second} over the range of GPU
prices.

The estimate rests on the following assumptions.
\begin{itemize}
    \item[--] \textit{Purchase price only}: the host server of the GPU, the rest of the
          CPU node, power, cooling and the lifetime of the hardware are not included.
    \item[--] \textit{List prices}: prices are those at the time of writing and vary
          between vendors and over time; for the GPU the range of prices is propagated
          to the time above.
    \item[--] \textit{Perfect scaling over cores}: every core is assumed to simulate at its
          single-core speed when all cores of the processor are busy, without losses to
          shared caches or memory bandwidth.
    \item[--] \textit{Memory per core}: the memory is set by the smallest standard
          configuration of the processor, which is more than a \textsc{Geant4} job needs.
\end{itemize}

\section{Scaling of the generation time with the shower size}\label{sec:apptiming}
\begin{figure}[htbp]  
  \centering
  \includegraphics[width=\timingpanelwidth]{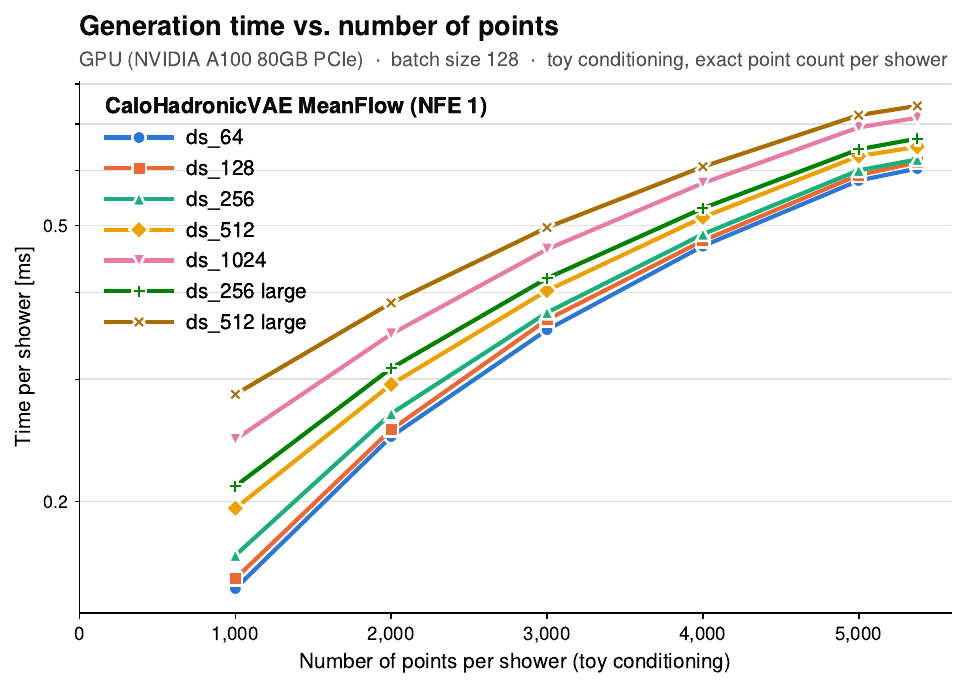}
  \caption{Generation time per shower of the full chain as a function of the number of
  points per shower (a single network evaluation, batch size 128, NVIDIA A100 GPU), for latent
  sizes $d_s$ from 64 to 1024 and for the large \LMF at $d_s = 256$ and 512 (``large'').}
  \label{fig:timing_ds}
  \end{figure}

Figure~\ref{fig:timing_ds} shows the generation time per shower of the full chain,
\layerFM, the single-step \LMF and the \VAEdec, on an NVIDIA A100 GPU at
batch size 128, as a function of the number of points per shower, for the five
latent sizes $d_s$ and for the large \LMF at $d_s = 256$ and 512. 

To isolate the
dependence on the shower size, every shower in a batch is given exactly the same number
of points; the conditioning is otherwise synthetic, so only the tensor shapes are
realistic. The time grows linearly with the number of points up to the maximum of 5376,
with the same slope in every configuration, about \SI{0.1}{\milli\second} per thousand
points. The latent size and the size of the \LMF only add a constant offset, which is largest,
about \SI{0.1}{\milli\second} compared with $d_s = 64$, for $d_s = 1024$ and for the
large \LMF at $d_s = 512$. The generation time is thus mainly determined by the number
of points in the shower, not by the length of the latent sequence or the size of the
\LMF.

\begin{figure}[htbp]  
  \centering
  \includegraphics[width=\timingpanelwidth]{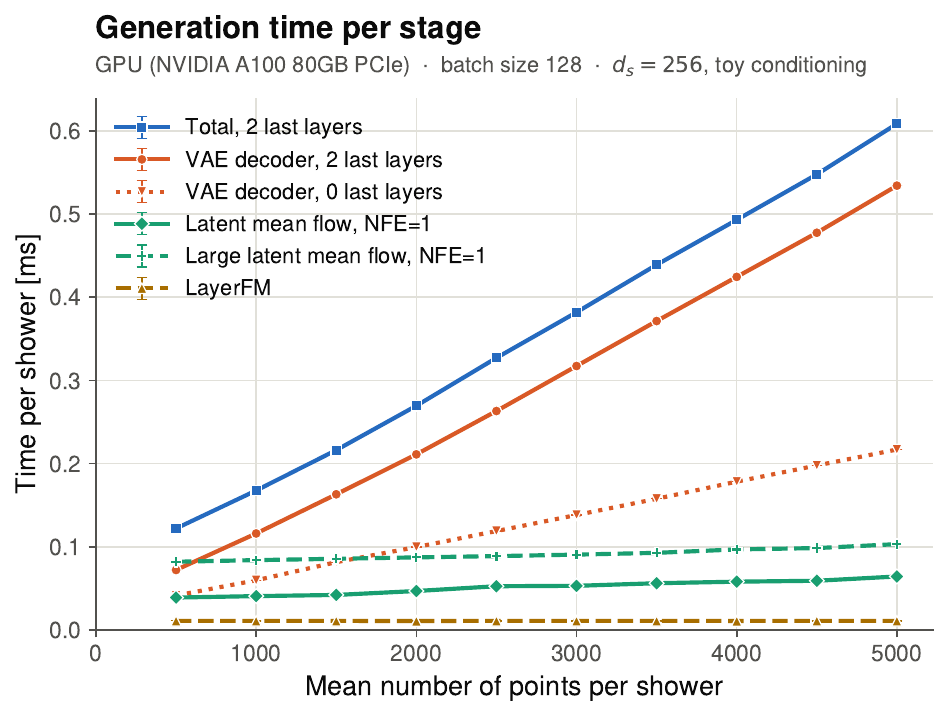}
  \caption{Generation time per shower of each stage of the chain for the \textsc{VAE} at
  $d_s = 256$ with the standard \LMF, and with the large \LMF as a dashed line, as a
  function of the number of points per shower (a single network evaluation, batch size 128,
  NVIDIA A100 GPU). The dotted line is the \VAEdec
  without its two refinement blocks, the self-attention blocks at full resolution that
  close it, labelled ``last layers'' in the legend.}
  \label{fig:breakdown}
  \end{figure}

\label{sec:appbreakdown}
Figure~\ref{fig:breakdown} splits the generation time per shower into its stages for the
\textsc{VAE} at $d_s = 256$ and the two \LMF models trained on its latent, the standard and
the large one, as a function of the number of points per shower, on an NVIDIA
A100 (\SI{80}{GB}, PCIe) GPU at batch size 128 (25 warm-up and 50 timed batches).

As above, every shower in a batch has exactly the same
number of points, here from 500 to 5000, and the conditioning is synthetic.
\layerFM distilled to a single step takes a constant \SI{0.011}{\milli\second}, the
single-step \LMF \SIrange{0.04}{0.06}{\milli\second} and the large \LMF
\SIrange{0.08}{0.10}{\milli\second}; the difference between the two, about
\SI{0.04}{\milli\second} at every shower size, is the offset of the large \LMF in the
left panel. The \VAEdec takes most of the time, and its share grows with the number of points. At
high multiplicity most of its time is spent in the two refinement blocks, and without
them the \VAEdec takes \SI{0.22}{\milli\second} instead of \SI{0.53}{\milli\second} at
5000 points. The remaining time still grows with the number of points, since every
output point is a query of the expansion step that cross-attends to the $d_s$ latent
tokens and passes through its own embeddings, feed-forward network and output
projection, with a cost proportional to the number of points times $d_s$. The
refinement blocks add self-attention between the points, which the sparse mask
restricts to a band of neighbours, so their cost also grows linearly with the number of
points, but with a larger slope. All times in this appendix are measured
with PyTorch 2.13.

\FloatBarrier  
\section{Further event displays}\label{sec:appevents}

\begin{figure}  
    \centering
    \includegraphics[width=\eventsthreedwidth]{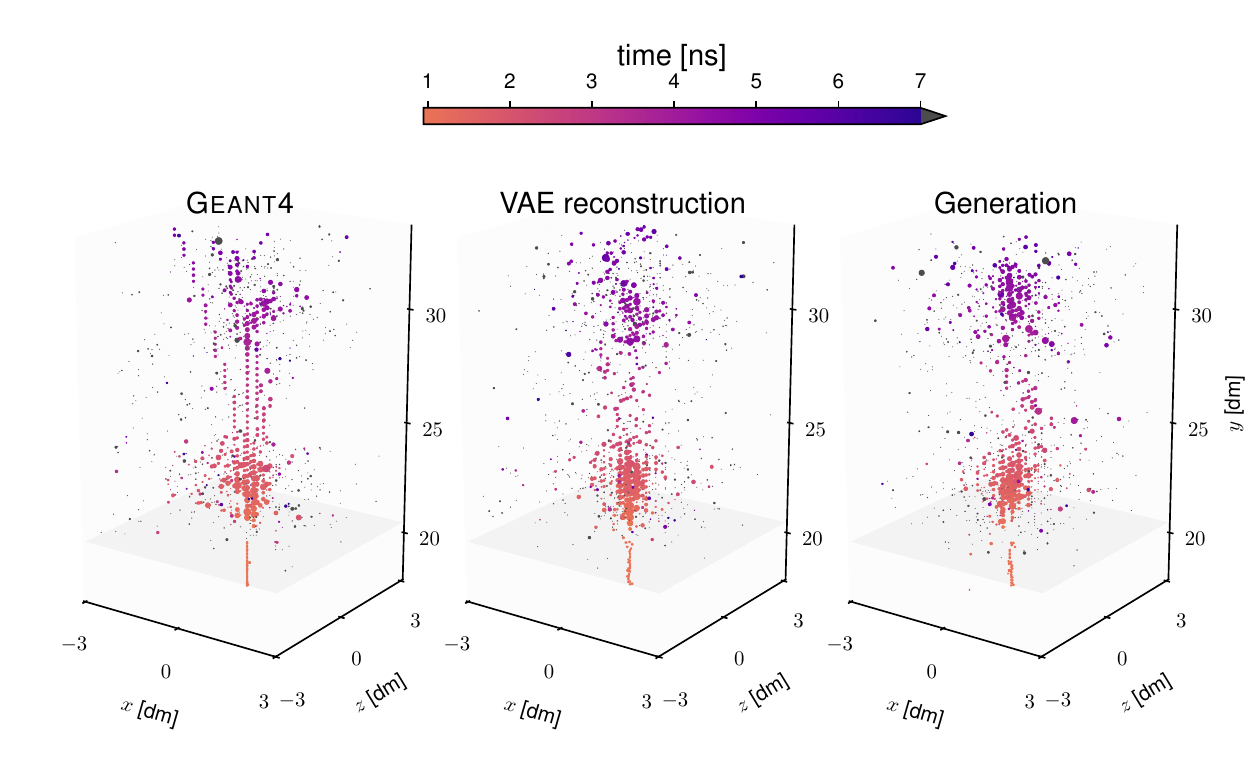}\\
    {\small (a) $\pi^{+}$, $E = \SI{36.0}{\giga\electronvolt}$}\\[8pt]
    \includegraphics[width=\eventsthreedwidth]{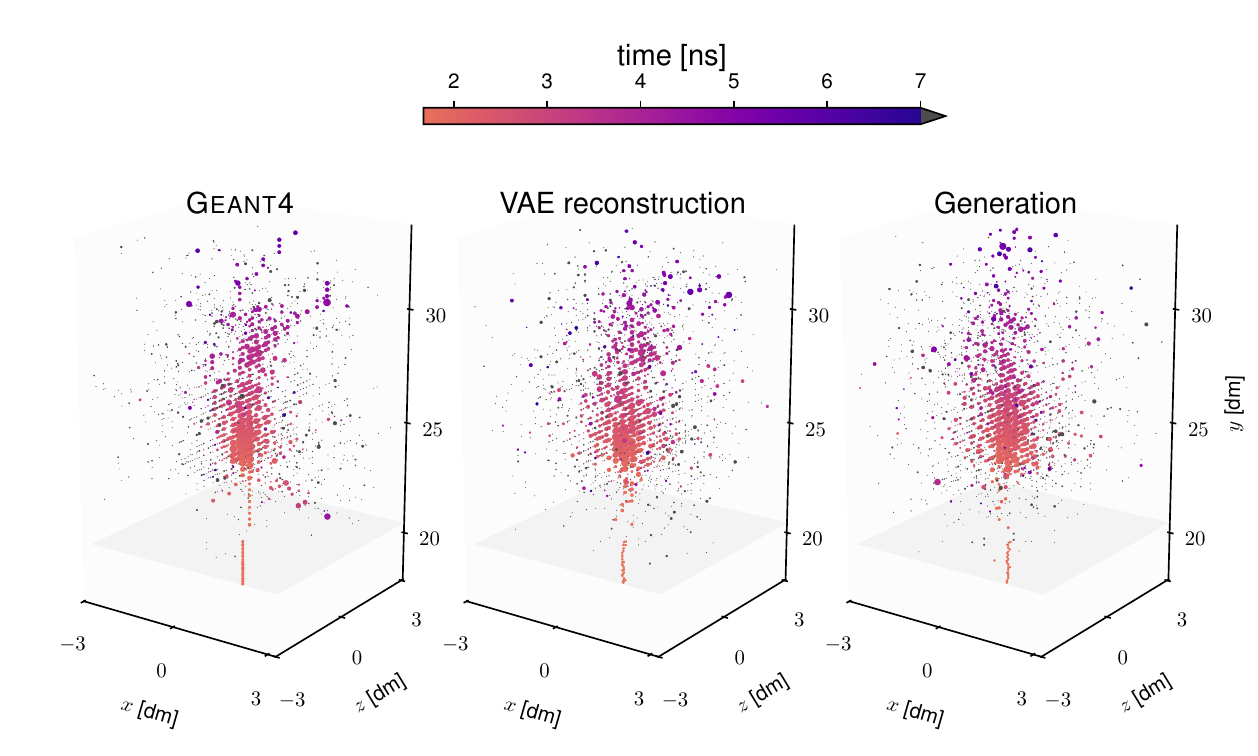}\\
    {\small (b) $\pi^{+}$, $E = \SI{64.6}{\giga\electronvolt}$}
    \caption{Two $\pi^{+}$ showers in three dimensions, \textsc{Geant4} (left), its
    reconstruction by the \textsc{VAE} (centre) and the shower generated from the same
    conditioning (right): (a) a shower with a second, later cluster higher in the HCal,
    connected to the core by tracks, and (b) a shower with secondary tracks leaving the
    core at large angles.}
    \label{fig:events3d_a}
\end{figure}

Figures~\ref{fig:events3d_a} and~\ref{fig:events3d_b} show five more showers in three
dimensions, in the format of figure~\ref{fig:event_display}: a \textsc{Geant4} shower,
its reconstruction by the \textsc{VAE}, and a shower generated by the large \LMF at
$d_s = 256$ with the incident energy and the per-layer numbers of points and earliest
times of the \textsc{Geant4} shower as conditioning. They are selected among \num{4000}
\textsc{Geant4} showers between \SI{30}{\giga\electronvolt} and
\SI{70}{\giga\electronvolt}. Every readout cell with a deposit is drawn at its centre, with $x$ and $z$ transverse
and the depth $y$ vertical, its size growing with the square root of its energy and its colour giving the
time of its earliest hit, from the start of the shower (red-orange) to
\SI{7}{\nano\second} (dark indigo); later cells are grey. The grey
plane is the boundary between the ECal and the HCal.

\ifdefined\prdlayout\begin{figure}[H]\else\begin{figure}\fi  
    \centering
    \includegraphics[width=\eventsthreedwidth]{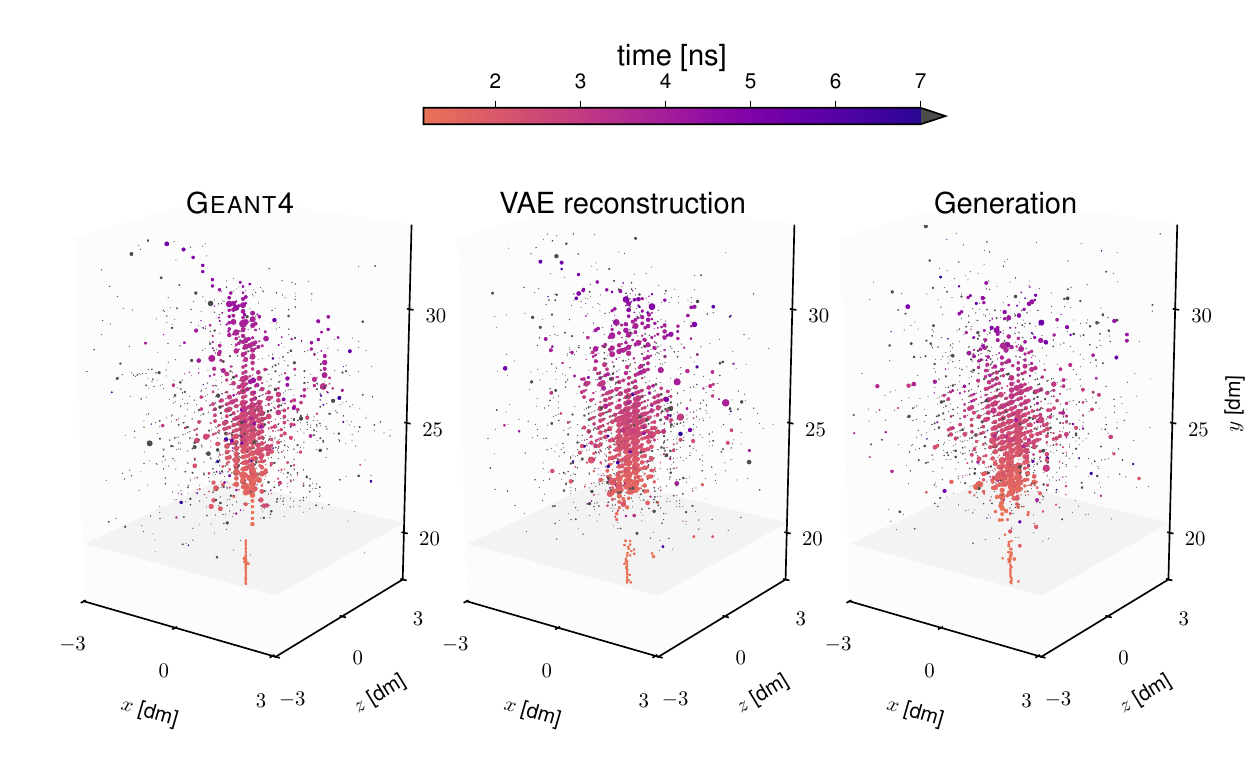}\\
    {\small (c) $\pi^{+}$, $E = \SI{48.5}{\giga\electronvolt}$}\\[8pt]
    \includegraphics[width=\eventsthreedwidth]{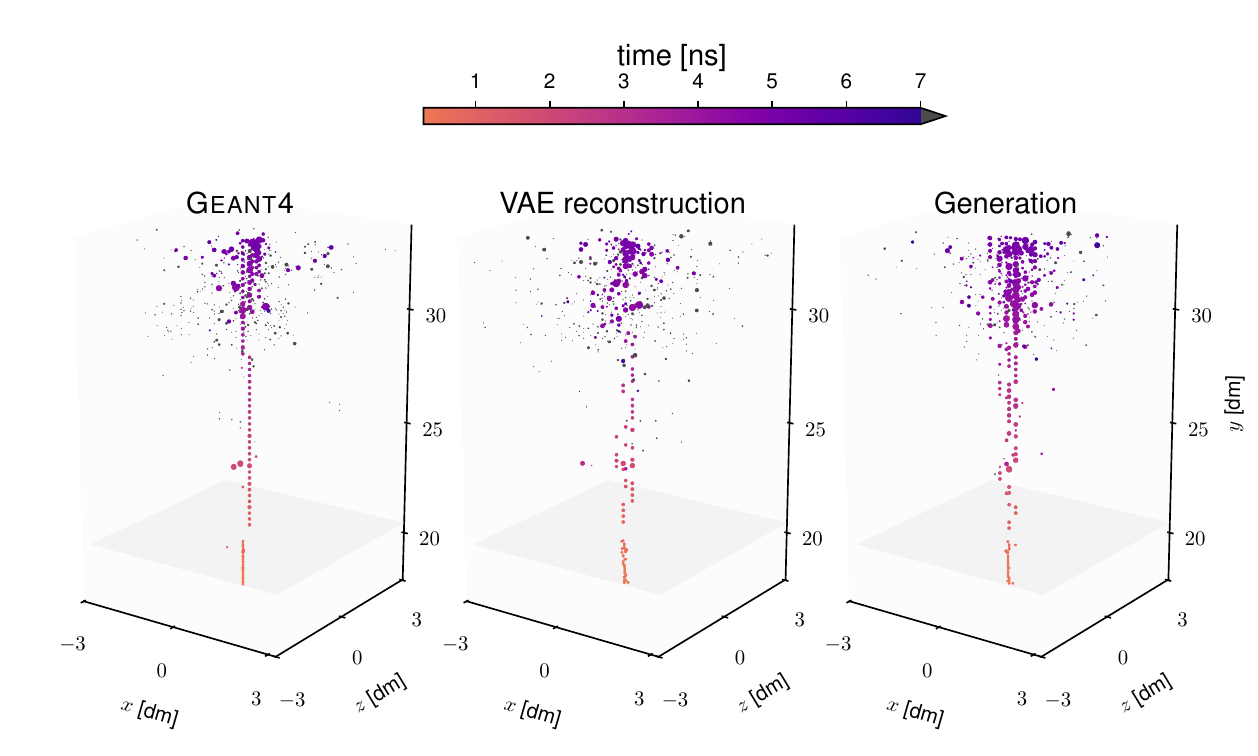}\\
    {\small (d) $\pi^{+}$, $E = \SI{36.8}{\giga\electronvolt}$}\\[8pt]
    \includegraphics[width=\eventsthreedwidth]{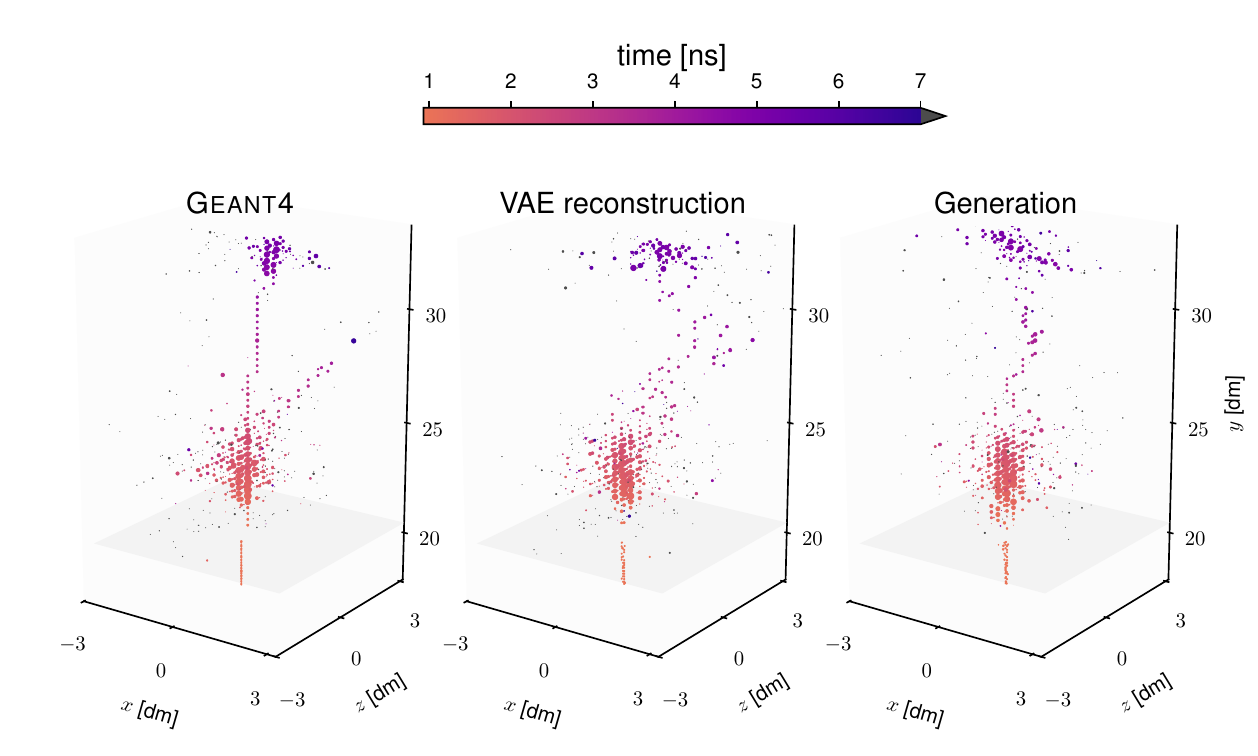}\\
    {\small (e) $\pi^{+}$, $E = \SI{35.5}{\giga\electronvolt}$}
    \caption{As figure~\ref{fig:events3d_a}: (c) a shower whose secondary tracks leave
    the core towards the back of the HCal, (d) a primary track that crosses almost the
    whole calorimeter before the shower starts, and (e) a shower in the HCal followed by
    a long secondary track and a second interaction at the back.}
    \label{fig:events3d_b}
\end{figure}

\end{document}